\documentclass{aa}
\usepackage{natbib}
\usepackage[FIGTOPCAP]{subfigure}

\usepackage{graphicx}
\usepackage{color}
\usepackage[table, dvipsnames]{xcolor}
\usepackage{hyperref}
\hypersetup{
    colorlinks=true,
    citecolor=blue,
    filecolor=black,
    linkcolor=blue,
    urlcolor=blue
}

\usepackage{ulem}

\definecolor{highlight}{rgb}{1, 0.8, 0.8}

\usepackage[flushleft]{threeparttable}
\usepackage{multirow}
\usepackage{tablefootnote}
\usepackage{pdflscape}
\usepackage{txfonts}
\usepackage[capitalise]{cleveref} 
\newcommand*{\rowstyle}[1]{
  \gdef\@rowstyle{#1}%
  \@rowstyle\ignorespaces%
  }

\usepackage{placeins}

\begin{document}

   \title{X-Shooting ULLYSES: Massive stars at low metallicity}

   \subtitle{VI. Atmosphere and mass-loss properties of O-type giants in the Small Magellanic Cloud}

   \author{F.\,Backs \inst{\ref{inst:kul}, \ref{inst:API}}
          \and S.\,A.\,Brands \inst{\ref{inst:API}}
          \and A.\,de\,Koter \inst{\ref{inst:API}, \ref{inst:kul}}
          \and L.\,Kaper \inst{\ref{inst:API}}
          \and J.\,S.\ Vink \inst{\ref{inst:armagh}}
          \and J.\,Puls \inst{\ref{inst:LMU}}
          \and J.\,Sundqvist \inst{\ref{inst:kul}}
          \and F.\,Tramper \inst{\ref{inst:CAB}, \ref{inst:kul}}
          \and H.\,Sana \inst{\ref{inst:kul}}          
         \and M.\,Bernini-Peron\inst{\ref{inst:ari}}
         \and J.\,M.\ Bestenlehner\inst{\ref{inst:sheffield}}
         \and P.\,A.\,Crowther\inst{\ref{inst:sheffield}}
         \and C.\,Hawcroft \inst{\ref{inst:stsci}}
         \and R.\,Ignace \inst{\ref{inst:ETSU}}
         \and R.\,Kuiper\inst{\ref{inst:ude}}
         \and J.\,Th.\,van\,Loon\inst{\ref{inst:keele}}
         \and L.\,Mahy \inst{\ref{inst:rob}}
         \and W.\,Marcolino\inst{\ref{inst:ufrj}}
         \and F.\,Najarro\inst{\ref{inst:CAB}} 
         \and L.\,M.\ Oskinova\inst{\ref{inst:up}}
         \and D.\,Pauli\inst{\ref{inst:up}}
         \and V.\,Ramachandran\inst{\ref{inst:ari}}
         \and A.\,A.\,C.\,Sander\inst{\ref{inst:ari}}
         \and O.\,Verhamme\inst{\ref{inst:kul}}
          }

   \institute{
        {Institute of Astronomy, KU Leuven, Celestijnenlaan 200D, 3001 Leuven, Belgium \label{inst:kul} \\
              \email{frank.backs@kuleuven.be}}
         \and
         {Anton Pannekoek Institute for Astronomy, Universiteit van Amsterdam, Science Park 904, 1098 XH Amsterdam, The Netherlands\label{inst:API}}
         \and 
         Armagh Observatory and Planetarium, College Hill, BT61 9DG Armagh, UK \label{inst:armagh}
         \and
         LMU München, Universitätssternwarte, Scheinerstr. 1, 81679 München, Germany \label{inst:LMU}
         \and
         Centro de Astrobiología (CAB), CSIC-INTA, Carretera de Ajalvir km 4, E-28850 Torrejón de Ardoz, Madrid, Spain \label{inst:CAB}
         \and
         Department of Physics \& Astronomy, Hounsfield Road, University of Sheffield, Sheffield, S3 7RH, United Kingdom\label{inst:sheffield}
         \and
         {Observatório do Valongo, Universidade Federal do Rio de Janeiro, Ladeira Pedro Antônio 43, Rio de Janeiro, CEP 20080-090, Brazil \label{inst:ufrj}}
         \and
          Department of Physics and Astronomy, East Tennessee State University, Johnson City, TN 37614, USA \label{inst:ETSU}
         \and
         {Institut f\"ur Physik und Astronomie, Universit\"at Potsdam, Karl-Liebknecht-Str. 24/25, 14476 Potsdam, Germany
         \label{inst:up}}
         \and
         {Lennard-Jones Laboratories, Keele University, ST5 5BG, UK\label{inst:keele}}
         \and
         {Zentrum f\"ur Astronomie der Universit\"at Heidelberg, Astronomisches Rechen-Institut, M\"onchhofstr. 12-14, 69120 Heidelberg, Germany\label{inst:ari}}
         \and
         {Fakult\"{a}t für Physik, Universit\"{a}t Duisburg-Essen, Lotharstraße 1, 47057 Duisburg, Germany \label{inst:ude}}
         \and
         {Royal Observatory of Belgium, Avenue Circulaire/Ringlaan 3, B-1180 Brussels, Belgium} \label{inst:rob}
         \and
         {Space Telescope Science Institute, 3700 San Martin Drive, Baltimore, MD 21218, USA \label{inst:stsci}}
         }
   \date{Received xxx xx, xxxx; accepted xxx xx, xxxx}

 
  \abstract
   {Mass loss through a stellar wind is an important physical process that steers the evolution of massive stars and controls the properties of their end-of-life products, such as the supernova type and the mass of compact remnants. To probe its role in stellar evolution over cosmic time, mass loss needs to be studied as function of metallicity. For mass loss to be accurately quantified, the wind structure needs to be established jointly with the characteristics of small-scale inhomogeneities in the outflow, which are known as wind clumping.   }
   {We aim to improve empirical estimates of mass loss and wind clumping for hot main-sequence massive stars, study the dependence of both properties on the metal content, and compare the theoretical predictions of mass loss as a function of metallicity to our findings.}
   {Using the model atmosphere code {\sc Fastwind} and the genetic algorithm fitting method {\sc Kiwi-GA}, we analyzed the optical and ultraviolet spectra of 13 O-type giant to supergiant stars in the Small Magellanic Cloud galaxy, which has a metallicity of approximately one-fifth of that of the Sun. We quantified the stellar global outflow properties, such as the mass-loss rate and terminal wind velocity, and the wind clumping properties. To probe the role of metallicity, our findings were compared to studies of Galactic and Large Magellanic Cloud samples that were analyzed with similar methods, including the description of clumping.}
   {We find significant variations in the wind clumping properties of the target stars, with clumping starting at flow velocities $0.01 - 0.23$ of the terminal wind velocity and reaching clumping factors $f_{\rm cl} = 2 - 30$. In the luminosity ($\log L / L_{\odot} = 5.0 - 6.0$) and metallicity ($Z/Z_{\odot} = 0.2 - 1$) range we considered, we find that the scaling of the mass loss $\dot{M}$ with metallicity $Z$ varies with luminosity. At $\log L/L_{\odot} = 5.75$, we find $\dot{M} \propto Z^m$ with $m = 1.02 \pm 0.30$, in agreement with pioneering work in the field within the uncertainties. For lower luminosities, however, we obtain a significantly steeper scaling of $m > 2$. }
    {The monotonically decreasing $m(L)$ behavior adds a complexity to the functional description of the mass-loss rate of hot massive stars. Although the trend is present in the predictions, it is much weaker than we found here. However, the luminosity range for which $m$ is significantly larger than previously assumed (at $\log L/L_{\odot} \lesssim 5.4$) is still poorly explored, and more studies are needed to thoroughly map the empirical behavior, in particular, at Galactic metallicity.} 


   \maketitle

\section{Introduction}

Feedback from massive stars plays an important role in the thermal and dynamical evolution of the interstellar medium and galaxy halos \citep[e.g.,][]{2011MNRAS.417..950H,2018MNRAS.477.1578H,2020MNRAS.494.3328A}. It is provided through jets, (ionizing) radiation, powerful stellar outflows, and, when their cores ultimately collapse, through supernovae \citep[e.g.,][]{2018A&A...616A.101K,2011MNRAS.414..321D,2016ApJ...824..125L,2020MNRAS.492..915G,2000MNRAS.317..697E,2019MNRAS.486.5263M}.

The mass and angular momentum that are lost through their stellar outflows (processes that persist throughout their lives) strongly influence the evolution of massive stars 
\citep[e.g.,][]{2008A&ARv..16..209P,2012ARA&A..50..107L,2017A&A...603A.118R,2022ARA&A..60..203V}.
At Galactic metallicities, mass loss causes the mass at the end of carbon burning to be 25 to 40\% of the initial mass for stars that started their lives with 20\,$M_{\odot}$ or more \citep{2012A&A...537A.146E}. The cumulative mass loss experienced by a massive star throughout its evolution is found to be a function of the initial stellar metal content (see below).

Metal-poor hot luminous stars have weaker winds, as has been shown empirically for the Large Magellanic Cloud \citep[LMC, e.g.,][]{2022A&A...663A..36B}, the Small Magellanic Clould \cite[SMC, e.g.,][]{2019A&A...625A.104R}, and for sub-SMC \citep[e.g.,][]{2014ApJ...788...64G,2015MNRAS.449.1545B}. One consequence of this is that properties of core-collapse supernovae are expected and are found to differ for galaxies with low and high metal content. Long-duration gamma-ray bursts \citep[e.g.,][]{2008AJ....135.1136M,2003ApJ...591..288H,2009ApJ...691..182S}, superluminous supernovae \citep[e.g.,][]{2017MNRAS.470.3566C}, and pair-instability supernovae \citep[e.g.,][]{2010A&A...512A..70Y}, for instance, favor lower-metallicity environments.
It also implies that the first stars to form in the Universe \citep[e.g.,][]{2015MNRAS.448..568H,2016ApJ...824..119H} must have lost relatively little mass \citep[e.g.,][]{2002ApJ...577..389K}, with important consequences for their end-of-life products \citep[e.g.,][]{2003A&A...399..617M}.

An accurate understanding of the mass loss from massive stars is therefore very important, including its dependence on the metal content. On the theory side, the hot-phase driving mechanism has been identified to be radiation pressure on the metal lines of mostly CNO and iron-group elements that are clustered in the ultraviolet, where the stars emit the bulk of their radiation \citep[e.g.,][]{1970ApJ...159..879L,1975ApJ...195..157C,1985ApJ...288..679A,1986A&A...164...86P,1996A&A...305..171P}. Combined stellar atmosphere and hydrodynamical methods that take into account several $10^{5}$ of these lines have been developed to make quantitative predictions \citep{1997ApJ...477..792D,2001A&A...369..574V,2017A&A...603A..86S,2018A&A...612A..20K,2021A&A...648A..36B,2022A&A...661A..51G}.

The behavior of mass loss versus metallicity was probed by studies of the solar metallicity and of the LMC ($Z = 0.5\,Z_{\odot}$), and SMC ($Z = 0.2\,Z_{\odot}$) metallicities. When they were based on relatively high-luminosity sources ($\log L/L_{\odot} \gtrsim 5.4$), where wind signatures in H$\alpha$ and \ion{He}{ii}\,4686 are clearly visible, the studies inferred a luminosity-independent behavior $\dot{M} \propto Z^{0.5-0.8}$ \citep{2007A&A...473..603M,2022MNRAS.511.5104M}. To constrain the mass-loss properties for lower-luminosity stars, the spectral analysis relies strongly on the ultraviolet spectral range, where UV resonance lines such as \ion{C}{iv} 1548,1550\,\AA\ provide more sensitive probes of the wind strength. \citet{2022MNRAS.511.5104M} used a compilation of UV and optical studies to first address this parameter space and tentatively found a weaker or even vanishing dependence on $Z$: $\dot{M} \propto Z^{0.1}$ at $\log L/L_{\odot} = 4.5$. They stressed that their findings should be tested with larger samples that include more accurate determinations of the terminal wind velocity.

The combination of the {\it Hubble} Space Telescope (HST) UV Legacy Library of
Young Stars as Essential Standards (ULLYSES) \citep{2020RNAAS...4..205R} and Very Large Telescope (VLT) X-Shooting ULLYSES (XShootU) \citep{2023A&A...675A.154V} programs provides high-quality spectra of OB stars in the Magellanic Clouds with an extensive wavelength coverage from the UV to the optical. We analyze a sample of O4 to O9.5 dwarfs, giants, bright giants, and supergiants in the SMC from these programs. The spectra allow us to constrain the photospheric conditions and a range of wind properties, including inhomogeneities in the outflow and the terminal velocity. To gain insight into the metallicity dependence of our findings, we compare them with similar studies of Galactic \citep{2021A&A...655A..67H} and LMC stars \citep[Brands et al. in prep.]{Hawcroft2024}.

A major challenge in the empirical determination of mass-loss rates is the inhomogeneity in the wind, which is also referred to as clumping. These clumps are overdense regions that are surrounded by lower density material. Clumps affect the strength of spectral lines both directly and indirectly. They affect the lines directly in the case of recombination lines, such as H$\alpha$ and \ion{He}{ii}\,4686, the strength of which depends on the mean of the square of the density of the medium; and they affect them indirectly in the case of lines that trace ionic species whose ionization balance shifted significantly because the different density. Scattering lines of abundant ionic species, such as \ion{C}{iv}\,1550\,\AA, may be unaffected by clumping depending on the stellar temperature, however. 

Because of these different effects, clumping has been suggested and used as solution for the discrepancies between inferred mass-loss rate from different diagnostic lines \citep[e.g.][]{2003ApJ...595.1182B,2006ApJ...637.1025F}.
Whether wind clumping itself is a function of metallicity and/or luminosity has been little explored so far. If this is the case, it may affect the derived mass-loss properties and should thus be accounted for when the empirical $\dot{M}(Z)$ or $\dot{M}(L,Z)$ relations are derived. This challenging problem is a goal of this study as well. So far, \citet{2022A&A...663A..40D} reported a weak relation between the metallicity and the clumping factor in 2D line-deshadowing instability simulations. \citet{2024MNRAS.52711422P} reported a similar trend for structures that appeared in single-epoch observations of wind lines in the UV spectra of B supergiants.

This paper is organized as follows. We describe the data and the sample in \cref{p2:sec:data}, along with the data preparation and normalization process. The model and fitting routine are presented in \cref{p2:sec:methods}. \cref{p2:sec:results} shows and describes the best-fit line profiles and the corresponding stellar and wind parameters. In \cref{p2:sec:discussion}, we briefly discuss the evolutionary stage of the sample stars and present the implications for $\dot{M}(L,Z)$, and the wind structure parameters. Finally, \cref{p2:sec:conclusion} lists the main conclusions.

\section{Data and sample} \label{p2:sec:data}

Our sample consists of 13 SMC O-type giants to supergiants that are available in ULLYSES DR5 \citep{2020RNAAS...4..205R} and XShootU eDR1 \citep[][XShootU I and II]{2023A&A...675A.154V,2024A&A...688A.104S}. The ULLYSES data consist of a mix of FUSE\footnote{Far Ultraviolet Spectroscopic Explorer}, HST/COS\footnote{HST Cosmic Origins Spectrograph}, and HST/STIS\footnote{HST Space Telescope Imaging Spectrograph} spectra, with both archival and new observations. The ULLYSES sample was selected to include presumably single stars \citep{2023A&A...675A.154V}. However, the presence of binaries in our sample cannot be excluded.  We used the high-level science products as reduced by \citet{2020RNAAS...4..205R}. The XShootU data consist of VLT/X-shooter spectra covering the UVB and VIS arms ($\sim$3100--10000\,\AA). 
The sample was selected based on data availability in the optical and UV wavelengths, where we required that the UV data cover at least the wavelength range 1150--1700\,\AA. 
The sample of stars with their spectral type and available UV observations is listed in \cref{p2:tab:targets}. The literature spectral types we used for target selection are included, as are updated spectral types using the XShootU data from Bestenlehner et al. in prep.   

All FUSE data were taken with the LWRS aperture, which has a spectral resolution of $\sim$17500 and covers 905--1180\,\AA. For AV~307 alone, the data were taken with HST/COS using the G130M/1096 grating. This grating covers the wavelength range from 940--1240\,\AA and has a resolution of $R\sim6000$ around $\lambda\sim1120$\,\AA. The HST/COS observations using the G130M/1291 and G160M/1611 gratings cover the wavelength range 1141--1783\,\AA\, and have a resolution from 11000 to 19000. 
The HST/STIS data that were obtained using the E140M and E230M gratings have a resolution of 45800 and 30000, respectively. The E140M grating covers the wavelength range from 1141 to 1708\,\AA\, and the E230M grating covers the wavelengths 1608--2366\,\AA.
When multiple data sets covered a modeled feature, the data were chosen based on the signal-to-noise ratio, resolving power, and possible systematic effects. The optical data consist of the UVB and VIS arms of VLT/X-shooter. The UVB arm covers 3100--5500\,\AA\ and has a resolution of 6700 for the chosen slit width of 0."8. The VIS arm covers 5500 to 8000\,\AA\ with a resolution of 11400 with a slit width of 0."7.

The signal-to-noise ratio ({S/N)} of the data varies depending on the observation and wavelength range. The optical data typically have a high signal-to-noise ratio of >100. The UV data have an {S/N} $\sim$ 20 on average. In some cases, a higher {S/N} was obtained. The lowest signal-to-noise ratio is found in the FUSE data, where it can be as low as $\sim$5.

\begin{table*}[]
    \centering
    \caption{Sample of SMC O giants, their spectral types, and the UV instruments.}
    \begin{tabular}{lllllccc} \hline \hline \\[-10pt]
    Target & Spectral type & Ref &  Updated Spectral type$^\dagger$ & UV observations & $V$  &  $A_V$& abs $K_s^\star$  \\ \hline \\[-10pt]
    \object{AV 80}       &   O4-6n(f)p       & a & O6 IIInn(f)p & FUSE, G130M, G160M & 13.38 &  0.47 & -5.41  \\        
    \object{AV 15}       &   O6.5 II(f)      & a & O6.5 III(f)  & FUSE, E140M        & 13.18 &  0.48 & -5.47  \\        
    \object{AV 95}       &   O7 III((f))     & a & O7.5 V((f))  & FUSE, E140M, E230M & 13.83 &  0.55 & -4.79  \\        
    \object{AV 207}      &   O7 III((f))     & b & O7 V((f))z   & FUSE, G130M, G160M & 14.35 &  0.28 & -4.07  \\        
    \object{AV 69}       &   OC7.5 III((f))  & a & OC7 III      & FUSE, E140M        & 13.33 &  0.44 & -5.28  \\        
    \object{AV 469}      &   O8.5 II((f))    & b & O9 Iab(f)    & FUSE, E140M        & 13.18 &  0.41 & -5.34  \\        
    \object{AV 479}      &   O9 Ib           & d & O9 Iab((f))  & FUSE, G130M, G160M & 12.42 &  0.74 & -6.10  \\        
    \object{AV 307}      &   O9 III          & e & B0.5 II      & G130M, G160M       & 14.02 &  0.28 & -4.39  \\        
    \object{AV 372}      &   O9.5 Iabw       & f & O9.2 Iab     & FUSE, E140M, E230M & 12.65 &  0.37 & -5.91  \\        
    \object{AV 327}      &   O9.5 II-Ibw     & a & O9.7 Ib      & FUSE, E140M        & 13.09 &  0.17 & -5.15  \\ \hline 
    \object{AV 83}       &   O7 Iaf$^+$      & a & O7 Iaf$^+$   & E140M              & 13.37 &  0.39 & -5.07  \\    
    \object{AV 70}       &   O9.5 Ibw        & f & O9.5 Iab     & FUSE, E140M, E230M & 12.31 &  0.52 & -6.16  \\       
    \object{2dFS 163}    &   O8 Ib(f)        & c & O7.5 Ib(f)   & G130M, G160M       & 15.11 &  0.57 & -3.82  \\  
    \hline
    \end{tabular}
    \tablefoot{We only found poor fits for the sources listed at the bottom. \\
     $^\dagger$ New spectral type determined using XShootU data (Bestenlehner et al. in prep.). 
     Spectral type reference: a: \citet{2000PASP..112.1243W}, 
          b: \citet{2016ApJ...817..113L}, 
          c: \citet{2004MNRAS.353..601E},
          d: \citet{1997A&A...317..871L},
          e: \citet{1987AJ.....93.1070G},
          f: \citet{2002ApJS..141..443W} \\
    $^\star$ The absolute $K_s$ magnitude was also corrected for line-of-sight extinction. }
    
    \label{p2:tab:targets}
\end{table*}

\subsection{Photometry}
Photometric information of the stars was used to determine the line-of-sight extinction. The extinction and the distance were used to determine the absolute $K_s$-band magnitude (listed in \cref{p2:tab:targets}), which was used as luminosity anchor in the analysis. All photometric data were taken from the \citet{2010AJ....140..416B} catalog. From this catalog, we used the $U, V, B,$ and $ I$ photometry of the Magellanic Cloud Photometric Survey \citep{2002AJ....123..855Z}, and the $J, H,$ and $K_s$ bands from the Two Micron All Sky Survey \citep[2MASS;][]{2006AJ....131.1163S} and the InfraRed Survey Facility \citep{2007PASJ...59..615K}. For the luminosity anchor (see \cref{p2:sec:data_prep}), we prioritized the 2MASS photometry. Finally, we used the extinction curves of \citet{1999PASP..111...63F} to parameterize the line-of-sight extinction.
We adopted $R_V=3.1$ and determined the $A_V$ by fitting the photometry to Castelli-Kurucz model spectral energy distributions (SED) \citep{2004A&A...419..725C}. We opted to fix $R_V$ as only marginal differences are found when it is made a free parameter \citep{2024A&A...689A..30S}. For this part of the analysis, the temperature of the model was selected based on the literature spectral type, as listed in Table~\ref{p2:tab:targets}. These spectral types were then used to obtain an approximate temperature following \citet{2005A&A...436.1049M}.

\subsection{Data preparation} \label{p2:sec:data_prep}
The spectrum-fitting routine employed in this work (see \cref{p2:sec:methods}) requires normalized line profiles. Therefore, the observed spectra need to be normalized. Incorrectly identifying the continuum may affect the strength and shape of spectral lines, so it is of great importance to determine it as accurately as possible. To do this, we attempted to improve on the normalization done by \citet[XShootU II]{2024A&A...688A.104S} in eDR1 by pursuing the following normalization routine. 

We divided the observed flux-calibrated spectrum by a normalized CMFGEN model \citep{1998ApJ...496..407H}. Ideally, that is, if the model fits perfectly, this results in a featureless pseudo-continuum with noise. This pseudo-continuum was fit with a polynomial. The normalized spectrum was then obtained by dividing the observed spectrum by the polynomial. The CMFGEN model parameters were selected from a limited grid of models, in which we used the model with the lowest $\chi^2$ when we compared the normalized data to the normalized model spectrum. In this process, we masked diverging features that are not properly covered in the model, such as wind-sensitive lines and interstellar features, so that they did not affect the polynomial fit or $\chi^2$ determination. The normalization procedure was done locally, that is, around spectral lines of interest, in the optical, and per grating in the UV spectral range (see below).

By using normalized spectra to fit our models and by using a polynomial to fit the pseudo-continuum, we minimized the possible effect of uncertainties in the extinction on the analysis. This was aided by choosing the $K_s$-band magnitude as the luminosity anchor, as the effect of extinction toward our targets is limited at longer wavelengths. 

For the UV data, the normalization was performed simultaneously with the determination of the radial velocity, using the same $\chi^2$ analysis as in the selection of the model. The optical X-shooter data were corrected for their radial velocity using a cross-correlation with a selection of hydrogen and helium lines. This should give corrections accurate to $\sim$10 km\,s$^{-1}$, depending on the signal-to-noise ratio and projected rotational velocities.
The UV spectrum of hot stars in the SMC is rich in Fe lines. These lines cover the whole spectral range, making it hard to identify the continuum between the lines. Therefore, the normalization was applied to the full spectrum separately for each grating. 

The differences between the normalization performed in XShootU eDR1 \citep[XShootU II]{2024A&A...688A.104S} and this work are typically small (< 1\%). We ascribe them to some broad local features that may have affected the global normalization of \citeauthor{2024A&A...688A.104S}.

Strong interstellar Ly$\alpha$ absorption is present in all spectra. The wings of this absorption overlap with the \ion{N}{v}~$\lambda1240$ feature and are often wide enough to affect the \ion{C}{iii}~$\lambda1176$ and \ion{C}{iv}~$\lambda1169$ features. Since interstellar Ly$\alpha$ is not part of our modeling, we corrected for it. We fit a Voigt-Hjerting function to the affected data \citep{2006MNRAS.369.2025T}. This resulted in good fits, which allowed us to set the continuum around these features at unity. Features in the spectral regimes of interest that were not modeled, such as missing lines or elements, diffuse interstellar bands, and interstellar absorption lines, were clipped from the observed spectrum. In this way, they did not affect the fitting efforts. 

\section{Methods} \label{p2:sec:methods}
We aim to determine detailed stellar and wind parameters of our targets by calculating model spectra and comparing those to the observations. The model spectra were calculated using the model atmosphere code {\sc Fastwind}. As the number of models required to fully explore the parameter space is substantial, we used a genetic algorithm to efficiently explore the parameter space and converge to the optimal parameters and their associated uncertainties. Sect.\,\ref{p2:sec:fastwind} describes {\sc Fastwind} in more detail. The genetic algorithm is covered in Sect.\,\ref{p2:sec:GA}. Finally, in Sect.\,\ref{p2:sec:fit_method}, we describe the fitting approach and free parameters.

\subsection{Fastwind} \label{p2:sec:fastwind}
{\sc Fastwind}\footnote{{\sc Fastwind} version 10.6.} is a stellar atmosphere and radiative transfer code that is optimized for hot stars and their winds \citep{1997A&A...323..488S,2005A&A...435..669P,2012A&A...537A..79R,2016A&A...590A..88C,2018A&A...619A..59S}. The model assumes nonlocal thermodynamic equilibrium (NLTE) and includes line-blanketing. {\sc Fastwind} aims to minimize computational cost. To this end, the elements are split into explicit and background species. The explicit elements are treated in detail in the comoving frame, while the background elements are only used to account for line-blocking and line-blanketing. For this reason, only explicit elements can be used to produce diagnostic line profiles. The explicit elements used here are H, He, C, N, O, Si, and P. 
{\sc Fastwind} uses a pseudo-hydrostatic photosphere that smoothly transitions into a trans-sonic wind. The wind is then described in a parameterized way, with a specified mass-loss rate and a radially increasing smooth-wind $\beta$-velocity law that asymptotically approaches the terminal velocity $\varv_\infty$. Additionally, we used the clumping prescription that we refer to as ``optically-thick'', as described in \citet{2018A&A...619A..59S}, to account for small-scale inhomogeneities in the outflow. In this prescription, the wind and clumps are not assumed to be optically thin, which allows for both optically thin and thick parts. The clumped wind is described by a set of six wind-structure parameters, $f_{\rm cl}, f_{\rm ic}, f_{\rm vel}, \varv_{\rm cl,start}, \varv_{\rm cl,max}$, and $\varv_{\rm windturb}$, which we introduce briefly (for more detailed descriptions, see \citet{2018A&A...619A..59S,2022A&A...663A..36B}).

The medium was assumed to consist of two components: regions in which the density $\rho_{\rm cl}$ is relatively high (clumps), and regions in which the density $\rho_{\rm ic}$ is relatively low (inter-clump medium). The clumps fill a fraction $f_{\rm vol}$ of the total volume, such that the mean density
\begin{eqnarray}
    \langle\rho\rangle = f_{\rm vol} \,\rho_{\rm cl} + (1 - f_{\rm vol}) \,\rho_{\rm ic}.
\end{eqnarray}
The density of the rarefied medium in between the clumps is set by the inter-clump density contrast,
\begin{equation}
    f_{\rm ic} \equiv \frac{\rho_{\rm ic}}{\langle \rho \rangle}.
\end{equation}
The clumping factor $f_{\rm cl}$ relates the mean density to the mean-square density as

\begin{equation}
    f_{\rm cl} \equiv \frac{\langle\rho^2\rangle}{\langle\rho\rangle^2} = \frac{f_{\rm vol} \,\rho_{\rm cl}^2 + (1 - f_{\rm vol}) \,\rho_{\rm ic}^2}{\left[ f_{\rm vol} \,\rho_{\rm cl} + (1 - f_{\rm vol}) \,\rho_{\rm ic}\right]^2},
\end{equation}
such that for a void inter-clump medium ($\rho_{\rm ic} = 0$), $f_{\rm cl} = 1/f_{\rm vol}$ or $f_{\rm vol} = 1/f_{\rm cl}$. For the more general case of a nonvoid inter-clump medium,
\begin{equation}
    f_{\rm vol} = \frac{(1 - f_{\rm ic})^2}{f_{\rm cl} - 2 f_{\rm ic} + f_{\rm ic}^2}.
\end{equation}

All the above parameters are a function of radial distance, which we left out for simplicity of notation.
The model inherently assumes the clumping to result from the line-deshadowing instability \citep{1988ApJ...335..914O}. Therefore, the clumping only starts when the wind starts to accelerate significantly. Here, the onset of clumping is a free parameter (see \cref{p2:sec:fit_method}), with the lowest allowed value $\varv_{\rm cl, start} = 0.01\varv_\infty$. From this starting velocity, the clumping increases linearly with velocity until it reaches its maximum value of $f_{\rm cl}$ at $\varv_{\rm cl,max}$. 

Each clump is assumed to have an internal velocity dispersion $\delta \varv$. To put this in perspective, as clumps have a physical size, it may be related to the velocity span of the clumps as a result of the underlying radially increasing smooth outflow velocity, $\delta \varv_{\rm sm}$. If the clumps have small internal $\delta \varv$ and are located relatively far from one another, the gas in optically thick clumps can only absorb a modest amount of light as Doppler shifts are small and do not spread out the line opacity over a wide velocity range. However, if $\delta \varv$ is larger than $\delta \varv_{\rm sm}$ , light will be much more effectively blocked. This effect of porosity in velocity space \citep[termed velocity-porosity or vorosity in][]{2008cihw.conf..121O} is quantified using a normalized velocity-filling factor $f_{\rm vel}$ that takes values between 0 and 1. Finally, $\varv_{\rm windturb}$ describes the turbulence in the wind.

The wind-structure parameters $f_{\rm ic}, f_{\rm vel}, \varv_{\rm cl, start},$ and $ \varv_{\rm windturb}$ are defined as a function of the clumping factor $f_{\rm cl}$. Therefore, if no inhomogeneities are present in the flow, that is, when the wind is truly smooth, the clumping factor $f_{\rm cl}$ equals unity. In this case, the other wind-structure parameters no longer have any effect or meaning. This implies that when the fitting procedure (see below) selects $f_{\rm cl} = 1$ for a model calculation, none of these five detailed structure parameters affects the line profiles.

\subsection{Genetic algorithm} \label{p2:sec:GA}
Given the large number of parameters, it was not feasible to fully explore the parameter space with a standard grid-based approach. However, it was essential that all parameters were explored simultaneously, as many parameters are connected and correlated. Determining parameter values sequentially would then likely result in an underestimation of the uncertainties and possibly in a suboptimal final fit. Therefore, an efficient fitting algorithm was required to fit all parameters simultaneously. A genetic algorithm (GA) was found to be  efficient, while remaining robust enough to not become stuck in local minima \citep[e.g.,][]{2005A&A...441..711M,2014A&A...572A..36T}. 
 We used the genetic algorithm Kiwi-GA\footnote{\url{https://github.com/sarahbrands/Kiwi-GA}} \citep{2022A&A...663A..36B} to find the optimal fit parameters, and we refer to \citet{2022A&A...663A..36B} for an in-depth discussion of the method. Kiwi-GA functions both as fitting routine and as {\sc Fastwind} wrapper. The algorithm starts with a set of models, the parameters of which result from a random uniform sampling of the parameter space. 

Subsequent generations of models were then generated by combining two models of the parent population. The parameters of these models were mixed, giving the new model the parameter value from either parent. Additionally, the parameters were allowed to mutate, which changed the value. The parent models were selected semi-randomly based on their $\chi^2$ value, with a lower $\chi^2$ giving a higher probability to be selected. The $\chi^2$ value was determined by comparing the normalized model spectra to the observed spectra and their uncertainty. Only a selected set of spectral features was compared (see Section\,\ref{p2:sec:fit_method}). 

We determined the confidence intervals of the fit parameters using the root mean square error of approximation \citep[RMSEA,][]{RMSEA_PAPER}. This statistics was chosen over a regular $\chi^2$ statistics because the latter requires the residuals of the data to follow a standard normal distribution. Because of the large number of high-quality (i.e., with a high signal-to-noise ratio) data points and inherently imperfect models, too many residuals diverged from a normal distribution. This resulted in high $\chi^2$ values and too large $\Delta \chi^2$ when it deviates from the best-fit model. The RMSEA corrects for this deviation from a normal distribution by rescaling the $\chi^2$ distribution as follows:
\begin{equation}
    {\rm RMSEA} = \sqrt{{\rm max} \left(\frac{\chi^2 - n_{\rm dof}}{n_{\rm dof}(N - 1)}, 0\right)},
\end{equation}
 with $N$ the number of data points, and $n_{\rm dof}$ the number of degrees of freedom. The best-fitting model is given by the lowest RMSEA value, the 1$\sigma$ confidence interval is given by the models that have ${\rm RMSEA} < 1.04 \times {\rm min(RMSEA)}$, and the 2$\sigma$ intervals are given by ${\rm RMSEA} < 1.09 \times {\rm min(RMSEA)}$. These values of 1.04 and 1.09 were calibrated such to obtain similar confidence intervals as the $\chi^2$ analysis performed in \citet{2022A&A...663A..36B}. The RMSEA statistics does not affect the sampling performed in the genetic algorithm.

Using Kiwi-GA, we were able to fit 15 free parameters and converged to a robust solution in 80 generations. Each generation consisted of 128 models, resulting in $\sim$10,000 models per star in total. For comparison, this is already less effort than calculating all parameter value permutations given only two values per parameter ($2^{15}$ = 32\,768). A grid consisting of five values per parameter, which is unrealistically coarse for most parameters, would constitute $\sim 3 \times 10^{10}$ models, and would only be computational-cost effective if more than $3 \times 10^{6}$ stars were scrutinized to the level pursued here.

\subsection{Fitting approach} \label{p2:sec:fit_method}
The parameters we aimed to constrain are listed in \cref{p2:tab:parameters}. We fit essential atmosphere parameters such as the effective temperature $T_{\rm eff}$, the surface gravity $\log g$, and the projected equatorial rotational velocity $\varv \sin i$, along with the surface abundance of helium, carbon, nitrogen, oxygen, silicon, and, when available, phosphorus. The wind (structure) parameters we fit are the mass-loss rate $\dot{M}$, the velocity law index $\beta$, the terminal velocity $\varv_\infty$, the (maximum) clumping factor $f_{\rm cl}$, the inter-clump density contrast $f_{\rm ic}$, the velocity porosity $f_{\rm vel}$, the onset velocity of clumping $\varv_{\rm cl,start}$, and the turbulent wind velocity $\varv_{\rm windturb}$. We assumed that maximum clumping was reached at $\varv_{\rm cl,max} = 2 \varv_{\rm cl,start}$ after increasing linearly from $\varv_{\rm cl,start}$. 

The analysis of each star was split into an optical-only fit to constrain the projected rotational velocity $\varv\sin i$ and the helium abundance $y_{\rm He}$ of the star. We did not distinguish between $\varv \sin i$ and additional broadening from macroturbulence. This parameterization of the broadening profile is not expected to critically affect other parameters such as the temperature and surface gravity \citep{2018A&A...613A..65H}. Then, another fit addressing both optical and UV diagnostics was preformed using the $\varv \sin i$ and $y_{\rm He}$ value from the optical fit. This two-step approach was used to ensure that we found the correct value of the rotation. From test calculations, we found that the $\varv\sin i$ value was often found to be degenerate with, notably, $\varv_{\rm windturb}$ in wind lines, which allows slightly better fits in the UV at the cost of worse fits to many optical lines. The helium abundance was fixed in the optical and UV fits as there was little to no sensitivity to this parameter in the UV and because it would lead to more uncertain and less accurate values. Additionally, the two-step approach made it possible to check for systematic changes or biases caused by including the UV diagnostics. 

In the optical-only fits, the only free wind parameter was the mass-loss rate. All other wind parameters were fixed to fiducial values of $\beta=1$, $f_{\rm cl}=10$, $f_{\rm ic}=0.1$, $f_{\rm vel} = 0.5$, $\varv_{\rm cl, start}=0.05\varv_\infty$, and $\varv_{\rm windturb}=0.1\varv_\infty$, and we used the $\varv_\infty$ from ULLYSES spectroscopy \citep[XShootU III]{2024A&A...688A.105H}. 

The spectral features scrutinized in this work are listed in \cref{p2:tab:diagnostics}.  
The top and bottom parts of the table list UV and optical diagnostics, respectively. All lines that are part of the same complex were fit together, and the formal solution of their radiative transfer was solved together. 

Wind-embedded shocks, caused by instabilities in the wind \citep[e.g.]{1988ApJ...335..914O}, can cause the emission of X-rays. The diagnostic features we selected to scrutinize have limited sensitivity to these X-rays. However, for completeness, we included the emission in our modeling with {\sc Fastwind} \citep{2016A&A...590A..88C}. Our implementation follows the prescription of Brands et al. in prep.

Some lines such as the \ion{O}{v}\,1371 lines and \ion{N}{iv}\,1718 lines were excluded from the analysis. These did not show clear wind signatures, and the photospheric absorption lines are blended with metal lines that are not included in the model. The \ion{N}{v}\,1240 line was excluded because it strongly depends on the assumptions regarding X-rays. This results in systematic changes compared to the optical-only fits (see also \cref{p2:sec:struggles}).

The mix of spectral features of different elements with varying ionization stages allowed us to accurately determine the temperature of these stars. The Balmer lines allowed us to infer the surface gravity.
The large selection of wind-sensitive lines, including resonance lines and recombination lines, allowed us to constrain the mass-loss rate and terminal velocity, and if possible, also the wind-clumping parameters. We found that the latter required a relatively strong outflow (see \cref{p2:sec:results}).

Additionally, we fit the CNO-cycle element abundances and silicon and phosphorus abundances, the latter of which was only fit in the optical and UV fits.  However, not many diagnostics were available for studying the Si and P abundances, with the added complexity that these diagnostics are also sensitive to wind.  

The radius of the stars, and therefore, their luminosity, was determined using the absolute $K_s$-band magnitude as an anchor. This means that the stellar parameters determine the shape of the SED, which was then scaled with the appropriate radius to match the observed absolute magnitude. The $K_s$-band magnitude was chosen as it is widely available and is not strongly affected by the interstellar extinction or the thermal radiation of dust. The uncertainty on the absolute magnitude is a result of the uncertainty on the photometry, the extinction toward the star, and the distance to the star. This uncertainty affects the radius, luminosity, spectroscopic mass, ionizing flux, and mass-loss rate. The latter is affected because we assumed $\dot{M} / R_\star^{3/2}$ to be constant \citep{1996A&A...305..171P}. However, intrinsic uncertainties on the mass-loss rate are typically significantly higher than the uncertainty on the radius. We assumed a distance of 62.44$\pm$2\,kpc to the sources in the SMC \citep{2020ApJ...904...13G,2009A&A...496..399S}. Here, the uncertainty is an estimate of the depth of the SMC, as it is unclear where the source is in the direction of the line of sight through the galaxy. 

\begin{table}[]
    \centering
    \caption{Model parameters that were fit in the optical and optical and UV fits. }
    \begin{tabular}{ll} \hline \hline \\[-10pt]
      Run            &  Free parameters \\ \hline \\[-10pt]
      Optical only   &  $T_{\rm eff}, g, \dot{M}, \varv\sin i, y_{\rm He}, \epsilon_{\rm C}, \epsilon_{\rm N}, \epsilon_{\rm O}, \epsilon_{\rm Si}$ \\
      Optical + UV   &  $T_{\rm eff}, g, \epsilon_{\rm C}, \epsilon_{\rm N}, \epsilon_{\rm O}, \epsilon_{\rm Si}, \epsilon_{\rm P}^{\dag}$ \\
                     &  $\dot{M}, \beta, \varv_\infty, f_{\rm cl}, f_{\rm ic}, f_{\rm vel}, \varv_{\rm cl,start}, \varv_{\rm windturb}$ \\ \hline
    \end{tabular}
    \begin{tablenotes}
    \item{\dag} Only if the \ion{P}{v} 1118 and 1128 lines are available.
    \end{tablenotes}
    \label{p2:tab:parameters}
\end{table}

\begin{table}[]
    \centering
    \caption{Diagnostic features used during the fitting.}
    \small 
    \begin{tabular}{lll} \hline \hline
    Ion             & Wavelength [\AA]          &  Part of complex          \\ \hline
    \ion{Si}{iii}   & 1113.2                    & \ion{Si}{iii} 1113        \\
    \ion{P}{v}      & 1118.0                    & \ion{P}{v} 1118           \\
    \ion{P}{v}      & 1128.0                    & \ion{P}{v} 1128           \\
    \ion{Si}{iv}    & 1128.3                    & \ion{P}{v} 1128           \\
    \ion{C}{iv}     & 1168.9, 1169.0            & \ion{C}{iv} 1169$^\dag$   \\
    \ion{C}{iii}    & 1174.9, 1175.3, 1175.6,   & \ion{C}{iii} 1176$^\dag$  \\
                    & 1175.7, 1176.4, 1177.0    &                           \\
    \ion{O}{iv}     & 1338.6, 1343.0, 1343.5    & \ion{O}{iv} 1340          \\
    \ion{Si}{iv}    & 1393.8, 1402.8            & \ion{Si}{iv} 1400         \\
    \ion{C}{iv}     & 1548.2, 1550.8            & \ion{C}{iv} 1550          \\
    \ion{He}{ii}    & 1640.4                    & \ion{He}{ii} 1640         \\
    \ion{N}{iii}    & 1747.9, 1751.2, 1751.7    & \ion{N}{iii} 1750         \\ \hline\\[-8pt]
    \ion{N}{iv}     & 3478.7, 3483.0, 3485.0    & \ion{N}{iv} 3480          \\
    \ion{O}{iii}    & 3961.6                    & H$\epsilon$               \\
    \ion{He}{i}     & 3964.7                    & H$\epsilon$               \\
    \ion{H}{i}      & 3970.1                    & H$\epsilon$               \\
    \ion{He}{ii}    & 4025.4                    & \ion{He}{i} 4026          \\
    \ion{He}{i}     & 4026.2                    & \ion{He}{i} 4026          \\
    \ion{C}{iii}    & 4068.9, 4070.3            & \ion{C}{iii} 4070         \\
    \ion{O}{ii}     & 4069.6, 4069.9, 4072.2    & \ion{C}{iii} 4070$^\ddag$ \\
                    & 4075.9                    &                           \\
    \ion{Si}{iv}    & 4088.9, 4116.1            & H$\delta$                 \\
    \ion{N}{iii}    & 4097.4, 4103.4            & H$\delta$                 \\
    \ion{He}{ii}    & 4099.9                    & H$\delta$                 \\
    \ion{H}{i}      & 4101.7                    & H$\delta$                 \\
    \ion{He}{i}     & 4143.8                    & \ion{He}{i} 4143           \\
    \ion{N}{iii}    & 4195.8, 4200.1            & \ion{He}{ii} 4200         \\
    \ion{He}{ii}    & 4199.6                    & \ion{He}{ii} 4200         \\
    \ion{He}{ii}    & 4338.67                   & H$\gamma$                 \\
    \ion{H}{i}      & 4340.5                    & H$\gamma$                 \\
    \ion{N}{iii}    & 4379.0, 4379.2            & \ion{He}{i} 4387          \\
    \ion{He}{i}     & 4387.9                    & \ion{He}{i} 4387          \\
    \ion{He}{i}     & 4471.5                    & \ion{He}{i} 4471          \\
    \ion{N}{iii}    & 4510.9, 4511.0, 4514.9    & \ion{N}{iii} qua          \\
                    & 4518.1                    &                           \\
    \ion{N}{iii}    & 4534.6                    & \ion{He}{ii} 4541         \\
    \ion{He}{ii}    & 4541.4                    & \ion{He}{ii} 4541         \\
    \ion{N}{iii}    & 4634.1, 4640.6, 4641.9    & \ion{C}{iii}\ion{N}{iii} 46 \\
    \ion{C}{iii}    & 4647.4, 4650.2, 4651.5    & \ion{C}{iii}\ion{N}{iii} 46 \\
    \ion{He}{ii}    & 4685.6                    & \ion{He}{ii} 4686$^\circ$ \\
    \ion{He}{i}     & 4713.1                    & \ion{He}{i} 4713$^*$      \\
    \ion{N}{iii}    & 4858.7, 4859.0, 4861.3    & H$\beta$                  \\
                    & 4867.1, 4867.2, 4873.6    &                           \\
    \ion{He}{ii}    & 4859.1                    & H$\beta$                  \\
    \ion{H}{i}      & 4861.4                    & H$\beta$                  \\
    \ion{He}{i}     & 4921.9                    & \ion{He}{i} 4922          \\
    \ion{He}{i}     & 5015.7                    & \ion{He}{i} 5015          \\
    \ion{He}{ii}    & 5411.3                    & \ion{He}{ii} 5411         \\
    \ion{O}{iii}    & 5592.4                    & \ion{O}{iii} 5592         \\
    \ion{C}{iii}    & 5695.9                    & \ion{C}{iii} 5696         \\
    \ion{C}{iv}     & 5801.3, 5812.0            & \ion{C}{iv} 5801          \\
    \ion{He}{i}     & 5875.6                    & \ion{He}{i} 5875          \\
    \ion{He}{ii}    & 6527.1                    & \ion{He}{ii} 6527         \\
    \ion{He}{ii}    & 6559.8                    & H$\alpha$                 \\
    \ion{H}{i}      & 6562.8                    & H$\alpha^{\dag\dag}$      \\
    \ion{He}{i}     & 6678.2                    & \ion{He}{ii} 6683         \\
    \ion{He}{ii}    & 6682.8                    & \ion{He}{ii} 6683         \\
    \ion{He}{i}     & 7065.2                    & \ion{He}{i} 7065          \\ \hline
    \end{tabular}
    \begin{tablenotes}
        \item \dag) \ion{C}{iv} 1169 and \ion{C}{iii} 1176 can be blended.
        \item \ddag) \ion{O}{ii} lines are only included in cooler stars. 
        \item $\circ$) \ion{He}{ii} 4685.6 is also included in the \ion{C}{iii}-\ion{N}{iii} 
               4634-4651 complex if the He feature shows strong emission.
        \item *) \ion{He}{i} 4713 is merged with \ion{He}{ii}\,4686 if the latter line shows strong emission. 
        \item \dag\dag) \ion{H}{i} at 6562.8 is also included in \ion{He}{ii} 6527. 
    \end{tablenotes}
    \label{p2:tab:diagnostics}
\end{table}

\subsection{Comparison with evolutionary models} \label{p2:sec:bonnsai}
We determined the initial mass, current evolutionary mass, and age using {\sc Bonnsai}\footnote{The BONNSAI web-service is available at \url{www.astro.uni-bonn.de/stars/bonnsai/index.php}.} \citep{2014A&A...570A..66S}. {\sc Bonnsai} is a Bayesian framework that can compare observed stellar properties with those of evolutionary models to obtain the posterior distribution of additional stellar properties. As input, we used the luminosity, temperature, and an upper limit of $\varv\sin i$ in combination with the SMC evolutionary tracks of \citet{2011A&A...530A.115B}. We used an upper limit for the rotation as we did not include macroturbulence in our modeling \citep[e.g.,][]{2017A&A...597A..22S}. Therefore, the rotational velocity could be lower than what we found. 

The inferred temperature and luminosity of AV\,15, AV\,70, and AV\,80 are not covered by the SMC model grid of \citet{2011A&A...530A.115B}, and we therefore used the LMC grids of \citet{2011A&A...530A.115B} and \citet{2015A&A...573A..71K} instead. We used the default {\sc Bonnsai} settings with the exception of the prior for the initial rotation velocity, for which we used the distribution of \citet{2013A&A...560A..29R}.  

\section{Results}
\label{p2:sec:results}

Tables\,\ref{p2:tab:fit_params} and \ref{p2:tab:derived_parameters} show the best-fit stellar atmosphere parameters and derived parameters along with their uncertainty. For completeness, we include the spectroscopic mass $M_{\rm spec}$, the number of \ion{H}{i} and \ion{He}{i} ionizing photons ($Q_0$ and $Q_1$), and the Eddington parameter for electron scattering $\Gamma_{\rm Edd,e}$ in \cref{p2:tab:derived_parameters}. We do not discuss these parameters further. 
The helium surface abundance $y_{\rm He}$ and the projected rotational velocity $\varv\sin i$ are the result of the optical-only fit. All other parameters are from the optical and UV fit. The remainder of the best-fit parameters of the optical-only fits, which are not the focus of this paper, can be found in Appendix\,\ref{p2:sec:optical_appendix}. Table\,\ref{p2:tab:wind_props} shows the best-fit mass-loss and wind-structure parameters and their uncertainties. Figs.\,\ref{p2:fig:wind_lines_overview} and \ref{p2:fig:photospheric_lines_overview} show an overview of the diagnostic lines and line complexes and their fits. More detailed figures of the line profiles and fit results can be found in Appendix\,\ref{p2:app:fit_summaries} and online\footnote{Available on \url{https://doi.org/10.5281/zenodo.13929105}}. 

Below, we describe the general fits for the UV and wind lines in Sect.\,\ref{p2:sec:results_wind}, and we describe the fit for the optical lines in Sect.\,\ref{p2:sec:results_optical}. In Sect.\,\ref{p2:sec:star_comments} we comment on some poorer fits. 

\subsection{UV and wind lines} \label{p2:sec:results_wind}
Overall, the UV lines are well reproduced by the models, with the exception of the phosphorus lines. The optical wind-sensitive lines are generally reproduced reasonably well. The fits are described per line group below. 

\paragraph{Phosphorus}
The P\,{\sc v} lines at 1118 and 1128\,\AA\ are in absorption in all of the sources. The two components of the doublet are very sensitive to the mass-loss properties and temperature of the star. For AV\,15, AV\,95, AV\,207, and AV\,307, the best-fit line profiles match the observations very well. For the other stars, the best-fitting results show a stronger in-filling of the line core or even P-Cygni profiles, while photospheric absorption profiles were observed.

\paragraph{Carbon}
The \ion{C}{iv}\,1550 doublet resonance lines are well reproduced by the models, with clear P-Cyngi wind profiles for most stars. The two components of the \ion{C}{iv}\,1169 and six components of the \ion{C}{iii}\,1176 lines are generally photospheric, with the \ion{C}{iv} lines typically being relatively weak. The \ion{C}{iii} line complex only shows clear wind signatures for AV\,83 and AV\,372. The photospheric profiles are typically well reproduced, with some exceptions where the model shows some emission from the wind. In these cases, notably AV\,479, AV\,70, and to a limited extent, AV\,469, the produced absorption feature is not strong enough.

\paragraph{Oxygen}
Only \ion{O}{iv}\,1340 was modeled in the UV. The \ion{O}{v}\,1371 line is either not present or too contaminated by blended metal lines to be included in our analysis. 
The \ion{O}{iv}\,1340 line, consisting of three transitions, is dominated by the photospheric component for all stars in the sample, with only some blueshifted absorption for some of the stars. Typically, the lines are well reproduced, but in some cases, the best-fit model results in a too weak absorption profile, particularly for the blue component.

\paragraph{Silicon}
The \ion{Si}{iii}\,1113 line is photospheric for all stars, and in most cases, it is well reproduced by the model. The \ion{Si}{iv}\,1400 doublets typically show both photospheric and wind features. The line profile resulting from the wind is well reproduced. However, the photospheric component is often underestimated by the model. This could be due to contamination by the superimposed interstellar absorption feature, which is difficult to distinguish from the stellar feature and was therefore not clipped from the data.

\paragraph{Helium}
The \ion{He}{ii}\,1640 line is blended with iron lines on the blue side of the profile, and therefore, only the red wing can be properly fit with {\sc Fastwind}. Only AV\,80 and AV\,83 show P-Cygni profiles for this line, and these are reasonably well reproduced. The stars showing a photospheric profile are also well reproduced.

\paragraph{Nitrogen}
When available, the \ion{N}{iii}\,1750 lines are photospheric in nature. They are typically well reproduced. For some of the cooler stars, we find slightly weaker lines in the models than in the observations.

\paragraph{Optical emission lines}
The main optical wind diagnostics are \ion{He}{ii}\,4686 and H$\alpha$. Whereas  the main UV wind lines are resonances lines, these are recombination lines. The combination of the two types of lines is essential for constraining the clumping parameters of the wind. The \ion{He}{ii}\,4686 line has proven hard to reproduce in this work. The exact strength of the line is often not recovered, with a slightly too strong central absorption component. The most extreme case of this is AV\,479, where we find a significantly stronger absorption component than is observed. 2dFS\,163 shows only a very weak and broad absorption, while the best-fit model features a significant emission profile (see also \cref{p2:sec:star_comments}). The H$\alpha$ lines are typically reproduced quite well within the uncertainties. We note that AV\,372 and AV\,70 show complex line shapes for the hydrogen and helium lines that are not fully recovered by the model, but the strength of the features are recovered within the uncertainty. The shape of these complex profiles can be reproduced by a 1D {\sc Fastwind} model, but they remain very sensitive to many parameters, as can also be seen in the large uncertainty region. The complex of \ion{C}{iii} and \ion{N}{iii} lines from 4630 to 4660\,\AA  is well reproduced. These lines are very sensitive to various other stellar parameters, such as temperature and gravity, and therefore, the uncertainties on the model profiles are substantial, as indicated by the shaded region. 

\subsection{Optical absorption lines} \label{p2:sec:results_optical}
Below, we describe the overall fit of several groups of absorption lines in the optical spectra. Their widths are generally well reproduced, suggesting that the $\varv\sin i$ was properly determined using the optical-only fit. 

\paragraph{Hydrogen}
The hydrogen lines H$\beta$ through H$\epsilon$ are accurately reproduced by the models, including the wings, which are sensitive to the surface gravity of the stars. Exceptions apply for stars with poor fits (see \cref{p2:sec:star_comments}). For some stars, the wind emission was strong enough to (partially) fill in the absorption profile. This was accurately reproduced by the models.

\paragraph{Helium}
The helium lines, which are good indicators of the accuracy of the temperature determination, fit the observations well, with the exception of 2dFS\,163 and AV\,327. The observations of both stars show stronger \ion{He}{ii} lines than the best-fit models, suggesting that the temperature is too low. However, the remaining diagnostics prevent a higher temperature from resulting in a better overall fit.

\paragraph{Nitrogen, carbon, oxygen, and silicon}
The optical nitrogen and carbon lines are typically well reproduced. The same holds for oxygen, where the \ion{O}{iii} 5592 line is consistent with the data, and the weak \ion{O}{iii} and \ion{O}{ii} lines in blends with other lines are reproduced within the uncertainties. These uncertainties are relatively large, however, which we tentatively ascribe to the strong temperature sensitivity of these lines. These considerations, combined with the fact that only a few lines could be used in the analysis, result in rather large uncertainties on the oxygen abundance.

The silicon lines, located in the wings of H$\delta$, are reproduced with varying success. The lines are very sensitive to temperature, gravity, and mass loss. As a result the lines may appear in as emission or absorption profiles depending on these parameter values. The observed strength of the line therefore varies significantly. This resulted in rather poor fits for AV\,83, AV\,207, and AV\,479.

\subsection{Anomalous fits} \label{p2:sec:star_comments}
{\sc Fastwind} successfully reproduced the observed spectra for most stars, but for a few stars, some spectral features could not be fully matched with the model. For three stars (AV\,70, AV\,83, and 2dFS\,163), the mismatch between model and observation was too significant for us to consider the inferred stellar parameters to be representative of the source. These stars were therefore left out in the further analysis, are colored gray in the tables, and have a red outline in the figures.  

\paragraph{2dFS 163} The observed \ion{N}{iv}\,3480 lines of 2dFS\,163 are significantly stronger than those calculated by {\sc Fastwind}. Additionally, the strength of the \ion{He}{ii} lines is systematically underpredicted by the model, except for \ion{He}{ii}\,4686. This wind-sensitive line is almost absent in the observed spectrum, but the model predicts a strong emission. The emission is the result of the high clumping factor and mass loss required to reproduce the other wind-sensitive lines. This likely also prevented the temperature from increasing further to allow the strength of the \ion{He}{ii} lines to match the observations, as a higher temperature would result in even stronger emission in \ion{He}{ii}\,4686. Furthermore, the hydrogen and \ion{He}{i} lines appear to be slightly blueshifted, while this does not seem to be the case for the other lines. The \ion{He}{i} lines also appear to be slightly narrower than the other lines, suggesting a different rotation rate. The H$\alpha$ line (see Fig.~\ref{p2:fig:wind_lines_overview} and \cref{p2:app:fit_summaries}. 
is double peaked, as are the Paschen lines, which we do not display here. Taken together, we suspect that 2dFS\,163 is a composite source. \citet[XShootU VIII]{Ramachandran2024} identified this star as a post-interaction binary system.
Therefore, this source was left out in the further analysis, is marked with a red outline in the figures, and is listed in gray in the tables.

\paragraph{AV 83}
The spectrum of AV\,83 is likely composite. It shows a feature in the red wing of the \ion{He}{ii}\,5411 line that is also visible in the spectrum used by \citet{2003ApJ...588.1039H}, but shifted in wavelength. This results in a broad line. The optical \ion{He}{ii}, \ion{C}{iv}, and \ion{N}{iv} lines all appear to be slightly redshifted compared to the other lines. Some \ion{He}{i} lines and the optical \ion{Si}{iv} lines show emission that is not reproduced by the model. As a result, the best-fit parameters of this star are likely not representative of its properties. This source was left out in the further analysis, is marked with a red outline in the figures, and is listed in gray in the tables.

\paragraph{AV 70}
The \ion{He}{i} lines in the spectrum of AV\,70 show a broad component that is not visible in other lines, nor is it reproduced by the model. The broad component could suggest the presence of a rapidly rotating cooler companion. This is backed by the apparent flat-bottomed \ion{C}{iv}\,1550 absorption trough of the P-Cygni profile, for which the flux remains at 30\% of the continuum at the bottom. For this reason, the star was left out in the further analysis, is marked red in the figures, and is listed in gray in the tables.

\paragraph{AV 307}
Additional absorption lines can be seen in the \ion{C}{iii}\,-\ion{N}{iii}\,4630-4655 complex of AV\,307. These additional lines coincide with the transitions of \ion{O}{ii}, but they are not observed in any of the other stars in the sample, with the exception of AV\,327. In this star, the \ion{O}{ii} lines are considerably weaker, however. A closer inspection shows the spectrum of AV\,307 to be rich in \ion{O}{ii} lines. AV\,307 and AV\,327 are the coolest stars in the sample, both with $T_{\rm eff} = 29\,500$\,K as the best-fit value, and they feature the lowest $\varv\sin i$, with 55 and 80 km\,s$^{-1}$. The abundance of oxygen is found to be higher in AV\,327, mainly based on the strength of \ion{O}{iv}\,1340.

\paragraph{AV 479}
The \ion{He}{i} and \ion{He}{ii} lines of AV\,479 fit well within the uncertainties, with the exception of \ion{He}{ii}\,4686, for which the best-fit model displays a significantly stronger absorption than the observations. The H$\alpha$ line shows a slightly too strong absorption feature in the model. 
A consistent spectral fit for both the UV and optical wind signatures therefore does not appear to be feasible: The \ion{Si}{iv}\,1400 and \ion{C}{iv}\,1550 lines are fit well, at the expense of H$\alpha$ and \ion{He}{ii}\,4686.
Therefore, a higher mass-loss rate would be at the expense of the fit quality of the two ultraviolet lines. Similarly, the fit to the optical wind lines could  benefit from a higher clumping factor. However, this would affect the ionization structure of Si and C, resulting in a worse fit for the \ion{Si}{iv}\,1400 and \ion{C}{iii}\,1176 lines. Alternatively, an earlier onset of clumping, that is, a lower value for $\varv_{\rm cl,start}$, could also improve the line fits of H$\alpha$ and \ion{He}{ii}\,4686. Again, this is disfavored for the ultraviolet wind lines.

\begin{table*}
\centering
\caption{Stellar atmosphere best-fit parameters and 1$\sigma$ uncertainties.  }
\label{p2:tab:fit_params}
\small
\def\arraystretch{1.5}
\begin{tabular}{lccccccccc} \hline \hline
Source & $T_{\rm eff}$ [K] & $\log g$ & $\varv\sin i$ [km s$^{-1}$]$^\dagger$ & $y_{\rm He}$ & $\epsilon_{\rm C}$ & $\epsilon_{\rm N}$ & $\epsilon_{\rm O}$ & $\epsilon_{\rm Si}$ & $\epsilon_{\rm P}$\\ \hline
AV 80 & $41500^{+1250}_{-750}$ & $3.88^{+0.12}_{-0.10}$ & $305^{+25}_{-20}$ & $0.08^{+0.02}_{-0.01}$ & $7.7^{+0.1}_{-0.6}$ & $7.5^{+0.5}_{-0.2}$ & $7.5^{+0.7}_{-0.5}$ & $6.7^{+0.5}_{-0.7}$ & $3.6^{+0.8}_{-0.2}$\\
AV 15 & $39750^{+1250}_{-1500}$ & $3.70^{+0.10}_{-0.12}$ & $110^{+10}_{-10}$ & $0.10^{+0.01}_{-0.01}$ & $7.8^{+0.3}_{-0.3}$ & $7.8^{+0.4}_{-0.1}$ & $7.5^{+0.6}_{-0.6}$ & $7.0^{+0.3}_{-0.9}$ & $4.3^{+0.2}_{-0.8}$\\
AV 95 & $38250^{+1000}_{-1000}$ & $3.64^{+0.08}_{-0.12}$ & $75^{+10}_{-10}$ & $0.14^{+0.01}_{-0.02}$ & $7.5^{+0.3}_{-0.4}$ & $7.8^{+0.3}_{-0.4}$ & $8.2^{+0.5}_{-0.4}$ & $6.6^{+0.6}_{-0.5}$ & $4.5^{+0.5}_{-0.6}$\\
AV 207 & $38000^{+1250}_{-750}$ & $3.82^{+0.28}_{-0.12}$ & $110^{+10}_{-15}$ & $0.12^{+0.01}_{-0.03}$ & $7.7^{+0.4}_{-0.3}$ & $7.7^{+0.7}_{-0.6}$ & $8.6^{+0.3}_{-0.7}$ & $8.1^{+0.4}_{-0.7}$ & $5.0^{+1.4}_{-1.9}$\\
AV 69 & $36750^{+1000}_{-1250}$ & $3.50^{+0.16}_{-0.08}$ & $100^{+10}_{-15}$ & $0.09^{+0.03}_{-0.01}$ & $7.6^{+0.2}_{-0.3}$ & $6.4^{+0.8}_{-0.4}$ & $8.2^{+0.2}_{-0.5}$ & $6.5^{+0.7}_{-0.5}$ & $4.1^{+0.3}_{-0.8}$\\
AV 469 & $34500^{+1000}_{-1000}$ & $3.36^{+0.08}_{-0.14}$ & $85^{+10}_{-10}$ & $0.20^{+0.03}_{-0.04}$ & $7.5^{+0.3}_{-0.2}$ & $8.2^{+0.3}_{-0.2}$ & $8.1^{+0.8}_{-0.2}$ & $7.8^{+0.3}_{-0.2}$ & $4.0^{+0.5}_{-0.9}$\\
AV 479 & $33250^{+1500}_{-1000}$ & $3.42^{+0.06}_{-0.24}$ & $90^{+15}_{-5}$ & $0.13^{+0.04}_{-0.01}$ & $7.5^{+0.2}_{-0.2}$ & $7.3^{+0.2}_{-0.4}$ & $8.2^{+0.4}_{-0.3}$ & $7.8^{+0.4}_{-0.1}$ & $4.0^{+0.5}_{-0.8}$\\
AV 307 & $29500^{+250}_{-250}$ & $3.46^{+0.16}_{-0.04}$ & $55^{+15}_{-10}$ & $0.11^{+0.01}_{-0.03}$ & $7.8^{+0.2}_{-0.1}$ & $7.7^{+0.3}_{-0.3}$ & $7.8^{+0.5}_{-0.1}$ & $7.1^{+0.1}_{-0.2}$ & $5.5^{+0.8}_{-1.4}$\\
AV 372 & $30750^{+1500}_{-1250}$ & $3.10^{+0.16}_{-0.12}$ & $155^{+15}_{-5}$ & $0.17^{+0.03}_{-0.03}$ & $7.7^{+0.2}_{-0.2}$ & $7.7^{+0.8}_{-0.9}$ & $8.2^{+0.8}_{-1.2}$ & $7.7^{+0.2}_{-0.9}$ & $3.2^{+0.9}_{-0.2}$\\
AV 327 & $29500^{+250}_{-1000}$ & $3.28^{+0.18}_{-0.08}$ & $80^{+10}_{-15}$ & $0.15^{+0.01}_{-0.04}$ & $8.0^{+0.2}_{-0.2}$ & $7.7^{+0.7}_{-0.3}$ & $8.8^{+0.2}_{-0.6}$ & $7.3^{+0.2}_{-0.2}$ & $5.0^{+0.8}_{-1.5}$\\
\hline
\rowstyle{\color{gray}}
AV 83 & \color{gray}$36000^{+1250}_{-1250}$ & \color{gray}$3.24^{+0.34}_{-0.10}$ & \color{gray}$80^{+45}_{-20}$ & \color{gray}$0.18^{+0.06}_{-0.05}$ & \color{gray}$7.8^{+0.5}_{-0.7}$ & \color{gray}$8.7^{+0.3}_{-0.3}$ & \color{gray}$7.8^{+0.3}_{-0.7}$ & \color{gray}$8.0^{+0.3}_{-0.5}$ & \color{gray}...\\
\rowstyle{\color{gray}}
2dFS 163 & \color{gray}$36500^{+250}_{-1250}$ & \color{gray}$4.14^{+0.02}_{-0.40}$ & \color{gray}$90^{+35}_{-35}$ & \color{gray}$0.08^{+0.05}_{-0.01}$ & \color{gray}$6.8^{+0.3}_{-0.2}$ & \color{gray}$8.0^{+0.3}_{-0.6}$ & \color{gray}$7.2^{+0.7}_{-0.6}$ & \color{gray}$6.1^{+0.5}_{-0.1}$& \color{gray}...\\
\rowstyle{\color{gray}}
AV 70 & \color{gray}$33750^{+500}_{-250}$ & \color{gray}$3.54^{+0.08}_{-0.06}$ & \color{gray}$120^{+15}_{-15}$ & \color{gray}$0.15^{+0.04}_{-0.01}$ & \color{gray}$7.2^{+0.1}_{-0.2}$ & \color{gray}$7.7^{+0.1}_{-0.8}$ & \color{gray}$7.8^{+0.1}_{-0.8}$ & \color{gray}$7.3^{+0.4}_{-0.1}$ & \color{gray}$3.5^{+1.1}_{-0.5}$\\
\hline 
\end{tabular}
\tablefoot{ The bottom rows with gray text indicate parameter values that are likely not representative of the stellar properties due to poor fits.
 \\ $\epsilon_{\rm x}$ refers to the number abundance relative to hydrogen as $\epsilon_{\rm x} = \log( n_{\rm x} / n_{\rm H}) + 12$. \\
$^\dagger$ The broadening profile of the rotation is also used to account for broadening macroturbulent velocities, and the listed $\varv\sin i$ may therefore be higher than the true value.
}
\end{table*}

\begin{table*}
\centering
\caption{Derived parameters based on the {\sc Fastwind}/GA fitting. }
\def\arraystretch{1.5}
\label{p2:tab:derived_parameters}
\small
\begin{tabular}{lcccccc|ccc} \hline \hline
Source & $\log L / {\rm L_\odot}$ & $R$\,[R$_\odot$] & $\log Q_0$ & $\log Q_1$ & $\Gamma_{\rm Edd, e}$ & $M_{\rm spec} $[M$_\odot$] & $M_{\rm evo} $[M$_\odot$] & $M_{\rm ini} $[M$_\odot$] & Age [Myr]\\ \hline
AV 80& $5.94^{+0.04}_{-0.03}$& $18.26^{+0.60}_{-0.63}$& $49.66^{+0.06}_{-0.05}$& $48.95^{+0.08}_{-0.05}$& $0.25^{+0.05}_{-0.04}$& $92.3^{+21.8}_{-16.6}$& $62.0^{+4.0}_{-2.8}$& $66.4^{+3.7}_{-3.4}$& $2.0^{+0.1}_{-0.2}$\\
AV 15& $5.91^{+0.04}_{-0.05}$& $19.20^{+0.70}_{-0.67}$& $49.62^{+0.05}_{-0.08}$& $48.83^{+0.07}_{-0.12}$& $0.32^{+0.07}_{-0.04}$& $67.4^{+12.2}_{-13.5}$& $57.8^{+3.3}_{-3.8}$& $61.6^{+3.8}_{-4.1}$& $2.3^{+0.2}_{-0.1}$\\
AV 95& $5.58^{+0.04}_{-0.04}$& $14.19^{+0.48}_{-0.48}$& $49.24^{+0.06}_{-0.07}$& $48.37^{+0.08}_{-0.11}$& $0.32^{+0.05}_{-0.04}$& $32.1^{+5.2}_{-6.4}$& $38.0^{+1.8}_{-1.6}$& $39.0^{+1.8}_{-1.8}$& $3.2^{+0.2}_{-0.2}$\\
AV 207& $5.29^{+0.04}_{-0.03}$& $10.23^{+0.34}_{-0.36}$& $48.88^{+0.08}_{-0.06}$& $47.98^{+0.13}_{-0.12}$& $0.21^{+0.05}_{-0.08}$& $25.2^{+19.5}_{-5.4}$& $29.2^{+1.2}_{-1.0}$& $29.6^{+1.2}_{-1.0}$& $3.7^{+0.2}_{-0.4}$\\
AV 69& $5.73^{+0.04}_{-0.05}$& $18.21^{+0.65}_{-0.62}$& $49.38^{+0.04}_{-0.07}$& $48.48^{+0.07}_{-0.13}$& $0.38^{+0.06}_{-0.08}$& $38.3^{+13.6}_{-5.5}$& $43.8^{+2.1}_{-1.0}$& $45.8^{+1.7}_{-1.7}$& $3.1^{+0.1}_{-0.1}$\\
AV 469& $5.67^{+0.04}_{-0.04}$& $19.39^{+0.66}_{-0.66}$& $49.23^{+0.05}_{-0.04}$& $48.04^{+0.13}_{-0.08}$& $0.40^{+0.08}_{-0.04}$& $31.4^{+4.3}_{-7.3}$& $39.6^{+2.5}_{-1.4}$& $41.2^{+2.3}_{-1.9}$& $3.5^{+0.1}_{-0.1}$\\
AV 479& $5.94^{+0.05}_{-0.04}$& $28.27^{+0.97}_{-1.07}$& $49.38^{+0.13}_{-0.03}$& $47.92^{+0.42}_{-0.03}$& $0.30^{+0.16}_{-0.00}$& $76.7^{+6.3}_{-29.7}$& $55.6^{+1.2}_{-2.4}$& $58.6^{+1.3}_{-2.6}$& $2.8^{+0.1}_{-0.1}$\\
AV 307& $5.11^{+0.03}_{-0.03}$& $13.93^{+0.45}_{-0.45}$& $47.95^{+0.03}_{-0.12}$& $45.35^{+0.05}_{-0.12}$& $0.17^{+0.01}_{-0.05}$& $20.4^{+7.9}_{-1.6}$& $21.6^{+0.6}_{-0.6}$& $21.8^{+0.6}_{-0.6}$& $6.5^{+0.2}_{-0.2}$\\
AV 372& $5.76^{+0.06}_{-0.05}$& $26.95^{+1.00}_{-1.05}$& $49.18^{+0.13}_{-0.09}$& $47.36^{+0.57}_{-0.44}$& $0.46^{+0.08}_{-0.07}$& $33.4^{+10.9}_{-6.7}$& $42.6^{+3.6}_{-2.4}$& $44.8^{+3.5}_{-3.0}$& $3.5^{+0.2}_{-0.2}$\\
AV 327& $5.41^{+0.03}_{-0.04}$& $19.60^{+0.69}_{-0.63}$& $48.40^{+0.07}_{-0.27}$& $45.77^{+0.07}_{-0.43}$& $0.26^{+0.04}_{-0.08}$& $26.7^{+12.0}_{-3.8}$& $28.4^{+0.9}_{-1.4}$& $29.0^{+1.0}_{-1.3}$& $5.1^{+0.2}_{-0.2}$\\
\hline
\color{gray} AV 83& \color{gray}$5.59^{+0.04}_{-0.05}$& \color{gray}$16.17^{+0.58}_{-0.57}$& \color{gray}$49.32^{+0.06}_{-0.10}$& \color{gray}$48.41^{+0.09}_{-0.16}$& \color{gray}$0.63^{+0.09}_{-0.32}$& \color{gray}$16.6^{+17.8}_{-2.8}$& \color{gray}$37.2^{+1.7}_{-2.1}$& \color{gray}$38.0^{+2.1}_{-2.0}$& \color{gray}$3.5^{+0.2}_{-0.2}$\\
\color{gray} 2dFS 163& \color{gray}$5.14^{+0.03}_{-0.05}$& \color{gray}$9.33^{+0.33}_{-0.30}$& \color{gray}$48.59^{+0.06}_{-0.08}$& \color{gray}$47.53^{+0.13}_{-0.16}$& \color{gray}$0.08^{+0.10}_{-0.00}$& \color{gray}$43.8^{+2.8}_{-25.1}$& \color{gray}$24.4^{+0.8}_{-1.0}$& \color{gray}$24.6^{+0.8}_{-1.0}$& \color{gray}$4.4^{+0.4}_{-0.1}$\\
\color{gray} AV 70& \color{gray}$5.98^{+0.03}_{-0.03}$& \color{gray}$28.95^{+0.93}_{-0.93}$& \color{gray}$49.43^{+0.04}_{-0.04}$& \color{gray}$48.10^{+0.10}_{-0.12}$& \color{gray}$0.24^{+0.02}_{-0.03}$& \color{gray}$106.0^{+17.1}_{-11.5}$& \color{gray}$60.0^{+2.6}_{-2.5}$& \color{gray}$65.2^{+3.2}_{-2.7}$& \color{gray}$2.5^{+0.1}_{-0.1}$\\
\hline
\end{tabular}
\tablefoot{The bottom rows with gray text indicate parameter values that are likely not representative of the stellar properties due to poor fits. The final three columns were derived using {\sc Bonnsai} (see text). Rows with gray text indicate parameter values that are likely not representative of the stellar properties.}
\end{table*}

\begin{table*}
\centering
\caption{Wind structure best-fit parameters and 1$\sigma$ uncertainties.}
\label{p2:tab:wind_props}
\small
\def\arraystretch{1.5}
\begin{tabular}{lccccccccc} \hline \hline
Source & $\log \dot{M}$ & $\varv_\infty$ [km s$^{-1}$] & $\beta$ & $f_{\rm clump}$ & $\log f_{\rm ic}$ & $f_{\rm vel}$ & $\varv_{\rm cl}$ & $\varv_{\rm windturb}$ & $\log D$\\ \hline
AV 80& $-6.24^{+0.05}_{-0.15}$& $1775^{+100}_{-50}$& $1.35^{+0.60}_{-0.20}$& $30^{+17}_{-8}$& $-1.6^{+0.3}_{-0.3}$& $0.70^{+0.16}_{-0.16}$& $0.14^{+0.06}_{-0.03}$& $0.01^{+0.07}_{-0.01}$& $28.44^{+0.10}_{-0.10}$\\
AV 15& $-6.17^{+0.20}_{-0.23}$& $2250^{+125}_{-100}$& $1.60^{+0.40}_{-0.35}$& $7^{+32}_{-6}$& $-2.0^{+0.6}_{-0.1}$& $0.68^{+0.10}_{-0.32}$& $0.16^{+0.10}_{-0.12}$& $0.04^{+0.03}_{-0.05}$& $28.63^{+0.22}_{-0.22}$\\
AV 95& $-7.12^{+0.20}_{-0.33}$& $1850^{+125}_{-75}$& $1.50^{+0.70}_{-0.35}$& $20^{+29}_{-12}$& $-1.4^{+0.7}_{-0.7}$& $0.70^{+0.20}_{-0.26}$& $0.01^{+0.03}_{-0.01}$& $0.03^{+0.07}_{-0.04}$& $27.52^{+0.26}_{-0.26}$\\
AV 207& $-7.65^{+0.15}_{-0.55}$& $1775^{+275}_{-550}$& $1.85^{+0.40}_{-1.10}$& $2^{+18}_{-2}$& $-1.1^{+0.4}_{-0.9}$& $0.22^{+0.22}_{-0.22}$& $0.01^{+0.23}_{-0.01}$& $0.26^{+0.04}_{-0.18}$& $26.91^{+0.37}_{-0.37}$\\
AV 69& $-6.72^{+0.25}_{-0.30}$& $1850^{+75}_{-25}$& $1.40^{+1.10}_{-0.25}$& $12^{+36}_{-3}$& $-1.1^{+0.6}_{-0.3}$& $0.46^{+0.24}_{-0.36}$& $0.01^{+0.23}_{-0.01}$& $0.05^{+0.03}_{-0.04}$& $27.98^{+0.28}_{-0.28}$\\
AV 469& $-6.27^{+0.25}_{-0.08}$& $2025^{+150}_{-50}$& $1.10^{+0.10}_{-0.35}$& $12^{+19}_{-1}$& $-1.4^{+0.3}_{-0.5}$& $0.54^{+0.04}_{-0.32}$& $0.01^{+0.04}_{-0.01}$& $0.08^{+0.07}_{-0.03}$& $28.48^{+0.17}_{-0.17}$\\
AV 479& $-6.27^{+0.25}_{-0.43}$& $1650^{+25}_{-175}$& $0.90^{+0.05}_{-0.30}$& $11^{+20}_{-8}$& $-1.3^{+0.4}_{-0.1}$& $0.06^{+0.24}_{-0.06}$& $0.09^{+0.01}_{-0.05}$& $0.15^{+0.09}_{-0.03}$& $28.48^{+0.34}_{-0.34}$\\
AV 307& $-8.21^{+0.13}_{-0.25}$& $2625^{+100}_{-400}$& $0.90^{+0.35}_{-0.25}$& $4^{+8}_{-3}$& $-1.2^{+0.7}_{-0.5}$& $0.04^{+0.80}_{-0.04}$& $0.12^{+0.04}_{-0.11}$& $0.29^{+0.01}_{-0.19}$& $26.58^{+0.19}_{-0.19}$\\
AV 372& $-6.02^{+0.30}_{-0.18}$& $1650^{+100}_{-100}$& $1.80^{+0.50}_{-0.45}$& $7^{+21}_{-3}$& $-1.1^{+0.4}_{-0.3}$& $0.74^{+0.10}_{-0.48}$& $0.12^{+0.11}_{-0.05}$& $0.24^{+0.06}_{-0.07}$& $28.71^{+0.24}_{-0.24}$\\
AV 327& $-7.41^{+0.15}_{-0.38}$& $1650^{+300}_{-175}$& $0.80^{+0.25}_{-0.25}$& $3^{+15}_{-1}$& $-1.0^{+0.5}_{-0.6}$& $0.50^{+0.18}_{-0.30}$& $0.23^{+0.07}_{-0.04}$& $0.29^{+0.01}_{-0.05}$& $27.25^{+0.27}_{-0.27}$\\ \hline
\color{gray} AV 83& \color{gray}$-6.12^{+0.28}_{-0.18}$& \color{gray}$1025^{+75}_{-125}$& \color{gray}$1.75^{+0.75}_{-0.40}$& \color{gray}$8^{+14}_{-5}$& \color{gray}$-1.1^{+0.8}_{-0.7}$& \color{gray}$0.82^{+0.14}_{-0.48}$& \color{gray}$0.08^{+0.10}_{-0.02}$& \color{gray}$0.13^{+0.09}_{-0.06}$& \color{gray}$28.30^{+0.23}_{-0.23}$\\
\color{gray} AV 70& \color{gray}$-5.59^{+0.13}_{-0.18}$& \color{gray}$1875^{+425}_{-50}$& \color{gray}$0.90^{+0.15}_{-0.20}$& \color{gray}$14^{+3}_{-9}$& \color{gray}$-1.9^{+0.1}_{-0.1}$& \color{gray}$0.02^{+0.22}_{-0.02}$& \color{gray}$0.07^{+0.01}_{-0.03}$& \color{gray}$0.18^{+0.08}_{-0.12}$& \color{gray}$29.21^{+0.16}_{-0.16}$\\
\color{gray} 2dFS 163& \color{gray}$-6.70^{+0.13}_{-0.03}$& \color{gray}$1025^{+175}_{-100}$& \color{gray}$1.70^{+0.55}_{-0.35}$& \color{gray}$50^{+1}_{-13}$& \color{gray}$-1.4^{+0.1}_{-0.5}$& \color{gray}$0.24^{+0.14}_{-0.12}$& \color{gray}$0.09^{+0.04}_{-0.03}$& \color{gray}$0.20^{+0.10}_{-0.06}$& \color{gray}$27.60^{+0.10}_{-0.10}$\\
\hline
\end{tabular}
\tablefoot{The bottom rows with gray text indicate parameter values that are likely not representative of the stellar properties due to poor fits.}
\end{table*}

\section{Discussion} \label{p2:sec:discussion}

The main part of this discussion focuses on the mass-loss properties of the SMC sample in contrast to comparable samples in the Milky Way and LMC, such as to probe the dependence of mass loss on metallicity. However, we start with a brief discussion of other derived properties to characterize the target stars and their evolutionary state better. 

Many of the stars in our sample have been the subject of other studies. We compare our findings to those studies in \cref{p2:sec:other_works_comparison}. Most parameters are consistent with previous findings, although we find slightly higher temperatures and luminosities. 

\subsection{Evolutionary state}

\begin{figure}
    \centering
    \includegraphics[width=\columnwidth]{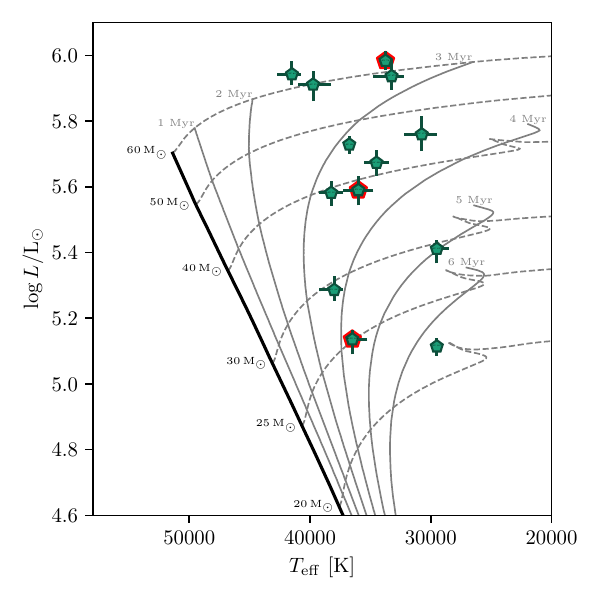}
    \caption{Hertzsprung-Russell diagram of our SMC sample of stars with temperatures and luminosities resulting from the fit. Points with a red border indicate stars with unreliable parameters. We overplot the evolutionary tracks (thin solid gray lines) and isochrones (thin dashed gray lines) of \citet{2011A&A...530A.115B}. The solid black line indicates the zero-age main sequence. The points are labeled with the name of the object in a small font that is visible when zoomed in. }
    \label{p2:fig:HRD}
\end{figure}

Figure~\ref{p2:fig:HRD} presents our sample of giants, bright giants, and supergiants in the Hertzsprung-Russell diagram (HRD). The figure shows that the sample covers a substantial part of the ($T_{\rm eff},L$) plane. All objects have moved away from the zero-age main sequence, but are still on the main sequence when considering single-star evolutionary tracks \citep{2011A&A...530A.115B}, and their ages range from approximately 2 to 6~Myr. For three stars (AV~83, 2dFS~163, and AV~70; see also \cref{p2:sec:star_comments}), we were not able to determine reliable parameter values. While we present their best-fit values, these sources were not considered in the mass-loss rate and clumping analysis that is the topic of the next subsections. In \cref{p2:fig:HRD} and also in \cref{p2:fig:mass_loss_predictions,p2:fig:wind_momentum_relation,p2:fig:DMZ} below, these sources are present and marked with a red border, but they are not included in any fits or analyses.

\begin{figure}
    \centering
    \includegraphics[width=\columnwidth]{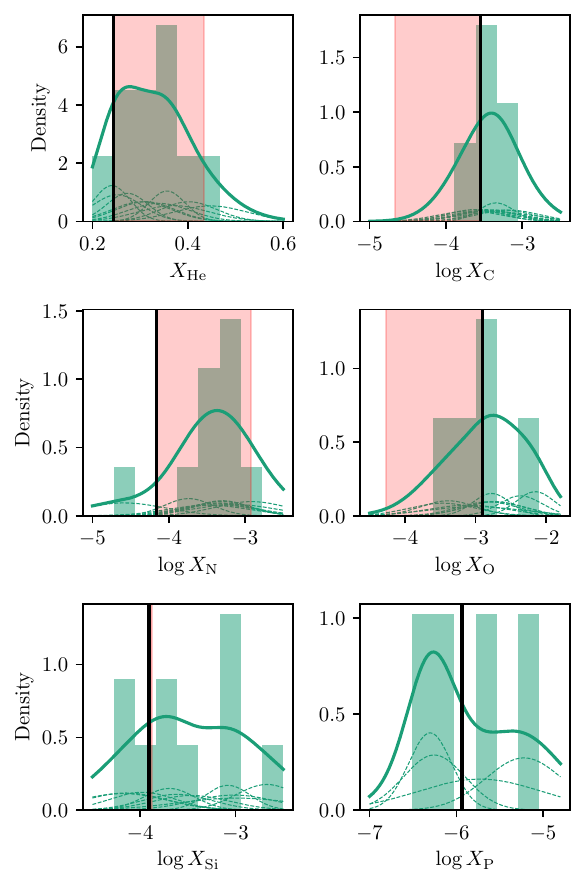}
    \caption{Distribution of the surface mass fractions of He, C, N, O, Si, and P from the {\sc Fastwind}/GA fitting of SMC stars. The solid green line indicates an approximate kernel density estimate of the distribution taking the uncertainties on the best-fit values into account. The thin dashed lines show the contribution of the individual sources. The vertical black line shows the baseline mass fraction from \citet{2019AJ....157...50D} for all elements except phosphorus, for which we show the scaled (by 0.2) solar abundance from \citet{2009ARA&A..47..481A}. The area marked in red shows the surface depletion or enrichment that can be expected due to mixing based on the evolution of a 40\,M$_\odot$ star with an initial rotation of 389\,km\,s$^{-1}$ \citep{2011A&A...530A.115B}. For phosphorus, only stars with a reasonable fit are included. The helium mass fractions are taken from the optical-only fit. All other fractions are from the optical and UV fit. }
    \label{p2:fig:abundaces}
\end{figure}

Figure\,\ref{p2:fig:abundaces} shows the derived surface mass fractions for He, C, N, O, Si, and P. We also plot the SMC baseline mass fraction following \citet{2019AJ....157...50D} for all elements save phosphorus, for which we scaled the \citet{2009ARA&A..47..481A} solar abundance by a factor of 1/5. We also show indications of the expected surface mass fractions, using an initially 40\,M$_{\odot}$ SMC model that rotates at the end of formation at 389\,km\,s$^{-1}$ and has a typical age of 3\,Myr \citep{2011A&A...530A.115B}. For the rotational velocity we chose, the range of helium mass fraction is well reproduced, but we remark that for a lower initial spin, rotation-induced mixing is less efficient. $X_{\rm He}$ is therefore only marginally enriched ($X_{\rm He} < 0.3$). The range of nitrogen mass fractions is well reproduced. The C and O abundances are slightly above the baseline, albeit with large uncertainties, while depletion is expected. In \cref{p2:sec:C_N_O_comp} we compare the total combined C, N, and O surface abundances to the baseline abundance and to the study of \citet[XShootU V][]{2024A&A...689A..31M}. Their sample overlaps with ours. They analyzed the same data with a different method that is more focused on surface abundances. 

Mixing is not expected to affect silicon and phosphorus. The surface abundances of these elements are poorly constrained; for silicon, they are typically above the baseline, and for phosphorous, they cluster around the baseline. 

The latter may be an effect of the sampling. Because the effect of the surface abundance on the strength of the \ion{P}{v} doublet is limited, the sampling is close to uniform. For example, AV\,307 has an excellent fit to the \ion{P}{v} lines, but the phosphorus abundance is unconstrained due to the dependence on temperature, among other parameters. This correlation might be too strong to determine the abundance reliably with only one available ionization stage.

We repeat that 2dFS\,163, AV\,83, and AV\,70 might be binaries (see Sect.~\ref{p2:sec:star_comments}) and that their best-fit parameter values are likely not representative of their true values. They were left out in the further analysis.

\subsection{Comparison of empirical and predicted mass-loss rates}

\begin{figure}
    \centering
    \includegraphics[width=\columnwidth]{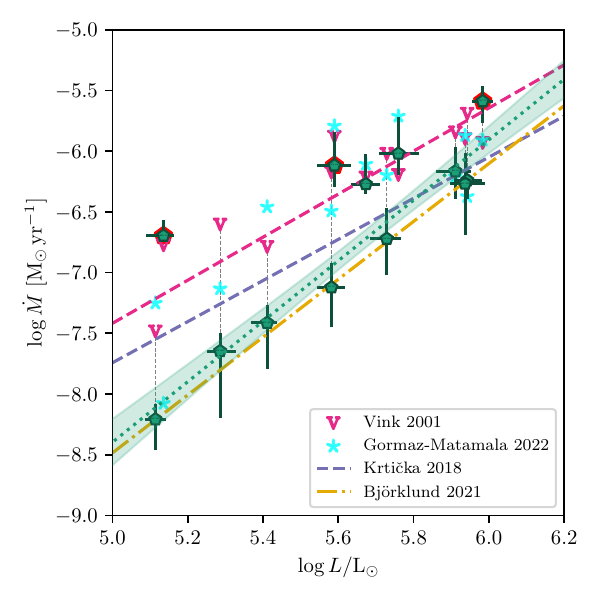}
    \caption{Mass-loss rate as a function of luminosity. The fit to the empirical mass-loss rates and its 1$\sigma$ uncertainty are indicated with the dotted green line and the shaded region, respectively. The mass-loss rate -- luminosity relations of \citet{2018A&A...612A..20K} and \citet{2021A&A...648A..36B} are shown with the dashed blue and yellow lines, respectively. For the mass-loss predictions of \citet{2001A&A...369..574V}, we used their equation 24 to obtain the predicted rates for each star based on their properties. The magenta V indicate these rates, and the dashed magenta line corresponds to the fit to these points. The light blue stars indicate the mass-loss rates of equation 7 from \citet{2022A&A...665A.133G}. Points with a red border are unreliable fits that are likely not representative of their physical properties; these are not included in the fit.}
    \label{p2:fig:mass_loss_predictions}
\end{figure}

We briefly show how the mass-loss rates in our sample compare to theoretical predictions for SMC stars. Figure\,\ref{p2:fig:mass_loss_predictions} shows the mass-loss rate as a function of luminosity, compared to the mass-loss rate predictions of \citet{2001A&A...369..574V,2018A&A...612A..20K,2021A&A...648A..36B}; for the SMC metallicity, we assumed 20\% of solar for all predictions, following \citet{2007A&A...473..603M}. The dotted line indicates a linear fit to the empirical mass-loss rates using orthogonal distance regression \citep[ODR][]{boggs1987odrpack}, which takes the uncertainties of all variables into account, that is here, mass-loss rate and luminosity, given by
\begin{equation}
    \log \dot{M} = (-20.81\pm1.48) + (2.48\pm0.26) \times \log (L/L_\odot).
\end{equation}
The luminosity dependence that we find is stronger than that of the theoretical predictions that are shown in the figure. Overall, the \citet{2021A&A...648A..36B} mass-loss rate--luminosity relation agrees best with the empirical mass-loss rate from our sample. \citet{2001A&A...369..574V}, \citet{2022A&A...665A.133G}, and \citet{2018A&A...612A..20K} predicted generally higher mass-loss rates at low luminosity, with the discrepancy decreasing toward higher luminosity. For the comparison to the prediction of \citet{2001A&A...369..574V}, we used the best-fit stellar properties and the spectroscopic mass. Slightly different results can be expected when the evolutionary mass as a discrepancy between the two masses remains (see \cref{p2:sec:mevo_mspec}).

\cref{p2:sec:more_SMC_Mdot} compares the mass-loss rates found here to those found in previous optical and UV stellar wind analyses in the SMC. We mostly find mass-loss rates consistent with those of previous studies.

For two stars, AV\,372 and AV\,469, a significantly higher mass-loss rate is found than for the rest of the sample given their luminosity. It is unclear why these stars require such a mass-loss rate for an optimal fit. Both stars agree within the uncertainties with the \citet{2001A&A...369..574V} and \citet{2022A&A...665A.133G} predictions, while the rest of the sample agrees with those of \citet{2021A&A...648A..36B}.
Both stars are classified as supergiants, but other supergiants in the sample follow the same trend as the rest of the sample. The He abundance for both stars is significantly enriched from the baseline, showing the highest abundances in the sample. Additionally, we find a strong nitrogen enrichment in AV\,469. The diverging values are still within the typical scatter found when determining mass-loss rates using optical and UV spectroscopy \citep[see e.g.,][]{2022A&A...663A..36B,2021A&A...647A.134B}. It remains possible, however, that these stars are physically different from the rest of the sample in ways in that are currently unknown, for instance, they may be post-interaction binaries. Alternatively, these stars could be in a transient episode of higher mass loss. 

\subsection{Dependence of the modified wind momentum on luminosity and metallicity}
\label{p2:sec:mass_loss_metallicity}

To empirically test the relation between mass loss, luminosity, and metallicity, we compared the mass-loss rates of samples of Galactic \citep{2021A&A...655A..67H}, LMC\footnote{In this study the potential binary stars from Brands et al. were left out and some poorer fits and the lowest luminosity star from \citeauthor{Hawcroft2024} were left out.} \citep[Brands et al. in prep.]{Hawcroft2024}, and SMC (this work) O stars. In these studies, the winds were scrutinized using a {\sc Fastwind} GA-analysis of both their optical and UV spectrum in which the inhomogeneous wind structure accounts for optically thick clumping. The uniform nature of these analyses -- they all applied the same formalism for the wind inhomogeneities -- allows for the best possible comparison of wind mass loss in different metallicity environments. However, it is important to note that the despite the similar approaches, the properties of the samples are not identical. \citet{2021A&A...655A..67H} focused on a set of Galactic supergiants with only one star with a spectral type later than O7 and all stars more luminous than 10$^{5.6}$\,L$_\odot$. The LMC stars are also on average of an earlier type than the SMC sample, with 14 stars having a spectral type earlier that O7, while only 2 such stars appear in the SMC sample.

\begin{figure}
    \centering
    \includegraphics[width=\columnwidth]{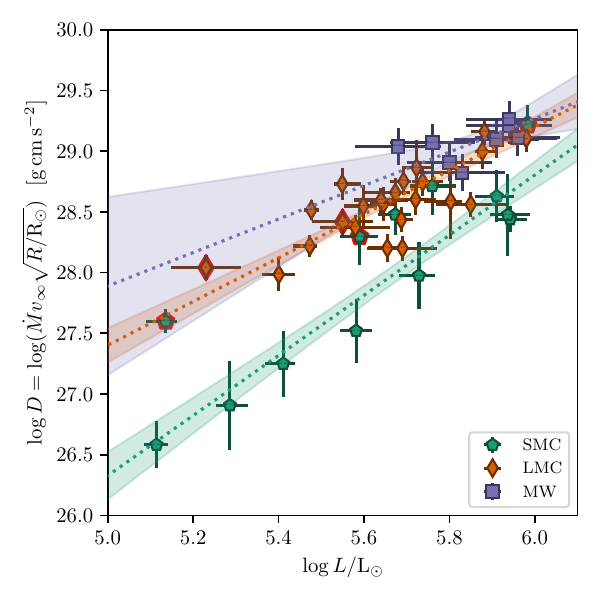}
    \caption{Modified wind momentum as function of luminosity for samples of SMC (this work), LMC \citep[Brands et al. in prep.]{Hawcroft2024}, and Galactic stars \citep{2021A&A...655A..67H}. The dotted lines indicate the linear fits to each subsample, and the shaded regions show their 1$\sigma$ uncertainties. The points with a red border represent unreliable fits that are likely not representative of their physical properties; these are not included in the fit.}
    \label{p2:fig:wind_momentum_relation}
\end{figure}

To isolate and quantify the impact of metallicity on mass loss from the limited size samples available, we resorted to the mechanical momentum of the stellar wind, modified by a term depending on stellar radius. This property, the so-called modified wind momentum,
\begin{equation}\label{eq:dmom-lum-slope-offset}
    D = \dot{M} \varv_\infty \sqrt{\frac{R_\star}{R_\odot}} ~~ [{\rm gr\,cm\,s^{-2}}],
\end{equation}
is  primarily a function of stellar luminosity and often predicted to be almost independent of stellar mass \citep{1995svlt.conf..246K,2000ARA&A..38..613K}. We note that for this to be valid, however, the CAK-$\alpha$ parameter (see below) needs to be constant at a value of 2/3. This may not be the case throughout our parameter space or everywhere in the wind \citep[e.g.,][]{2012A&A...537A..37M}. To compare our sample with other observations and theoretical predictions, we fit the following relation through the observed values: 
\begin{equation}\label{eq:dmom-lum-slope-offset}
    \log D = \log D_0 + x \log L/L_\odot.
\end{equation}
From now on, we refer to $x$ as the slope, and to $\log D_0$ as the offset of the modified wind momentum--luminosity relation.
\cref{p2:fig:wind_momentum_relation} shows the relation for the three samples, using ODR to constrain the fit variables $x$ and $\log D_0$. The empirical relations are distinguished by color: purple for Galactic stars, orange for LMC stars, and green for SMC stars. Although the uncertainties for the Galactic sample are sizeable, the overall trend is indeed that the wind momentum increases with metallicity. Furthermore, we find a tentative trend in the slope of the modified wind momentum--luminosity relation, with steeper slopes toward lower metallicities. Our fit values are presented in Table\,\ref{p2:tab:linear_fits}. 

\begin{table}[]
    \centering
    \caption{Slopes, $x$, and offsets, $\log D_{0}$, of the linear fits to the modified wind momentum (Equation~\ref{eq:dmom-lum-slope-offset}) of this work, \citet{2007A&A...473..603M}, and \citet{2022MNRAS.511.5104M}.}
    \label{p2:tab:linear_fits}
    \begin{tabular}{ccc} \hline \hline \\[-10pt]
    Galaxy      &  $x$               & $\log D_{0}$     \\ \hline \\[-10pt]
        \multicolumn{3}{c}{This work}                   \\ \hline \\[-10pt]
    Milky Way   &  1.38$\pm$0.86     & 20.99$\pm$5.01   \\
    LMC         &  1.87$\pm$0.21     & 18.03$\pm$1.19   \\
    SMC         &  2.46$\pm$0.26     & 13.99$\pm$1.51   \\ \hline \\[-10pt]
        \multicolumn{3}{c}{Mokiem (2007)} \\ \hline \\[-10pt]
    Milky Way   &  1.84$\pm$0.17     & 18.87$\pm$0.98   \\
    LMC         &  1.96$\pm$0.16     & 17.88$\pm$0.91   \\
    SMC         &  1.84$\pm$0.19     & 18.20$\pm$1.09   \\ \hline \\[-10pt]
        \multicolumn{3}{c}{Marcolino (2022)} \\ \hline \\[-10pt] 
    Milky Way   &  4.16$\pm$0.23     & 5.43$\pm$1.28    \\
    SMC         &  3.85$\pm$0.29     & 6.67$\pm$1.52    \\ \hline
    \end{tabular}
\end{table}

\begin{figure}
    \centering
    \includegraphics[width=\columnwidth]{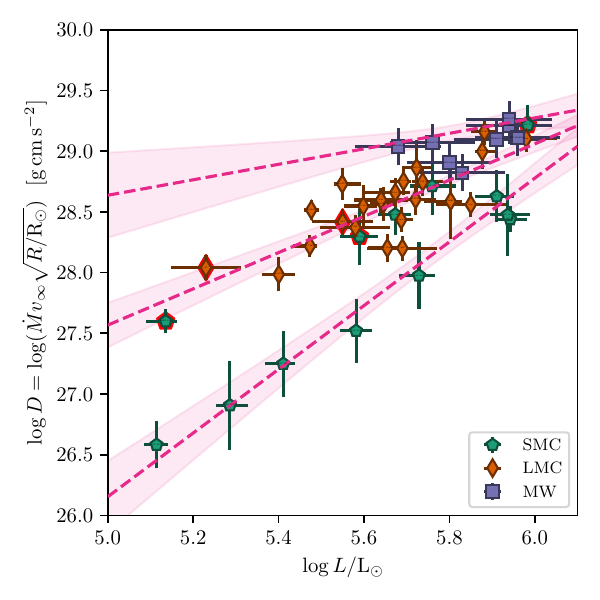}
    \caption{Modified wind momentum as function of luminosity and metallicity, as in Figure\,\ref{p2:fig:wind_momentum_relation}, but now fit with Equation\,\ref{p2:eq:DMZ} to the three samples, instead of separate linear fits for each sample. The dashed magenta lines and pink shaded regions indicate the best fit and the  1$\sigma$ confidence interval, respectively. Equation\,\ref{p2:eq:DMZ} has been evaluated at $Z = 1.0, 0.5,$ and 0.2 Z$_\odot$. The points with a red border are unreliable fits that are likely not representative of their physical properties; these are not included in the fit.}
    \label{p2:fig:DMZ}
\end{figure}

To account for a metallicity dependence of the modified wind momentum, we fit all three samples simultaneously with the function
\begin{equation}
    \log D(L, Z) = \left(a + b \log \frac{Z}{Z_\odot}\right) \log \frac{L}{10^6\,L_\odot} + c \log \frac{Z}{Z_\odot} + d, 
    \label{p2:eq:DMZ}
\end{equation}
following \citet{2018A&A...612A..20K} and \cite{2021A&A...648A..36B}, where $a$ through $d$ are the fitting parameters. We assumed $Z_{\rm MW} = {\rm Z_\odot}$, $Z_{\rm LMC} = 0.5\,{\rm Z_\odot}$, and $Z_{\rm SMC} = 0.2\,{\rm Z_\odot}$, consistent with \citet{2007A&A...473..603M}, \citet{2018A&A...612A..20K}, \citet{2021A&A...648A..36B}, and \citet{2022MNRAS.511.5104M}. We took the uncertainty in the wind momentum, luminosity, and metallicity into account. For the latter, we assumed the uncertainty to be $\Delta \log Z / {\rm Z_\odot} = 0.1$, which is conservative considering the typical uncertainties on the measured abundances \citep[e.g.][]{2019AJ....157...50D}. Figure\,\ref{p2:fig:DMZ} shows the best fit of Equation\,\ref{p2:eq:DMZ} to the data evaluated at $Z = 1, 0.5$ and 0.2\,Z$_\odot$. The fit shows that over the wide range in luminosity probed by the samples, the modified wind momentum is only poorly represented by a luminosity-independent power-law dependence on metallicity. For relatively dim stars, the metallicity dependence is stronger than for relatively bright stars. Although the fit agrees nicely with the LMC and SMC data points, it slightly overestimates the wind momentum of the MW stars. 
The best-fit values of the parameters $a$, $b$, $c$, and $d$ in \cref{p2:eq:DMZ} are listed in \cref{p2:tab:logD-LZ_params}. These parameters are strongly correlated, and therefore, their individual variances are a poor representation of their true uncertainty. The full covariance matrix of the fit can be found in Appendix\,\ref{p2:app:logD_cov}.

Comparing the two fitting procedures, we note that the gradient in slope between the direct linear fits to the individual samples is smaller than in the combined fit, as can be seen by comparing Figures~\ref{p2:fig:wind_momentum_relation} and \ref{p2:fig:DMZ}. 
This is a result of the difference in wind momentum between the Milky Way and the LMC, which is smaller than the same difference between the LMC and SMC stars, while in both cases, the difference in metallicity is similar. As a result, the metallicity-offset term, $c$ in Equation~\ref{p2:eq:DMZ}, cannot accommodate both differences at once. To mitigate this, the metallicity slope, $b$ in Equation~\ref{p2:eq:DMZ}, was adjusted. This works because of the lack of relatively low-luminosity points in the Milky Way sample. Therefore, caution is advised: Our fit result for Equation~\ref{p2:eq:DMZ} may overestimate the slope in the metallicity dependence. Additional studies using similar methods to model the clumped winds of, in particular, Galactic stars with luminosities $\log L/L_{\odot} \leq 5.6$ are needed to scrutinize this further. Alternatively, the discrepancy between the two fitting results may signal that the parameterization in Equation~\ref{p2:eq:DMZ} is not sufficiently suitable. This might be mitigated by adding a nonlinear (in log-space) metallicity term.

\subsubsection{Comparison to other empirical studies}

\begin{figure}
    \centering
    \includegraphics[width=\columnwidth]{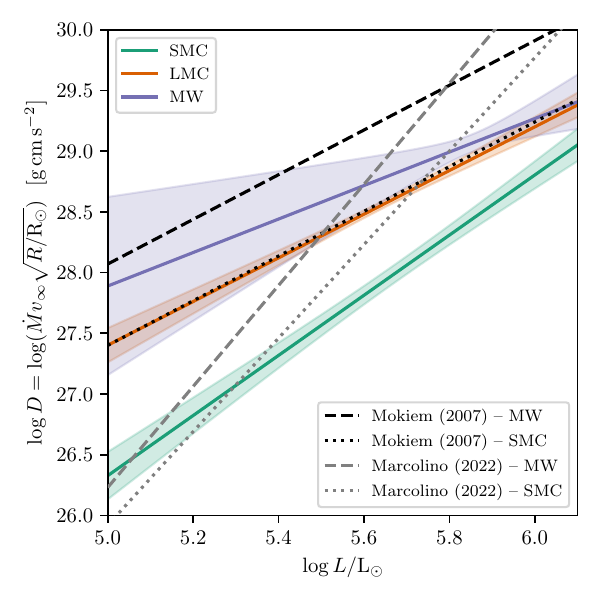}
    \caption{Fits to the empirically determined modified wind momentum as a function of luminosity. The solid colored lines show the fits to the MW, LMC, and SMC samples presented here. The dashed and dash-dotted black and gray lines indicate the modified wind momentum relations from \citet{2007A&A...473..603M} and \citet{2022MNRAS.511.5104M}, respectively. The values representing the fits are listed in Table~\ref{p2:tab:linear_fits}.}
    \label{p2:fig:empirical_relations}
\end{figure}

Figure~\ref{p2:fig:empirical_relations} repeats the fitting results from Fig.~\ref{p2:fig:wind_momentum_relation} along with the best relations from \citet{2007A&A...473..603M} and \citet{2022MNRAS.511.5104M}, who also empirically investigated the metallicity dependence of mass loss. 
\citet{2007A&A...473..603M} used data from optical studies of Galactic and Magellanic Clouds stars. Mass-loss rates follow from {\sc Fastwind} analyses without considering clumping effects. Because ultraviolet diagnostics are not considered, quite a few of their relatively low-luminosity sources yield only upper limits for $\dot{M}$. \citet{2022MNRAS.511.5104M} collected results from {\sc Cmfgen} \citep{1998ApJ...496..407H} studies that relied on both ultraviolet and optical spectra. In these MW and SMC studies, an optically thin clumping prescription was used that sometimes only yielded rough estimates of the clumping factor.

\citet{2007A&A...473..603M} estimated the metallicity dependence at only a single luminosity, that is,
$\log L / L_{\odot} = 5.75$, a typical mean for their sample. Near this luminosity, their slope is not too different from ours. Their wind momentum is higher than ours, however, which likely is a result of \citeauthor{2007A&A...473..603M} not considering clumping.
\citet{2022MNRAS.511.5104M} reported a significantly steeper slope in the $D(L)$ diagram for both the MW and SMC. Interestingly, their MW and SMC curves diverge for higher luminosities, while our curves converge. The luminosity range considered in \citet{2022MNRAS.511.5104M} extends to $\sim$$10^{4.6}$\,L$_\odot$, which is significantly lower than our sample. This might lead to a different slope.

We quantitatively compared our results to the pioneering work of \citet{2007A&A...473..603M} by evaluating Equation\,\ref{eq:dmom-lum-slope-offset} at $\log L/L_\odot = 5.75$ with the SMC, LMC, and MW coefficients as listed in Table\,\ref{p2:tab:linear_fits}.  
Under the assumptions that the mass-loss rate and terminal wind velocity scale as
\begin{equation}
    \dot{M} \propto Z^m
    \label{eq:m-slope}
\end{equation}
and 
\begin{equation}
    \varv_\infty \propto Z^n,
\end{equation}
such that
\begin{equation}
    m + n = \frac{d\log D_{\rm mom}}{d \log Z},
\end{equation}
with $n=0.13$ \citep{1992ApJ...401..596L}, we find from an ODR fitting of the three points on the $D(Z, L=10^{5.75}\,$L$_\odot)$-plane

\begin{equation}
    m = \frac{d\log \dot{M}}{d \log Z} = 1.02\pm0.30.
\end{equation}
This is slightly higher than but consistent with the value of \citeauthor{2007A&A...473..603M} ($m = 0.83\pm 0.16$).

\begin{figure}
    \centering
    \includegraphics[width=\columnwidth]{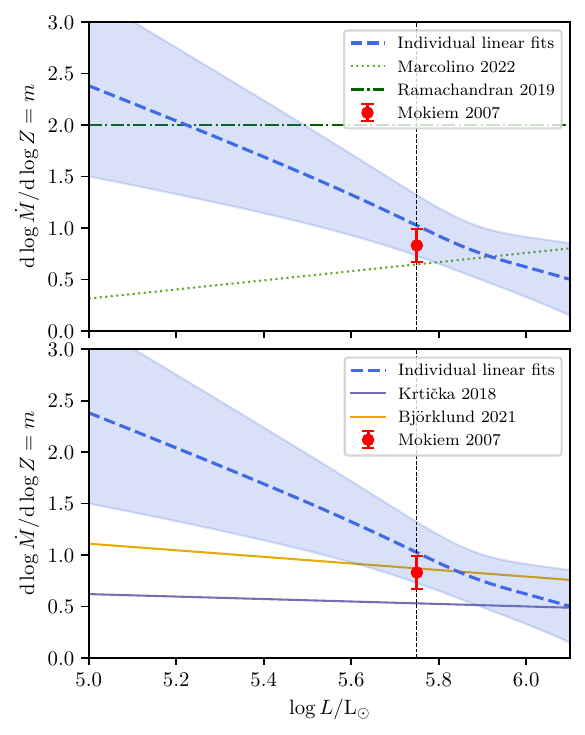}
    \caption{Slope of the mass-loss rate as a function of metallicity, $m$, for different values of the luminosity. The \textit{top} panel shows the empirical determinations of \citet{2007A&A...473..603M}, \citet{2019A&A...625A.104R}, and \citet{2022MNRAS.511.5104M}, along with our determination based on the linear fits as shown in \cref{p2:fig:wind_momentum_relation} and \cref{p2:fig:empirical_relations}. The empirical values of $m$ from \citet{2022A&A...666A.189R} lie above the vertical range of this figure. The \textit{bottom} panel shows the predicted values of $m$ of \citet{2018A&A...612A..20K} and \citet{2021A&A...648A..36B}, along with our empirical determination. The value of $m = 0.83$ from \citet{2007A&A...473..603M} is also shown for reference. In all cases, ${\rm d}\log \varv_\infty/{\rm d}\log Z = n = 0.13$ is assumed \citep{1992ApJ...401..596L}.}
    \label{p2:fig:MZ_slope_windmomentum}
\end{figure}

In the top panel of Figure\,\ref{p2:fig:MZ_slope_windmomentum}, we repeat this procedure for all luminosities (dashed blue line and light blue uncertainty region). At $\log L/L_{\odot} = 5.75$, $m = 1.02$, as discussed above. However, we do not find a constant value for $m$, but higher $m$ for lower luminosities. The luminosity-independent result $m = 0.83$ of \citeauthor{2007A&A...473..603M} is plotted as the red point. We plot with a dotted green line the results of \citet{2022MNRAS.511.5104M}. As their metallicity dependence only relies on MW and SMC results, they did not require a three-point fitting in order to obtain $m$. At the reference luminosity $\log L/L_{\odot} = 5.75$, they found $m \sim 0.6$, which is lower than \citet{2007A&A...473..603M} and just outside our 1$\sigma$ confidence interval. However, \citet{2022MNRAS.511.5104M} reported a positive trend of $m$ with luminosity, which is opposite to our findings and not in line with theoretical predictions and expectations (see \cref{p2:sec:wind_theory}). It is important to note, however, that the sample of \citet{2022MNRAS.511.5104M} extends to luminosities of $\sim$10$^{4.6}$\,L$_\odot$, where the winds are significantly weaker for both Galactic and SMC stars.
\citet{2019A&A...625A.104R} and \citet{2022A&A...666A.189R} also studied the atmosphere properties and mass-loss rates of massive stars in the SMC. Their samples consisted predominantly of dwarf stars and mostly covered lower luminosities. \citet{2019A&A...625A.104R} reported a value of $\sim$2, as shown in \cref{p2:fig:MZ_slope_windmomentum} with the dash-dotted dark green line, which is consistent with our findings at lower luminosities. \citet{2022A&A...666A.189R} found a similar trend with luminosity, that is, higher values of $m$ at lower luminosity, with the metallicity dependence scaling as $m(L) = -1.2 \log(L / \text{L}_\odot) + 10.3$. Their value of $m$ would range from 4.3 to 3 in our luminosity span. This significantly higher metallicity dependence might be caused by the so-called weak-wind problem in SMC dwarf stars.

\subsubsection{Comparison to predictions} \label{p2:sec:wind_theory} 

\begin{table}[]
    \centering
    \caption{Fit parameters of Equation\,\ref{p2:eq:DMZ} to the observed data, compared to \citet{2018A&A...612A..20K} and \citet{2021A&A...648A..36B}.}
    \label{p2:tab:logD-LZ_params}
    \begin{tabular}{c|ccc} \hline \hline \\[-10pt]
    Parameter & This work & Krti\v{c}ka (2018) & Bj\"orklund (2021) \\ \\[-11pt] \hline \\[-10pt]
       $a$  &  0.64 &  1.64 &  2.07 \\
       $b$  & $-2.84$ & $-0.32$ & $-0.73$ \\
       $c$  &  0.71 &  0.36 &  0.46 \\
       $d$  & 29.27 & 28.88 & 29.25$^\dagger$ \\ \hline
\end{tabular}
\begin{tablenotes}
    \item $^\dagger$ An offset of 30.80 is added to match the units of this work. 
\end{tablenotes}
\end{table}

Table\,\ref{p2:tab:logD-LZ_params} lists our fit parameters $a, b, c$, and $d$ of Equation~\ref{p2:eq:DMZ}, along with the predicted values from \citet{2018A&A...612A..20K} and \citet{2021A&A...648A..36B}. Additionally, these relations are shown graphically in Appendix\,\ref{p2:app:logD_cov}. In order to obtain the values of \citeauthor{2018A&A...612A..20K}, we calculated the modified wind momentum rates using the values from Tables 1 from both \citet{2017A&A...606A..31K} and \citet{2018A&A...612A..20K} up to 42\,500\,K. We find a significantly stronger metallicity dependence on the modified wind momentum--luminosity slope (parameter $b$) than both of the two predictions. Consequently, this implies a stronger metallicity dependence of the mass loss at low luminosity.

The bottom panel of Figure\,\ref{p2:fig:MZ_slope_windmomentum} again shows our empirical value of $m$ as a function of luminosity and the result from \citet{2005A&A...441..711M} of $m = 0.83$, but now compared to the two sets of theoretical results. Both predictions show a decreasing trend of $m$ with luminosity as in our empirical findings, albeit much weaker, and opposite to that of the work of \citet{2022MNRAS.511.5104M}. We note that  especially the \citet{2021A&A...648A..36B} result is close to the single $m$ value of \citeauthor{2007A&A...473..603M}. 

In passing, we note that we performed the same analysis for the derived mass-loss rates directly, instead of using the modified wind momentum. This can be found in Appendix~\ref{p2:sec:mdot_analysis}. The results from that analysis are consistent with the wind momentum results presented here. This suggests that the value of ${\rm d}\log \varv_\infty/{\rm d}\log Z = n = 0.13$ from \citet{1992ApJ...401..596L} that we adopted is reasonable, and a more recent determination of $n = 0.19$ by \citet{2021MNRAS.504.2051V} would also be consistent with our findings. A weak dependence of the terminal wind speed on metallicity was also reported by \citet[Paper III]{2024A&A...688A.105H}.

\subsubsection{Line-driven wind theory considerations}

The question arises whether the stronger dependence of the mass loss on metallicity for lower-luminosity stars can be explained by line-driven wind theory. The original theory by \citet[][henceforth CAK]{1975ApJ...195..157C}, and subsequent CAK-like theories such as \cite{2000A&AS..141...23P,1995ApJ...454..410G}, and re-formulations of CAK-like theories, such as \cite{1999isw..book.....L}, essentially have a principal scaling of
\begin{eqnarray}
    \dot{M} \propto \bar{Q}^{(1-\alpha)/\alpha} \cdot L^{1/\alpha} \cdot \left[ M(1-\Gamma) \right]^{-(1-\alpha)/\alpha},
\end{eqnarray}
where $M$ is the stellar mass, $\Gamma$ is the Eddington parameter for Thomson scattering, $\bar{Q}$ represents the maximum value of the multiplication factor by which the total continuum opacity in the outflow is to be multiplied to also account for line opacities \citep{1995ApJ...454..410G}, and $\alpha$ may be interpreted as the ratio of the line force from optically thick lines to the total force \citep{2000A&AS..141...23P}; hence $0 < \alpha < 1$. For stars that lie not too close to the Eddington limit, it thus holds that $\dot{M} \propto L^{1/\alpha}$. Assuming no direct dependence of terminal velocity on luminosity, we arrive at 
\begin{eqnarray}
   D \propto L^{1/\alpha} = L^{x},
\end{eqnarray}
where the last equality is what we used in \cref{eq:dmom-lum-slope-offset}.

The results in Table~\ref{p2:tab:linear_fits} would thus be matched by a decreasing value of $\alpha$ with metallicity: $\alpha = 0.72, 0.53, 0.41$ for $Z = 1, 0.5, 0.2~Z_{\odot}$, respectively. \citet{2000A&AS..141...23P} explored the metallicity (and effective temperature and radial depth) dependence of $\alpha$ in OBA-star outflows in the regime 0.1$-$3\,$Z_{\odot}$ (their figure 27, where $\hat{\alpha} \sim \alpha$). A CAK-optical depth parameter $t \sim 10^{-5}$ indeed shows a roughly similar behavior to our findings. However, for higher values of optical depth (i.e., closer to the stellar surface), this dependence disappears. The overall effect may therefore be modest \citep[such as found by][]{2018A&A...612A..20K,2021A&A...648A..36B}, at least in the metallicity regime probed here. Still, further theoretical studies, probing a wider metallicity range, would be extremely helpful in assessing the reasons for the luminosity dependence of the modified wind momentum versus metallicity dependence reported here and by \citet{2022MNRAS.511.5104M}. These studies may also address the weak empirical dependence of $D(Z)$ at high luminosities found here. This behavior might indicate that for very strong winds, the line force is dominated by optically thick lines, such that a $Z$-dependence is lost. Alternatively, at high luminosity, the Eddington limit may be exceeded already quite deep down in the photosphere, initiating a mass loss that the star is simply not able to cope with, leading to fall-backs and a maximum sustainable mass loss.

In addition to theoretical studies, empirical studies should be pursued to improve the coverage of the parameter space (especially in the low-luminosity regime of Galactic stars). Alternatives to the parameterization \cref{p2:eq:DMZ} should be explored.

\subsection{Wind structure and metallicity}

The dependence of the mass-loss properties on metallicity as described above is based on analyses that accounted for the presence of (optically thick and thin) clumps in the outflow as well as for the associated vorosity. The derived wind structure parameters for our target SMC stars are listed in Table~\ref{p2:tab:wind_props}. They show a wide range of clumping factors $f_{\rm cl}$, of 2--30, and a density of the inter-clump medium that is relatively low, with $f_{\rm ic} \sim 10^{-2}-10^{-1}$. The velocity-porosity effect, characterized by $f_{\rm vel}$, shows a wide range of values, from 0.04 to 0.74. The wind begins to form inhomogeneities close to the onset of the wind at $\varv_{\rm cl,start} \sim 0.01\,\varv_\infty$ to $0.23\,\varv_\infty$.
For the turbulent velocity, our analysis only probes the range of values from 0$-$0.3 $\varv_\infty$, following earlier empirical results by \citet{1989A&A...221...78G}. We find that the results fully span this allowed range, with 
 $\varv_{\rm windturb} \sim 0.01 - 0.29 \,\varv_\infty$.

The obtained clumping constraints allowed us to probe for the first time for O-type stars whether the wind-clumping properties themselves are metallicity dependent and thus contribute to the empirical behavior of $D(L,Z)$. If this dependence were strong, an in-depth treatment of clumping would be a prerogative for any empirical study of the dependence of mass loss on metal content. We continue here with caution as these first results should be considered as only exploratory for three important reasons. 
First, a comparison of clumping properties at different metallicities requires a similar treatment of the wind structure to within great detail. For this reason, we opted to only discuss the analyses of the sample of SMC stars described here and of LMC stars described in Brands et al. in prep., where very similar methods are applied to very similar data sets. We thus excluded the study by \citet{2021A&A...655A..67H}, who also accounted for optically thick clumping, but where slightly different assumptions regarding its treatment were adopted. 
Second, the clumping prescription adopted here still lacks a quantitative comparison to 2D or 3D hydro-simulations of radiation-driven outflow, in which wind structure naturally develops \citep[e.g.,][]{2018A&A...611A..17S}. This comparison would help us to assess its appropriateness.
Third, the O-star SMC and LMC samples scrutinized here are still relatively small and show some differences in intrinsic properties, with generally hotter and brighter stars in the LMC sample. The latter therefore implies overall somewhat more massive stars, with stronger stellar winds.

\begin{figure}
    \centering
    \includegraphics[width=\columnwidth]{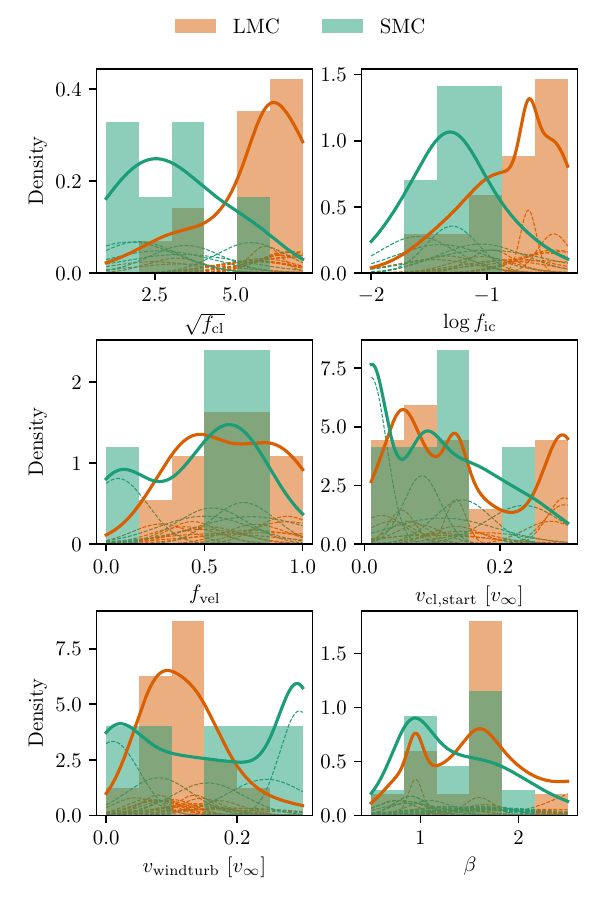}
    \caption{Distributions of the wind parameters of LMC and SMC XShootU samples (this work and Brands et al. in prep.). The solid lines indicate the approximate KDE based on the uncertainty on the parameter values. The thin dashed lines indicate the contribution of the individual objects to the total KDE. Objects for which the parameters are unconstrained are excluded.}
    \label{p2:fig:clump_hists}
\end{figure}

Figure~\ref{p2:fig:clump_hists} shows the distribution of the five wind-structure parameters ($f_{\rm cl}, f_{\rm ic}, f_{\rm vel}, \varv_{\rm cl,start}, \varv_{\rm windturb}$) and $\beta$, describing the acceleration rate of the smooth wind flow at LMC and SMC metallicities. The histograms also show the approximate kernel density estimation (KDE) based on the (often sizable) uncertainty on the parameter values. For the clumping factor, the square root of its value is displayed, as for optically thin clumping and a void inter-clump medium the optical wind-diagnostic lines are essentially invariant for the property $\dot{M} \cdot \sqrt{f_{\rm cl}}$. 

To assess the significance of the differences between the two distributions, we performed a two-sample Kuiper test. We found that the differences are consistent with resulting from a random sampling for $f_{\rm vel}, \varv_{\rm cl,start}, \varv_{\rm windturb}$ and $\beta$. For $f_{\rm cl}$ and $f_{\rm ic}$, we found that the probabilities of the difference result purely from random sampling of 0.08 and 0.11, suggesting that the underlying distributions may differ. We note, however, that the Kuiper test does not take the uncertainty on the parameter values into account, which may make the difference less significant. 

$f_{\rm cl}$ and $f_{\rm ic}$ appear to show higher values at higher metallicity. At face value, this would suggest that the inter-clump medium in the LMC is relatively dense (i.e., high $f_{\rm ic}$) and that the clumps have a high density (i.e., high $f_{\rm cl}$), but only fill a relatively small fraction of the wind volume. In the wind medium of the SMC stars, the inter-clump medium is relatively rarefied, the clumps have a modest density, but fill a relatively large fraction of the wind volume. \citet{2022A&A...663A..40D} reported a similar trend in the clumping factor with metallicity. They reported that the clumping factor scales with metallicity as $f_{\rm cl} \propto Z^{0.15}$. This would imply an increase in $f_{\rm cl}$ of $\sim$15\%, however, we find a difference of more than a factor of 5. Whether our analysis indeed implies these differences between the properties of the wind structure in the LMC and SMC sample requires additional studies (of larger samples). The differences might be associated with correlations between the clumping properties and other stellar parameters (e.g., mass loss). However, a search for these dependences did not yield any convincing correlation. As the clumping properties impact the strength and shape of spectral lines in complex ways, which differ from line to line and vary as a function of stellar properties, we cannot exclude possible degeneracies between the parameters in some parts of the parameter space. We conclude that the wind properties probed here do not show convincing evidence for a dependence on metallicity.

\section{Conclusion}
\label{p2:sec:conclusion}

We have analyzed ultraviolet ULLYSES and optical X-Shooting ULLYSES spectra of 13 massive O4$-$O9.5 stars in the SMC using the model atmosphere code {\sc Fastwind} and the genetic algorithm-fitting approach {\sc Kiwi-GA}. The {\sc Kiwi-GA} approach is the only way in which as much as 15 stellar and wind properties, of which 8 characterize the outflow, may be constrained simultaneously. The study accounted in detail for clumping, porosity, and velocity-porosity properties of the wind. These processes may impact the mass-loss-sensitive spectral-line diagnostics.

Assuming single-star evolution, we found that the sample stars originate from $\sim$20$-$60\,$M_{\odot}$ stars. They are somewhat evolved, with ages of about 2$-$7\,Myr, but all still reside on the main sequence. 

Although the ULLYSES sample was selected on the basis of an absence of indications of binarity (in existing spectra), we suspect three targets in our sample (2dFS\,163, AV\,83, and AV\,70) to have composite spectra. Additionally, two stars, AV\,372 and AV\,469, show enhanced mass-loss rates compared to the rest of the sample. These two stars also show enhanced He abundances. These stars may be physically different from the rest of the sample. They might also be post-interaction binaries. 

The main focus of the study lay on the stellar wind properties. Excluding the three possible binaries, we found mass-loss rates ranging from $\log \dot{M} \,[M_{\odot}\,{\rm yr}^{-1}] = -7.65$ to $-6.02$ and terminal velocities of $\varv_{\infty} = 1650$ to 2625\,km\,s$^{-1}$. We accounted for inhomogeneities in the outflows, which we characterized with a medium that partly consisted of a spectrum of clumps that may be either optically thick or thin and that may have varying degrees of an internal velocity dispersion. Our main findings are listed below.

\begin{enumerate}
    \item[$\bullet$] The mass-loss rates of our SMC stars are consistent with the predictions of \citet{2021A&A...648A..36B} to within 0.1$-$0.2 dex. The theoretical rates of \citet{2001A&A...369..574V}, \citet{2018A&A...612A..20K}, and \citet{2022A&A...665A.133G} produce a higher mass loss, especially at a relatively low luminosity. Care should be taken with this result. 
    Brands et al. in prep., using the same modeling approach as we did for a set of LMC stars, find mass-loss rates that lie between the predictions of \citet{2001A&A...369..574V} and \citet{2018A&A...612A..20K}. This underlines that the good agreement between empirical and predicted mass loss in one part of the parameter space is not necessarily valid for the full luminosity and metallicity range.
    
    \item[$\bullet$]  Incorporating the results from \citet{2021A&A...655A..67H} and \citet[Brands et al. in prep.]{Hawcroft2024} for Galactic and LMC stars, we found that in the luminosity interval $\log L / L_{\odot} = 5.0-6.0$, the modified wind momentum $D \propto \dot{M} \varv_{\infty} R^{1/2}$ is not merely a function of metallicity, but also of luminosity. Relatively low-luminosity stars show a stronger dependence on $Z$ than high-luminosity stars. This contradicts the recent finding of \citet{2022MNRAS.511.5104M}, who reported the opposite behavior. 
    
    \item[$\bullet$] Eliminating the dependence of the terminal velocity on metallicity, we found at a luminosity $\log L / L_{\odot} = 5.75$ the same $\dot{M} \propto Z^{m}$ as in the pioneering study by \citet{2007A&A...473..603M}: $m = 1.02 \pm 0.30$ in our study versus $m = 0.83 \pm 0.16$ in theirs.
    
    \item[$\bullet$] We used a mathematical formalism to describe $D(L,Z)$ for O-stars in both the Galaxy, LMC and SMC. This recipe does not capture the intricate dependences fully and may need revision in the future, but it reveals that in particular, effort is needed to better populate the sample of low-luminosity ($\log L/L_{\odot} \la 5.6$) Milky Way stars.
    
    \item[$\bullet$] We did not identify trends of wind inhomogeneity properties with metallicity, except for a tentative finding that the clumping factor $f_{\rm cl}$ and the
    contrast between the inter-clump density and the mean density $f_{\rm ic}$ show higher values with higher metallicity. Although this is interesting, firm conclusions require larger samples that are studied in identical ways. 
\end{enumerate}

This study is part of the analysis of the ULLYSES and X-Shooting ULLYSES datasets. A full analysis of the ultraviolet and optical spectra of the $\sim$250 targets in this program will greatly further our understanding of the outflowing atmospheres of massive stars. In parallel to these studies, time-dependent hydrodynamical simulations that resolve the clumping properties are needed to assess whether our analytical prescriptions of the wind inhomogeneity properties capture the essential characteristics of the wind structure. Such simulations are now becoming available in 2D \citep{2024A&A...684A.177D} and 3D \citep{2022A&A...665A..42M}.

\begin{acknowledgements}
This publication is part of the project ‘Massive stars in low-metallicity environments: the progenitors of massive black holes’ with project number OND1362707 of the research TOP-programme, which is (partly) financed by the Dutch Research Council (NWO). FB and JS acknowledge the support of the European Research Council (ERC) Horizon Europe grant under grant agreement number 101044048. Based on observations obtained with the NASA/ESA Hubble Space Telescope, retrieved from the Mikulski Archive for Space Telescopes (MAST) at the Space Telescope Science Institute (STScI). STScI is operated by the Association of Universities for Research in Astronomy, Inc. under NASA contract NAS 5-26555. We thank SURF (www.surf.nl) for the support in using the National Supercomputer Snellius.
DP acknowledges financial support by the Deutsches Zentrum f\"ur Luft und Raumfahrt (DLR) grant FKZ 50 OR 2005. 
AACS acknowledges support by the Deutsche Forschungsgemeinschaft (DFG, German Research Foundation) in the form of an Emmy Noether Research Group -- Project-ID 445674056 (SA4064/1-1, PI Sander). AACS further acknowledge support from the Federal Ministry of Education and Research (BMBF) and the Baden-Württemberg Ministry of Science as part of the Excellence Strategy of the German Federal and State Governments.
OV acknowledges support from the KU Leuven Research Council (grant C16/17/007: MAESTRO).

\end{acknowledgements}

\bibliography{aanda} 

\appendix

\FloatBarrier

\section{Comparison with other works} \label{p2:sec:other_works_comparison}

\begin{figure}
    \centering
    \includegraphics[width=\columnwidth]{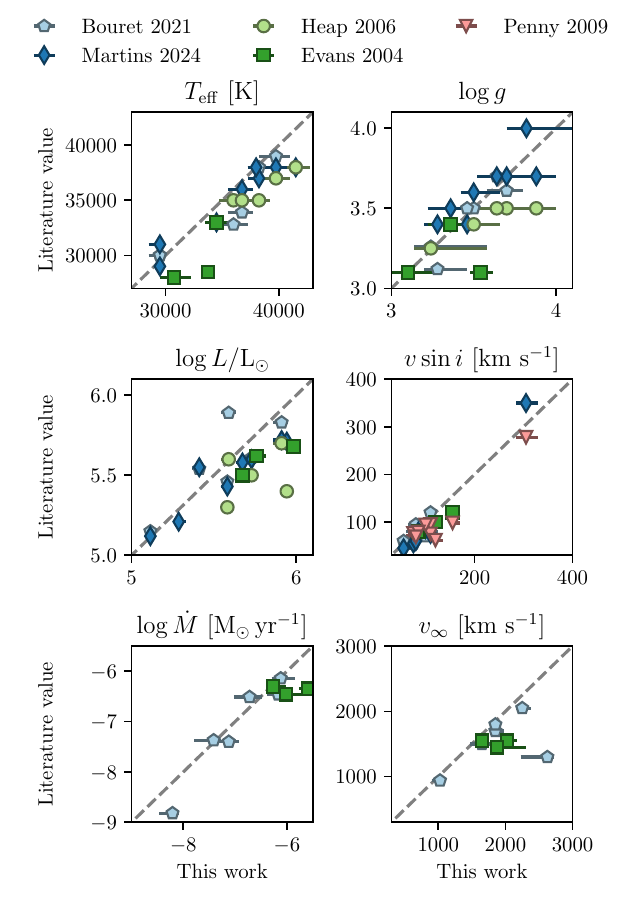}
    \caption{Best fit parameter values compared to \citet{2021A&A...647A.134B,2024A&A...689A..31M,2004ApJ...610.1021E,2006ApJ...638..409H,2009ApJ...700..844P}. The horizontal axis shows the values found in this work, and the vertical axis the parameter values found in the literature. }
    \label{p2:fig:parameter_comparison_literature}
\end{figure}

Here we compare the best fit parameter values of individual stars to those found by \citet{2021A&A...647A.134B,2024A&A...689A..31M,2004ApJ...610.1021E,2006ApJ...638..409H,2009ApJ...700..844P}. 
\cref{p2:fig:parameter_comparison_literature} shows the parameter values they found against the ones found here. All studies, except for \citet{2009ApJ...700..844P}, use optical and UV spectroscopy. \citet[XShootU V]{2024A&A...689A..31M} study a large sample of ULLYSES targets in order to determine their surface abundances. Their sample and our sample overlap with 8 stars, AV\,80, AV\,15, AV\,95, AV\,207, AV\,69, AV\,469, AV\,307 and AV\,327. Since the focus lies on stellar abundances their study does not include mass-loss properties, but does include effective temperature, surface gravity, luminosity, rotation, and abundances. Additionally, \citet{2021A&A...647A.134B} studies the fundamental stellar parameters of a sample of O-type giants and super giants including mass-loss rate. In a similar study \citet{2004ApJ...610.1021E} determine the atmosphere properties OB supergiants in the SMC including winds. Both take clumping into account adopting a volume filling fraction of 0.1. \citet{2006ApJ...638..409H} derive stellar parameters, but exclude mass loss. \citet{2009ApJ...700..844P} determined rotation velocities of O-type stars using FUSE data. 
In general \cref{p2:fig:parameter_comparison_literature} shows that we find slightly higher temperatures than the other studies. This is consistent with the results of \citet{2024A&A...689A..30S} who compare different analysis approaches to the same data. They find the same method as applied here to result in higher temperatures than other combined optical and UV analyses. These higher temperatures then also gives rise to a higher luminosity, however, a larger scatter is visible. The derived surface gravity values are mostly consistent between the different works. The projected rotation velocities are mostly consistent, however a slight trend with higher values in this work. This could be the result of excluding a separate macro turbulent velocity in our broadening. The mass-loss rates determined here are consistent with those found by \citet{2021A&A...647A.134B} and \citet{2004ApJ...610.1021E}, except for AV\,70 and AV\,307. The latter has a very low mass-loss rate, possibly hindering accurate determination. AV\,70, for which we find the highest mass-loss rate in the sample, is considered a poor fit and likely does not represent the physical properties of the star. We find terminal velocities consistent with or higher than the other studies. The most significant outlier, AV\,307, has a low mass-loss rate and therefore more challenging terminal velocity determination.

\subsection{SMC mass-loss rate} \label{p2:sec:more_SMC_Mdot}

\begin{figure}
    \centering
    \includegraphics[width=\columnwidth]{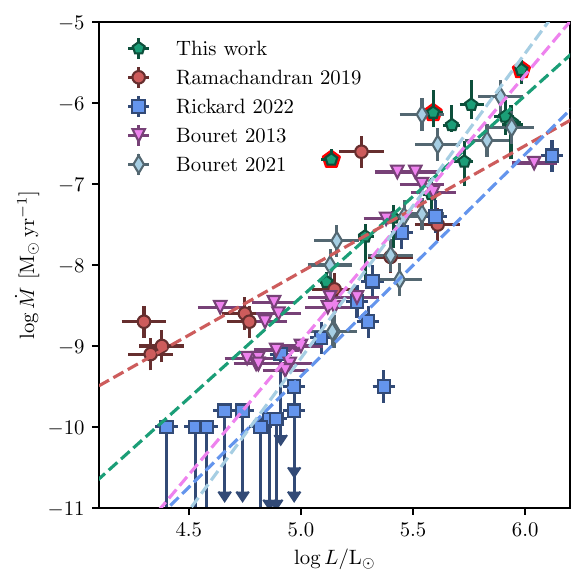}
    \caption{Mass-loss rate plotted against luminosity for SMC stars from \citet{2019A&A...625A.104R}, \citet{2022A&A...666A.189R}, and this work. The dotted line indicates a linear fit to the data points. The upper limits from \citet{2022A&A...666A.189R} are not considered in the linear fit, neither are stars with a luminosity $\log L/$L$_\odot < 5$ from \citet{2013A&A...555A...1B}. Stars with poor fits, outlined in red here, are not included in the linear fits. }
    \label{p2:more_SMC_stars}
\end{figure}

Here we compare the mass-loss rates of our sample of SMC stars and compare them to other optical and UV mass-loss rate determinations. \cref{p2:more_SMC_stars} shows the mass-loss rate as function of luminosity from \citet{2013A&A...555A...1B,2019A&A...625A.104R,2021A&A...647A.134B} and \citet{2022A&A...666A.189R}, as well as this work. \citet{2019A&A...625A.104R} and \citet{2022A&A...666A.189R} studied the mass-loss rate of stars in the SMC wing and in SMC cluster NGC 346. Their samples are diverse, but mostly dominated by dwarf stars and stars with lower luminosities than in the sample studied here. \citet{2013A&A...555A...1B} studied a sample of exclusively O-type dwarf stars. In their sample some stars with low luminosity have no significant wind features. For those stars, the values are to be considered upper limits. Similar to this work, \citet{2021A&A...647A.134B} studied a sample of O-type giants and supergiants. \citeauthor{2019A&A...625A.104R} finds a significantly more shallow mass-loss luminosity relation, which is mainly determined by the low-luminosity stars in their sample. \citeauthor{2022A&A...666A.189R} found a very similar slope of the mass-loss rate luminosity relation, but at lower mass-loss rate for at any luminosity. They found two stars with significantly lower mass-loss rates than expected given their luminosity. The bulk of their sample lies just below the mass-loss rates inferred here. \citeauthor{2013A&A...555A...1B} find a similar slopes in both works, which are steeper than the one found here. In the luminosity range studied in this work, the mass-loss rates are comparable to the ones studied in this sample. Overall the mass-loss rates appear consistent, but with a scatter of $\sim$1 dex, at $\log L / $L$_\odot > 5$, at lower luminosity mainly upper limits are found, with a larger scatter. 

\FloatBarrier

\section{CNO abundances} \label{p2:sec:C_N_O_comp}

\begin{figure}
    \centering
    \includegraphics[width=\columnwidth]{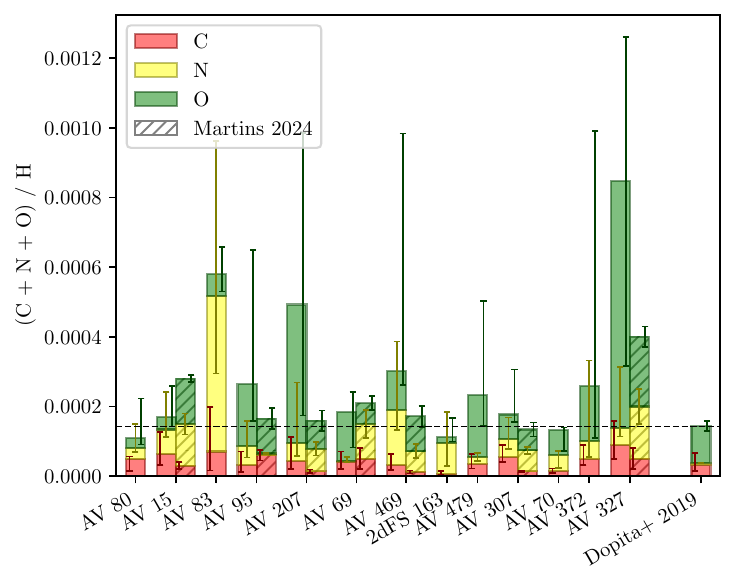}
    \caption{Total carbon, nitrogen, and oxygen abundances for each star, broken down in the separate components. In each bar the uncertainties of the carbon (left), nitrogen (middle), and oxygen abundances are shown. Stars from \citet{2024A&A...689A..31M} that overlap with our sample are shown in the hatched bars. The horizontal dashed line shows the total baseline abundance from \citet{2019AJ....157...50D}, which is also broken down in the separate components on the right. }
    \label{p2:fig:CNO_abundances}
\end{figure}

During the evolution of a star the surface abundances of carbon, nitrogen, and oxygen may change due to the CNO-cycle and internal mixing. However, in this process the total amount of C, N, and O is expected to remain constant at the total initial value. \cref{p2:fig:CNO_abundances} shows the total C + N + O surface abundance relative to hydrogen, broken down in their separate components. Additionally, the derived abundances of \citet[XShootU V]{2024A&A...689A..31M} are included, as well as the baseline abundances from \citet{2019AJ....157...50D}. Here, \citeauthor{2024A&A...689A..31M} used the same dataset from XshootU. Generally, our total abundance lies above the baseline abundance. This is mostly dominated by the high and uncertain oxygen abundances we find. Additionally, we assume a fixed micro turbulent velocity of $\varv_{\rm mic} = 15$\,km\,s$^{-1}$, which may affect the abundance determination, including the He abundance. In all stars we find a nitrogen enhancement, except for AV\,69 which is consistent with the baseline abundance. Within uncertainties our total abundance agrees with the values found by \citet{2024A&A...689A..31M}, however, we do find larger scatter and uncertainties. This is likely due to our analysis not being tailored to abundance determination and the possible correlation between abundances and other stellar parameters.

\section{Spectroscopic and evolutionary mass} \label{p2:sec:mevo_mspec}

\begin{figure}
    \centering
    \includegraphics[width=\columnwidth]{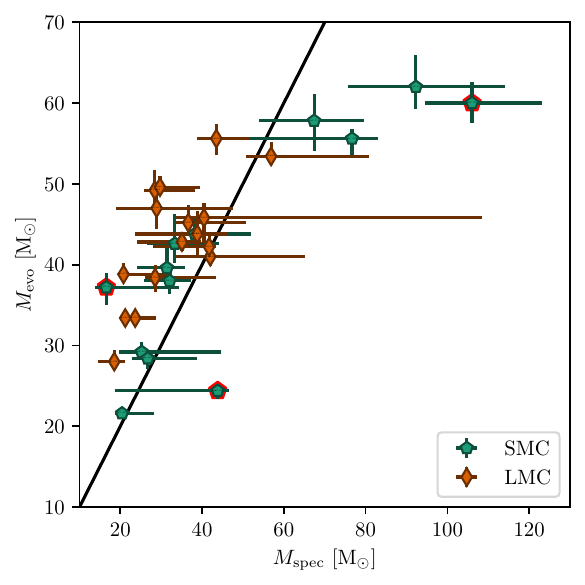}
    \caption{Spectroscopic mass plotted against the evolutionary mass. The solid black line indicates the where the two masses are equal. The LMC stars are from Brands et al. in prep. Stars with unreliable stellar parameters are marked with a red border. }
    \label{p2:fig:mass_discrepancy}
\end{figure}

The mass of a star can be determined based on their surface gravity and radius. This is referred to as the spectroscopic mass, $M_{\rm spec}$. Another way of determining the mass is though their position on the HRD and comparing that to evolutionary models. This gives the evolutionary mass, $M_{\rm evo}$. A discrepancy between these two types of masses has been observed \citep[e.g.,][]{1992A&A...261..209H}. \cref{p2:fig:mass_discrepancy} shows these masses for our sample and the sample of Brands et al. in prep. A clear discrepancy between the two masses remains visible, with lower spectroscopic masses below $\sim$50\,M$_\odot$, and higher spectroscopic masses above that threshold. This is consistent with the discrepancy found by \citet{2018A&A...613A..12M}, who report a similar trend, but a lower "transition" mass of $\sim$40\,M$_\odot$. We do note that our evolutionary masses, determined with {\sc Bonnsai}, see \cref{p2:sec:bonnsai}, are limited to $M < 60$\,M$_\odot$, and that for stars with higher masses evolutionary tracks of LMC metallicity are used. This could potentially affect the evolutionary masses, increasing the discrepancy.

\section{Challenges in the spectral fitting} \label{p2:sec:struggles}

Our spectral fitting relies on 55 spectral lines and line complexes, an overview of which is given in Table~\ref{p2:tab:diagnostics}. To add to our understanding of quantitative spectroscopy of hot massive stars, we discuss some notable complications. 

\paragraph{\ion{N}{v} 1240} 
The \ion{N}{V}\,1240 resonance line, located in the wing of the Ly$\alpha$ absorption, is strong and shows a P-Cygni profile for most stars. 
However, it requires X-ray emission from wind-embedded shocks to have \ion{N}{V} abundant enough to reproduce the observed profiles in our sample.
X-ray emission is implemented in {\sc Fastwind} \citep{2016A&A...590A..88C}, however the exact amount of emergent emission as well as the shock temperature remain uncertain \citep{2022MNRAS.515.4130C}. As \ion{N}{v} is a trace ion (and the diagnostic a resonance line), the line profile is highly sensitive to the X-ray radiation field and, hence, difficult to predict reliably. 
Test calculations in which we included the line in the analysis resulted in systematically higher nitrogen abundances (for more details see Brands et al. in prep.). We therefore opted to exclude it.

\paragraph{\ion{He}{i} singlets}

In our modeling, we consider both singlet and triplet transitions of \ion{He}{i}.
\citet{2006A&A...456..659N} point out the importance of accurately predicting the radiation field near the reasonance transition $1s^{2}\,^{1}$S$-2p\,^{1}P^{\circ}$ (at 584\,\AA) in order to correctly reproduce \ion{He}{i} 4387, 4922 and 6678\,\AA, lines that have $2p\,^{1}P^{\circ}$ as their lower level. In modeling efforts, however, the radiation
field near the \ion{He}{i} resonance transition may be uncertain due to assumptions on line-blanketing and turbulent velocity. Specifically, the \ion{Fe}{iv} ion has two transitions near to 584\,\AA\ that may affect the pumping efficiency of the resonance line, hence may impact the diagnostic \ion{He}{i} singlet lines.
Despite this, we find that in almost all cases the singlet lines are well reproduced. For AV\,80 and AV\,83 the singlet fits are of relatively poor quality; the triplet lines in these sources fit somewhat better.

\paragraph{The \ion{P}{v} problem}
\citet{2006ApJ...637.1025F} signal a discrepancy between mass-loss rates determined from the \ion{P}{v}\,1118,1128 resonance doublet and those determined from H$\alpha$, termed the '\ion{P}{v} problem'.
The \ion{P}{v} doublet would systematically result in lower mass-loss rates. The authors suggest the discrepancy can be solved by introducing clumping in the outflow. Such clumps increase the strength of the H$\alpha$ recombination line while the \ion{P}{v} line is not affected by changes in density directly, though it may be through changes in ionization as a result of the higher density in the clumps \citep[see][]{2007A&A...476.1331O}.
\citet{2013A&A...559A.130S} successfully reproduce both the \ion{P}{v} lines and H$\alpha$ line with the same mass-loss rate by including micro clumping (analogous to $f_{\rm cl}$) and macro clumping, the latter being a measure for the porosity of the wind.

Despite our relatively sophisticated and flexible prescription of wind inhomogeneities and leaving the phosphorus abundance a free parameter, we do not get satisfactory fits to the \ion{P}{v} lines for part of our sample. This may suggest the applied clumping prescription fails to capture intricacies of wind structure key for the formation of \ion{P}{v} or some other cause. Though, often we find wind features in the model, while for all stars we observe photospheric line profiles. Issues with photospheric line profiles are unlikely to be resolved with wind clumping properties. Possibly the problem is connected to the poor constraints on the phosphorous abundance.

\section{Optical results} \label{p2:sec:optical_appendix}

Here we briefly discuss the optical-only fits and how they compare to the optical + UV fits. Table\,\ref{p2:tab:optical_only} lists the best fit values of the free parameters used in the optical-only fits. Here we fixed $\beta=1, f_{\rm cl}=10, f_{\rm ic}=0.1, f_{\rm vel} = 0.5, \varv_{\rm cl, start}=0.05, \varv_{\rm windturb}=0.1$, and $\varv_\infty$ from \citet{2024A&A...688A.105H}. Generally, the optical-only fits agree well with the fits using the full set of diagnostics. The most significant deviations are found in the lower mass-loss rates, with the optical-only fits resulting in significantly higher values at the low mass loss end. The most extreme example of this is AV\,307, where the mass-loss rate is 1.3 dex higher in the optical only fit. The available diagnostics in the optical are not well suited for constraining low rates, as the sensitivity of the line profiles to the mass-loss rate drops significantly. The mass-loss rate needs to increase a significant amount to compensate a small filling in of the H$\alpha$ line profile. The wind diagnostics in the UV remain sensitive to lower values of the mass-loss rate. This indicates the importance of the UV diagnostic features in constraining low mass-loss rates.

\begin{table*}
\centering
\caption{Best fit parameters and uncertainties of optical-only fits. Rows with gray text indicate parameter values that are likely not representative of the stellar properties. }
\small
\def\arraystretch{1.5}
\label{p2:tab:optical_only}
\begin{tabular}{lccccccccc} \hline \hline
Source & $T_{\rm eff}$ [K] & $\log g$ & $\varv\sin i$ [km s$^{-1}$] & $\log \dot{M}$ & $y_{\rm He}$ & $\epsilon_{\rm C}$ & $\epsilon_{\rm N}$ & $\epsilon_{\rm O}$ & $\epsilon_{\rm Si}$\\ \hline
AV 80 & $43250^{+500}_{-250}$ & $3.84^{+0.12}_{-0.06}$ & $305^{+25}_{-20}$ & $-6.14^{+0.08}_{-0.05}$ & $0.08^{+0.02}_{-0.01}$ & $8.2^{+0.1}_{-0.8}$ & $7.5^{+0.4}_{-0.4}$ & $7.0^{+0.7}_{-0.8}$ & $6.4^{+0.7}_{-0.4}$\\
AV 15 & $40500^{+750}_{-750}$ & $3.64^{+0.08}_{-0.06}$ & $110^{+10}_{-10}$ & $-6.07^{+0.08}_{-0.05}$ & $0.10^{+0.01}_{-0.01}$ & $7.5^{+0.3}_{-0.6}$ & $7.8^{+0.2}_{-0.2}$ & $6.1^{+1.1}_{-0.1}$ & $6.3^{+0.5}_{-0.3}$\\
AV 95 & $37750^{+1000}_{-250}$ & $3.60^{+0.08}_{-0.04}$ & $75^{+10}_{-10}$ & $-6.50^{+0.03}_{-0.08}$ & $0.14^{+0.01}_{-0.02}$ & $7.4^{+0.1}_{-0.5}$ & $7.8^{+0.1}_{-0.1}$ & $8.2^{+0.2}_{-0.8}$ & $6.7^{+0.2}_{-0.7}$\\
AV 207 & $38250^{+750}_{-750}$ & $3.86^{+0.06}_{-0.06}$ & $110^{+10}_{-15}$ & $-7.02^{+0.13}_{-0.10}$ & $0.12^{+0.01}_{-0.03}$ & $7.5^{+0.2}_{-0.8}$ & $7.9^{+0.1}_{-0.3}$ & $8.4^{+0.4}_{-0.6}$ & $6.5^{+0.4}_{-0.5}$\\
AV 69 & $36750^{+1500}_{-250}$ & $3.48^{+0.12}_{-0.06}$ & $100^{+10}_{-15}$ & $-6.47^{+0.08}_{-0.05}$ & $0.09^{+0.03}_{-0.01}$ & $7.5^{+0.4}_{-0.5}$ & $6.7^{+0.2}_{-0.7}$ & $6.2^{+1.1}_{-0.3}$ & $6.8^{+0.3}_{-0.8}$\\
AV 469 & $33750^{+1000}_{-1000}$ & $3.24^{+0.12}_{-0.08}$ & $85^{+10}_{-10}$ & $-6.00^{+0.05}_{-0.08}$ & $0.20^{+0.03}_{-0.04}$ & $7.3^{+0.4}_{-0.4}$ & $8.3^{+0.1}_{-0.3}$ & $8.0^{+0.7}_{-0.2}$ & $7.2^{+0.4}_{-0.3}$\\
AV 479 & $33750^{+250}_{-750}$ & $3.34^{+0.04}_{-0.10}$ & $90^{+15}_{-5}$ & $-6.04^{+0.05}_{-0.08}$ & $0.13^{+0.04}_{-0.01}$ & $7.5^{+0.2}_{-0.2}$ & $7.5^{+0.4}_{-0.2}$ & $7.8^{+0.5}_{-1.0}$ & $6.9^{+0.4}_{-0.2}$\\
AV 307 & $29250^{+250}_{-1750}$ & $3.42^{+0.06}_{-0.18}$ & $55^{+15}_{-10}$ & $-6.88^{+0.08}_{-0.28}$ & $0.11^{+0.01}_{-0.03}$ & $7.5^{+0.4}_{-0.3}$ & $7.8^{+0.5}_{-0.3}$ & $8.0^{+0.2}_{-0.5}$ & $6.7^{+0.5}_{-0.3}$\\
AV 372 & $29750^{+1250}_{-500}$ & $2.88^{+0.14}_{-0.06}$ & $155^{+15}_{-5}$ & $-5.91^{+0.05}_{-0.05}$ & $0.17^{+0.03}_{-0.03}$ & $7.7^{+0.3}_{-0.3}$ & $7.6^{+0.3}_{-0.2}$ & $8.3^{+0.7}_{-2.2}$ & $7.3^{+0.2}_{-0.5}$\\
AV 327 & $31750^{+750}_{-1000}$ & $3.26^{+0.10}_{-0.08}$ & $80^{+10}_{-15}$ & $-6.47^{+0.10}_{-0.18}$ & $0.15^{+0.01}_{-0.04}$ & $7.2^{+0.2}_{-0.3}$ & $7.8^{+0.2}_{-0.4}$ & $7.8^{+0.5}_{-1.7}$ & $6.8^{+0.3}_{-0.2}$\\
\hline
\color{gray} AV 83 & \color{gray} $37000^{+1750}_{-1750}$ & \color{gray} $3.24^{+0.24}_{-0.12}$ & \color{gray} $80^{+45}_{-20}$ & \color{gray} $-5.87^{+0.05}_{-0.05}$ & \color{gray} $0.18^{+0.06}_{-0.05}$ & \color{gray} $7.4^{+0.3}_{-1.4}$ & \color{gray} $8.6^{+0.4}_{-0.3}$ & \color{gray} $7.0^{+1.8}_{-1.1}$ & \color{gray} $6.5^{+1.4}_{-0.6}$\\
\color{gray} 2dFS 163 & \color{gray} $33500^{+750}_{-750}$ & \color{gray} $3.54^{+0.24}_{-0.18}$ & \color{gray} $90^{+35}_{-35}$ & \color{gray} $-6.20^{+0.08}_{-0.05}$ & \color{gray} $0.08^{+0.05}_{-0.01}$ & \color{gray} $7.2^{+0.5}_{-1.2}$ & \color{gray} $7.8^{+0.6}_{-1.3}$ & \color{gray} $6.2^{+2.0}_{-0.3}$ & \color{gray} $6.2^{+0.7}_{-0.2}$\\
\color{gray} AV 70 & \color{gray} $30750^{+750}_{-1000}$ & \color{gray} $3.02^{+0.06}_{-0.08}$ & \color{gray} $120^{+15}_{-15}$ & \color{gray} $-5.75^{+0.08}_{-0.05}$ & \color{gray} $0.15^{+0.04}_{-0.01}$ & \color{gray} $7.3^{+0.3}_{-0.2}$ & \color{gray} $7.7^{+0.2}_{-0.2}$ & \color{gray} $7.0^{+1.6}_{-1.0}$ & \color{gray} $6.6^{+0.4}_{-0.2}$\\
\hline
\end{tabular}
\tablefoot{The bottom rows with gray text indicate parameter values that are likely not representative of the stellar properties due to poor fits.}
\end{table*}

\FloatBarrier

\section{Mass loss relations} \label{p2:sec:mdot_analysis}

Here we show the same analysis as Section~\ref{p2:sec:mass_loss_metallicity}, but now fitting directly the mass-loss rates, rather than the modified wind momentum. 

\begin{figure}
    \centering
    \includegraphics[width=\columnwidth]{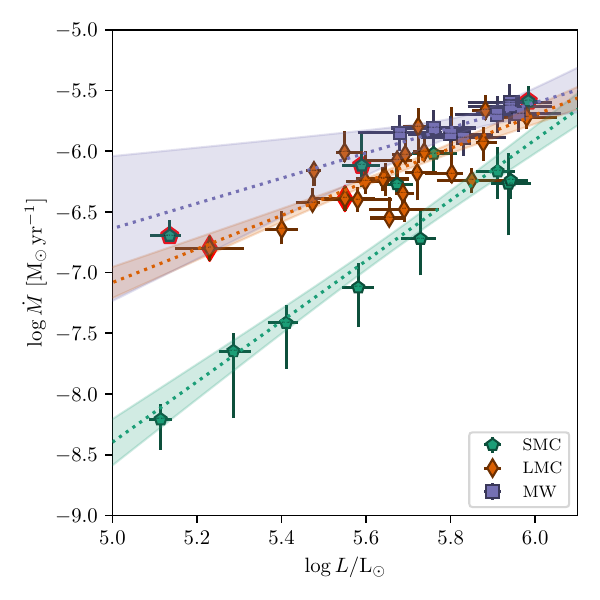}
    \caption{Mass-loss rates as function of luminosity for samples of SMC (this work), LMC \citep[Brands et al. in prep.]{Hawcroft2024}, and MW stars \citep{2021A&A...655A..67H}. The dotted lines indicate the linear fits to the data for each of the metallicities, with the shaded region the 1$\sigma$ confidence interval on the fit. Stars with poor spectral fits, marked with a red border, are left out the linear fit here. The parameters of the fit are listed in Table\,\ref{p2:tab:linear_fits_massloss}.}
    \label{p2:fig:mass_loss_relation}
\end{figure}

Figure~\ref{p2:fig:mass_loss_relation} shows the mass-loss rate as function of the luminosity for each of the samples. The dotted lines indicate linear fits to the individual samples. The slope and offsets of these fits are listed in Table~\ref{p2:tab:linear_fits_massloss}. A general trend of increased mass-loss rate for increased metallicity is visible. We fitted the linear fits at $\log L/{\rm L}_\odot = 5.75$ to determine the mass loss metallicity relation, and find $d\log \dot{M} / d \log Z = m = 1.01\pm0.26$, consistent with the wind momentum results. 

\begin{table}[]
    \centering
    \caption{Slopes and offsets of the linear fits to the mass-loss rate as function of luminosity.}
    \label{p2:tab:linear_fits_massloss}
    \begin{tabular}{ccc} \hline \hline 
    Galaxy      &   Slope            & Offset           \\ \hline
    Milky Way   &  1.04$\pm$0.70     & $-11.82\pm$4.07   \\
    LMC         &  1.49$\pm$0.19     & $-14.60\pm$1.05   \\
    SMC         &  2.48$\pm$0.26     & $-20.81\pm$1.48   \\ \hline
    \end{tabular}
\end{table}

\section{Combined metallicity and luminosity dependence} \label{p2:app:logD_cov}

\begin{figure}
    \centering
    \includegraphics[width=\columnwidth]{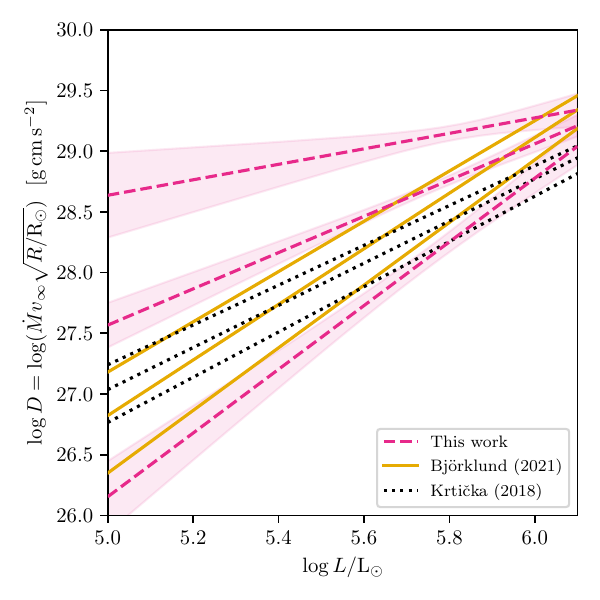}
    \caption{Empirical wind momentum relation with luminosity and metallicity compared to the theoretical predictions. The displayed metallicities are Solar, half Solar, and one fifth Solar, from top to bottom for each relation. }
    \label{p2:fig:MLZ_with_predictions}
\end{figure}

Here we supply additional information on the combined fit to the modified wind momentum using both a metallicity and luminosity dependence simultaneously. \cref{p2:fig:MLZ_with_predictions} shows the fit as displayed in \cref{p2:fig:DMZ} together with the theoretical predictions of \citep{2018A&A...612A..20K} and \citep{2021A&A...648A..36B}. The coefficients of wind momentum relations are listed in \cref{p2:tab:logD-LZ_params}. Most notably the theoretical predictions lie much closer together at low luminosity. The Solar metallicity wind momentum we find at low metalicity is significantly higher than the predictions. This is likely due to a lack of low luminosity stars in our sample. Towards higher luminosities of 10$^6$\,L$_\odot$ the predictions and observations converge.

The covariance matrix of the fit of Equation~\ref{p2:eq:DMZ} to the empirically determined modified wind momentum rates is
\begin{equation}
Cov(a, b, c, d) = 
\begin{bmatrix}
  0.18 & 0.34 & 0.06 & 0.03\\
  0.34 & 0.90 & 0.17 & 0.06\\
  0.06 & 0.17 & 0.06 & 0.02\\
  0.03 & 0.06 & 0.02 & 0.01\\
\end{bmatrix}.
\end{equation}

\section{Fit overview}

 \begin{sidewaysfigure*}
  \centering
  \includegraphics[width=1\textwidth]{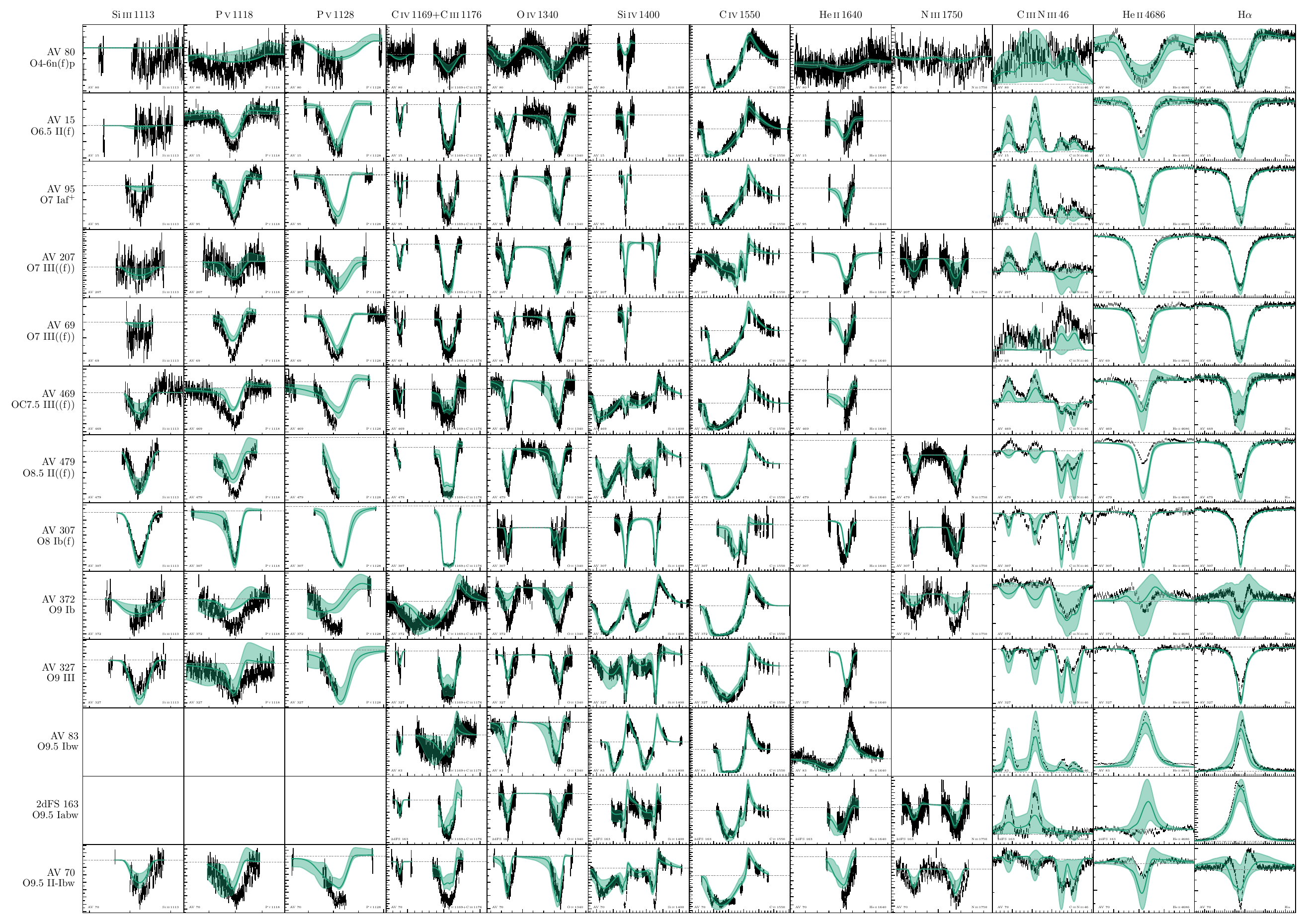}
  \caption{Overview of the best fits of the UV and wind sensitive optical lines. The normalized observed data is plotted as black vertical bars with the size of the bar indicating the 1~$\sigma$ uncertainty on the data. The green line indicates the best-fit model and the shaded region is the 1~$\sigma$ confidence interval of the models. The thin dashed grey lines are at unity. The major ticks on the x-axis indicate steps of 5\,\AA, and the minor ticks indicate steps of 1\,\AA. On the y-axis the major ticks indicate steps of 10\% of the continuum and the minor ticks indicate 5\%. A more detailed figure of the line profiles can be found online \url{https://doi.org/10.5281/zenodo.13929105} and in Appendix\,\ref{p2:app:fit_summaries}. }
  \label{p2:fig:wind_lines_overview}
 \end{sidewaysfigure*}

 \begin{sidewaysfigure*}
  \centering
  \includegraphics[width=1\textwidth]{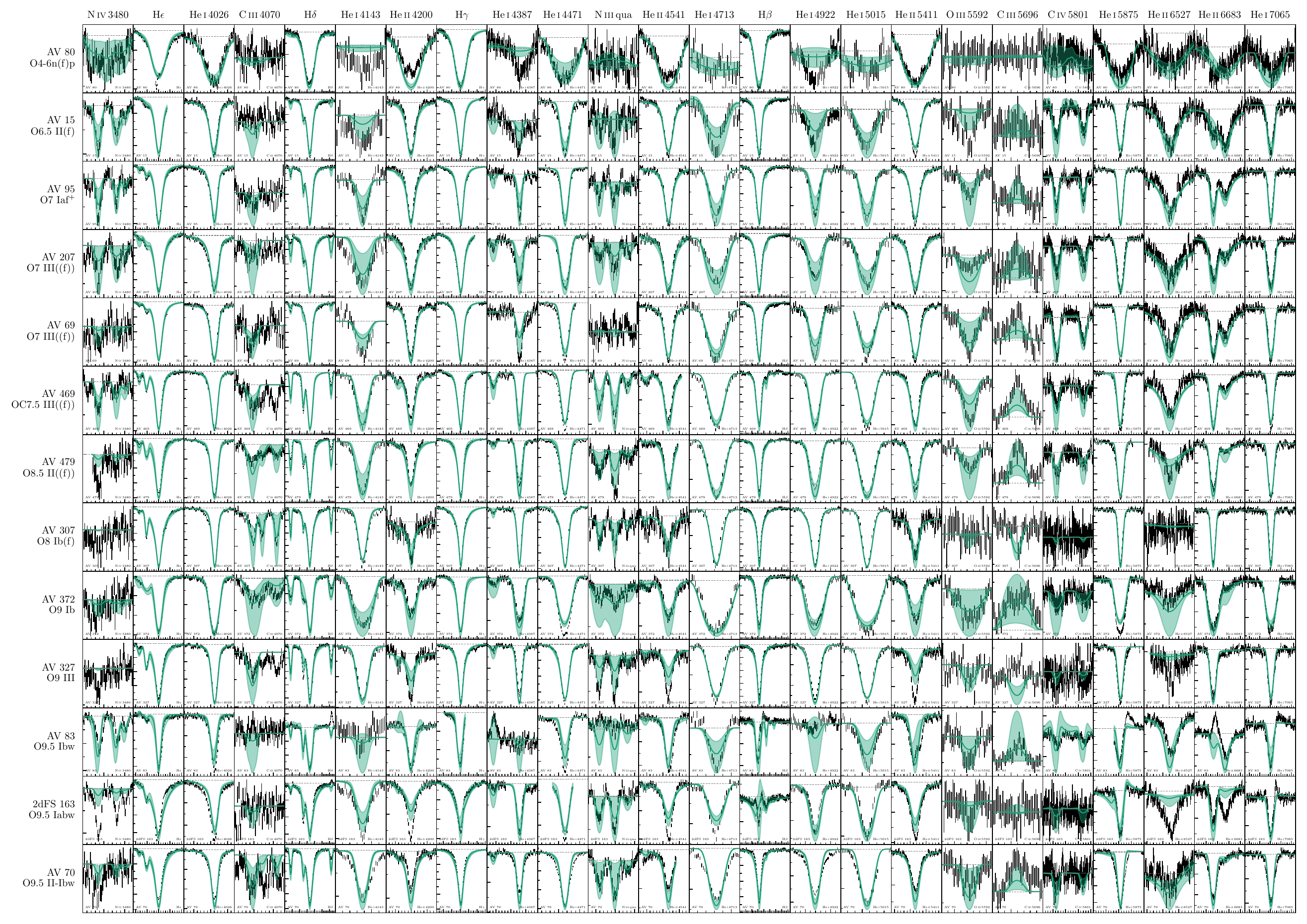}
  \caption{
  As Figure\,\ref{p2:fig:wind_lines_overview}, but for optical lines of mostly photospheric origin.}
  \label{p2:fig:photospheric_lines_overview}
 \end{sidewaysfigure*}

\section{Fit summaries}\label{p2:app:fit_summaries}

\begin{figure*}
   \centering
   \includegraphics[width=0.85\textwidth]{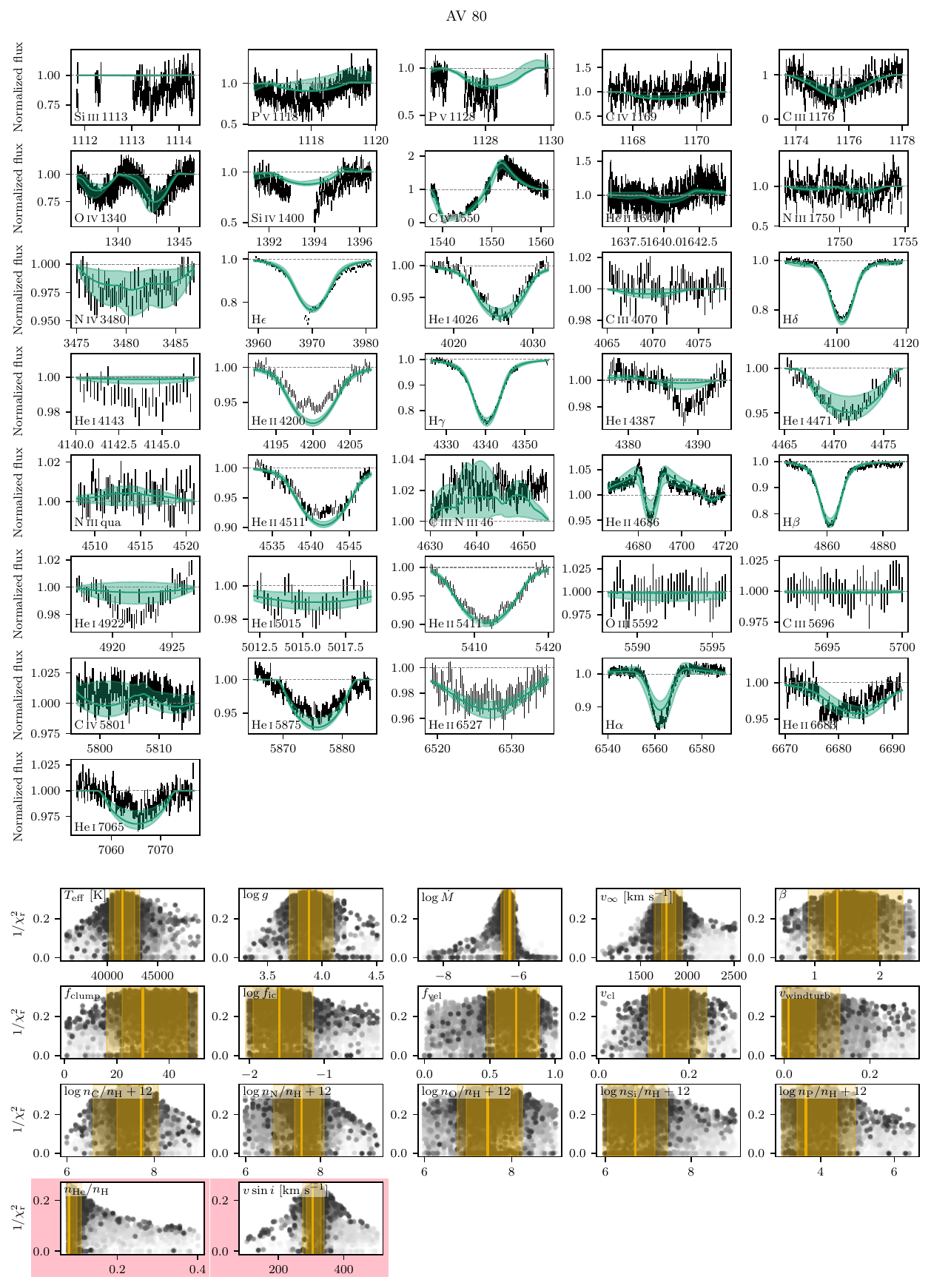}
   \caption{Overview of the fitting results of AV\,80. The \textit{top} part shows the line profiles included in the fitting process, with the name of each feature indicated in the bottom left. The green line shows the best fit, with the shaded region showing the 1$\sigma$ uncertainty on the model fit. The black vertical bars indicate the observed flux, with the length indicating the uncertainty. The horizontal axis shows the wavelength in \AA. The \textit{bottom} part shows the distribution of $1/\chi^2_{\rm r}$ for each parameter (indicated on the top left), with $\chi^2_{\rm r}$ the reduced $\chi^2$ value. Each scatter point indicates one {\sc Fastwind} model calculated by Kiwi-GA. The color indicates the generation in which the model was computed, with light gray the first generation and black the last. The vertical yellow line indicates the best fit value for each parameter, and the shaded regions are the 1 and 2 $\sigma$ confidence intervals. The distributions in the red shaded area are from the optical only GA fit. }
   \label{p2:fig:fit_summary}
\end{figure*}

\begin{figure*}
    \centering
    \includegraphics[width=0.95\textwidth]{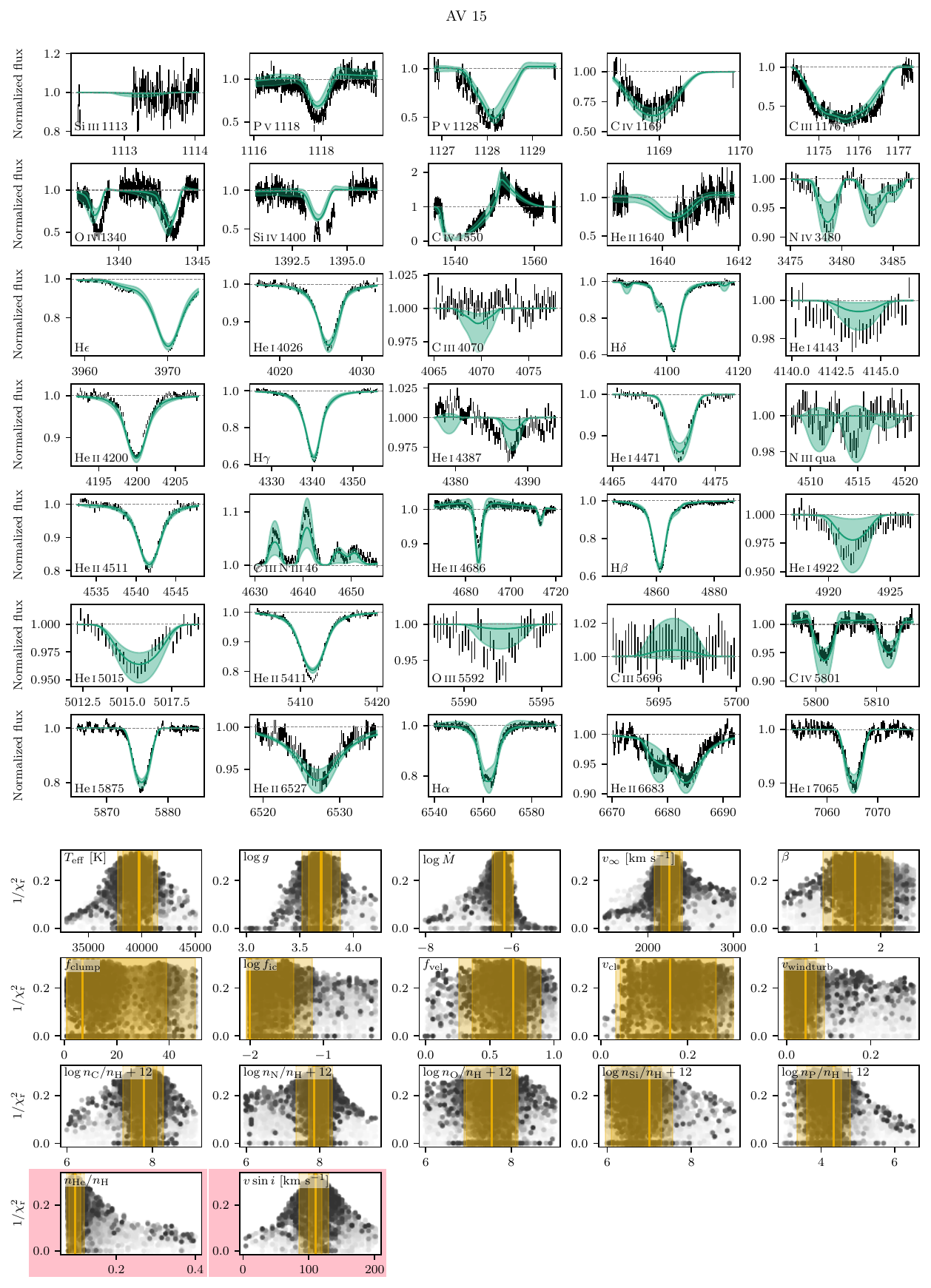}
    \caption{Same as Figure\,\ref{p2:fig:fit_summary}, but for AV\,15.}
\end{figure*}

\begin{figure*}
    \centering
    \includegraphics[width=0.95\textwidth]{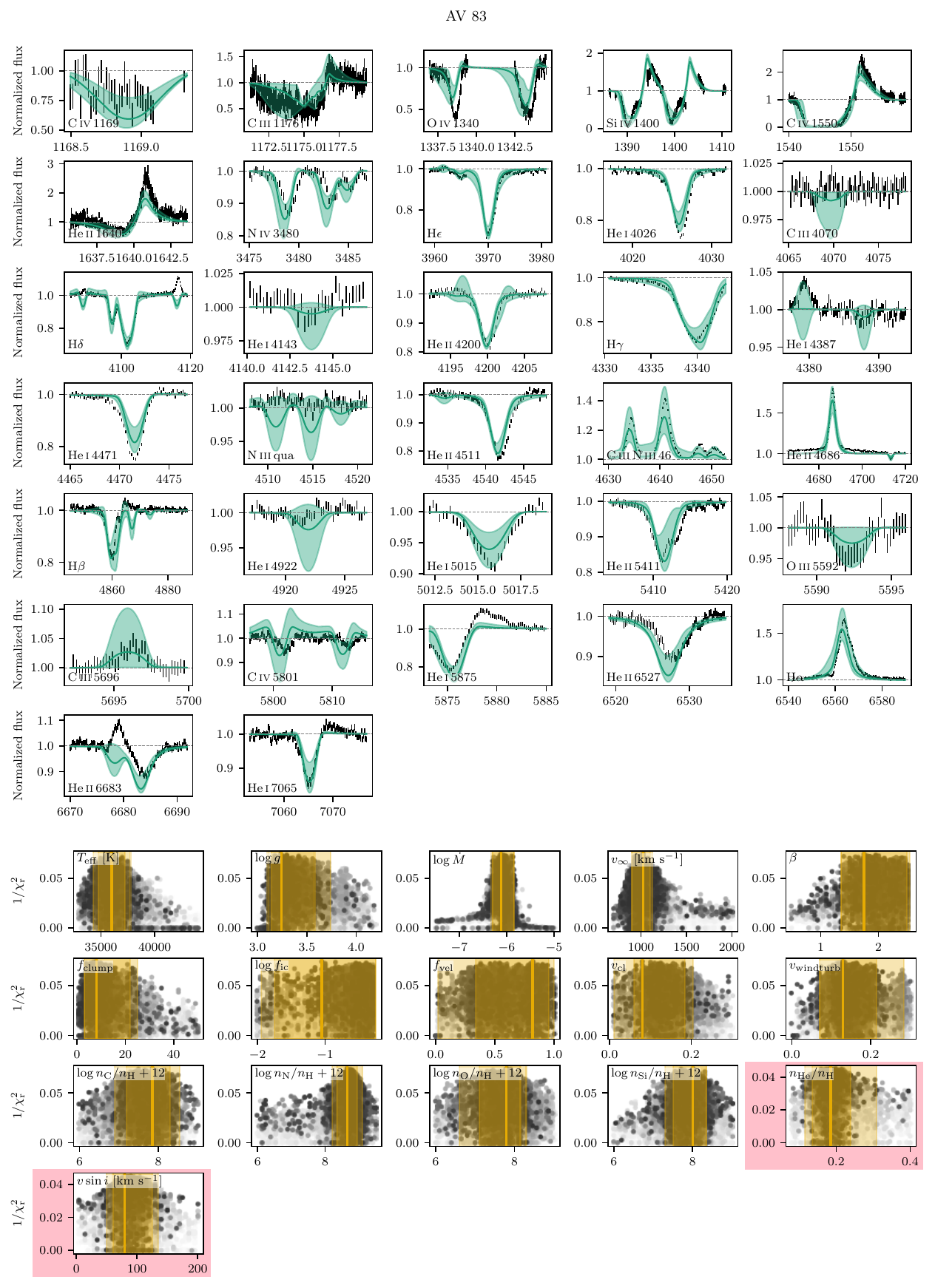}
    \caption{Same as Figure\,\ref{p2:fig:fit_summary}, but for AV\,83.}
\end{figure*}

\begin{figure*}
    \centering
    \includegraphics[width=0.95\textwidth]{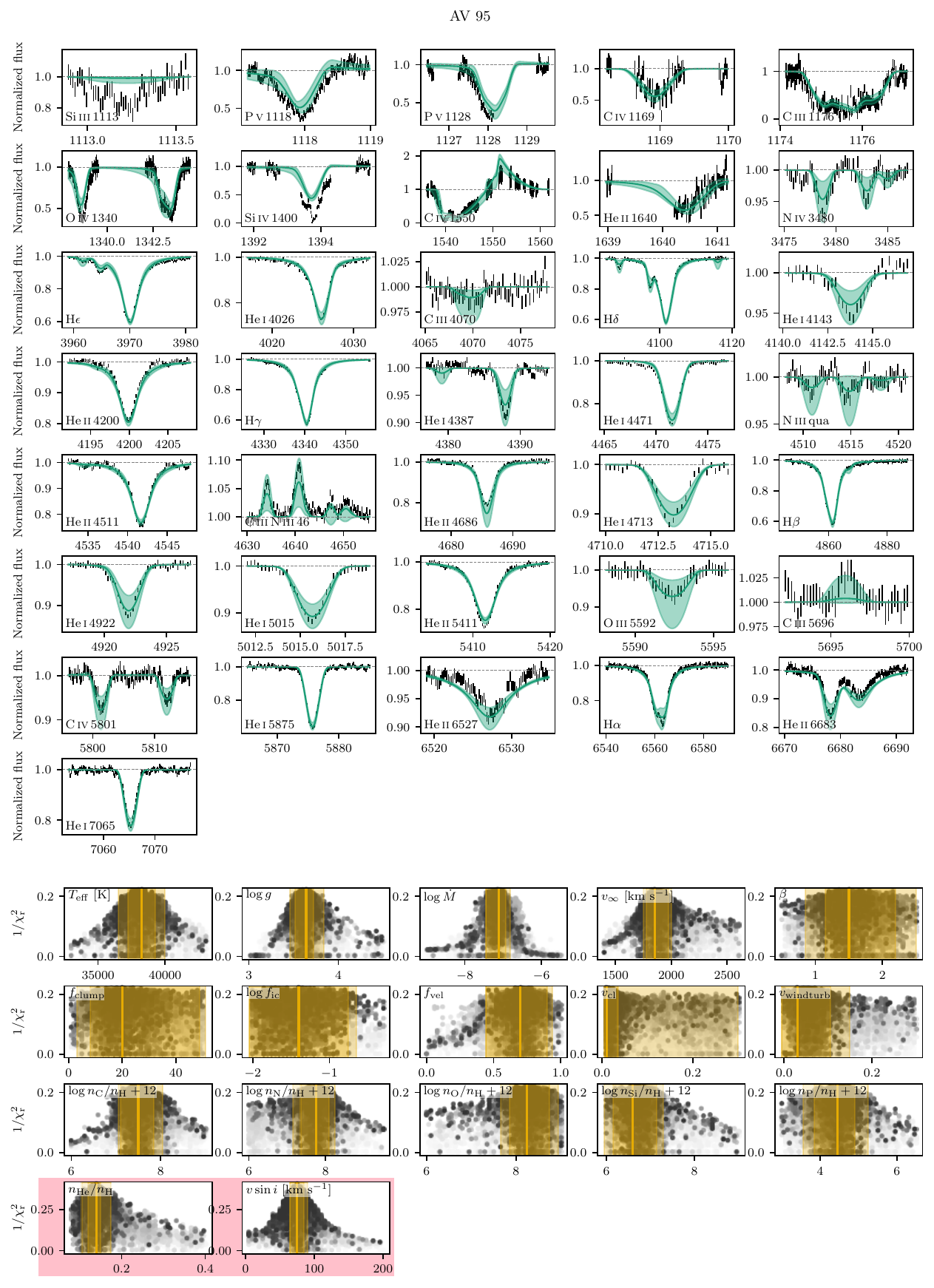}
    \caption{Same as Figure\,\ref{p2:fig:fit_summary}, but for AV\,95.}
\end{figure*}

\begin{figure*}
    \centering
    \includegraphics[width=0.95\textwidth]{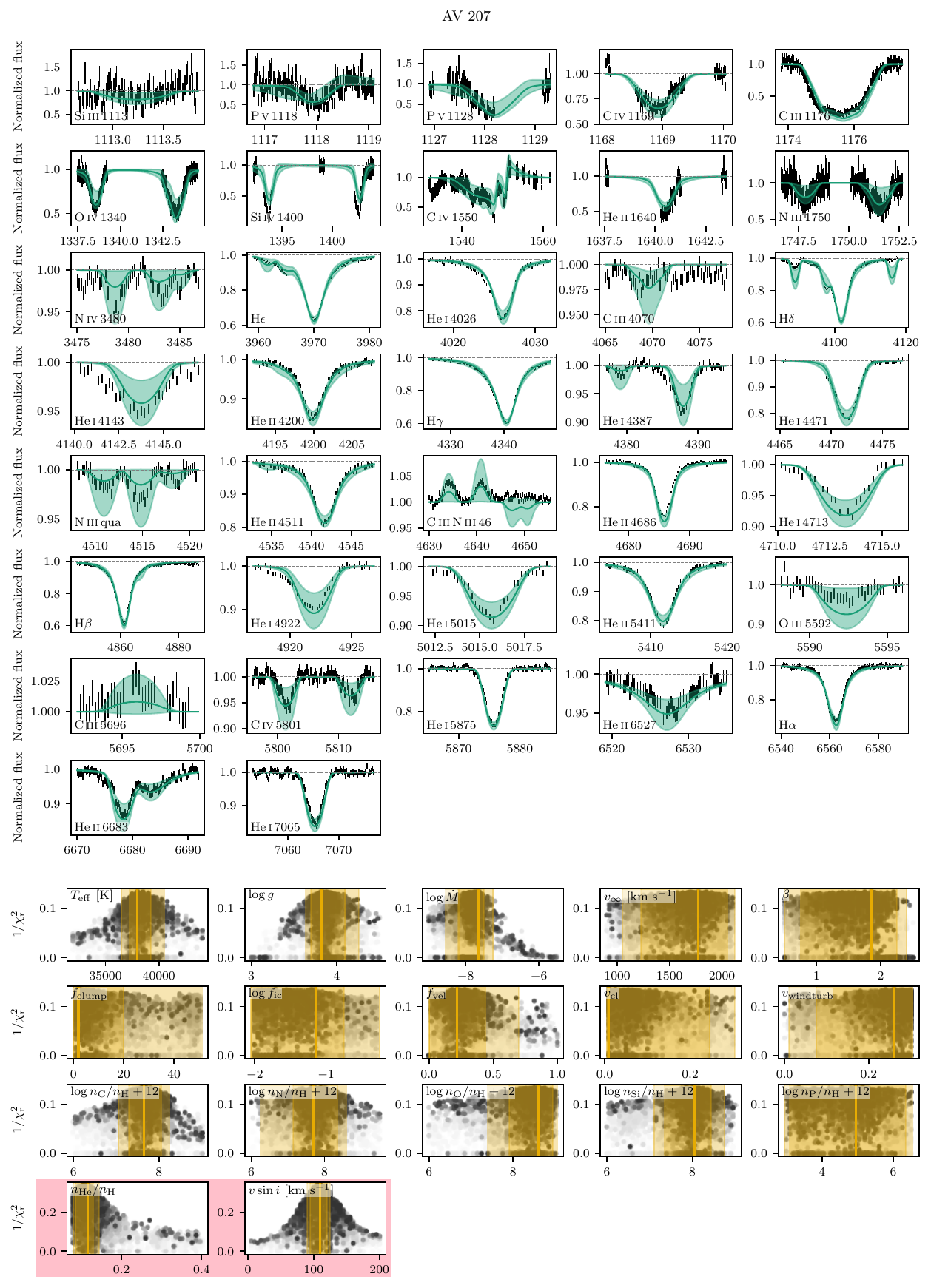}
    \caption{Same as Figure\,\ref{p2:fig:fit_summary}, but for AV\,207.}
\end{figure*}

\begin{figure*}
    \centering
    \includegraphics[width=0.95\textwidth]{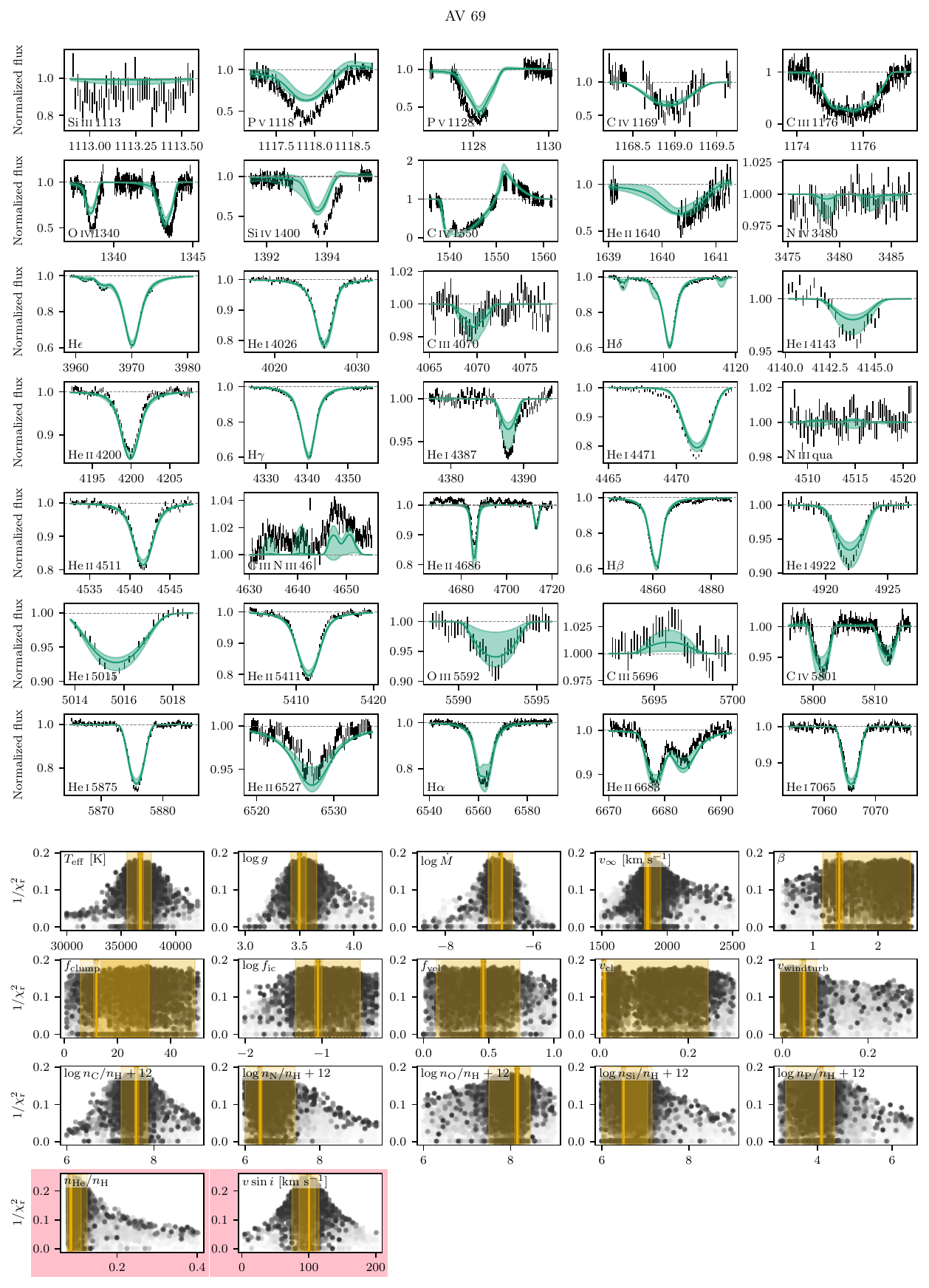}
    \caption{Same as Figure\,\ref{p2:fig:fit_summary}, but for AV\,69.}
\end{figure*}

\begin{figure*}
    \centering
    \includegraphics[width=0.95\textwidth]{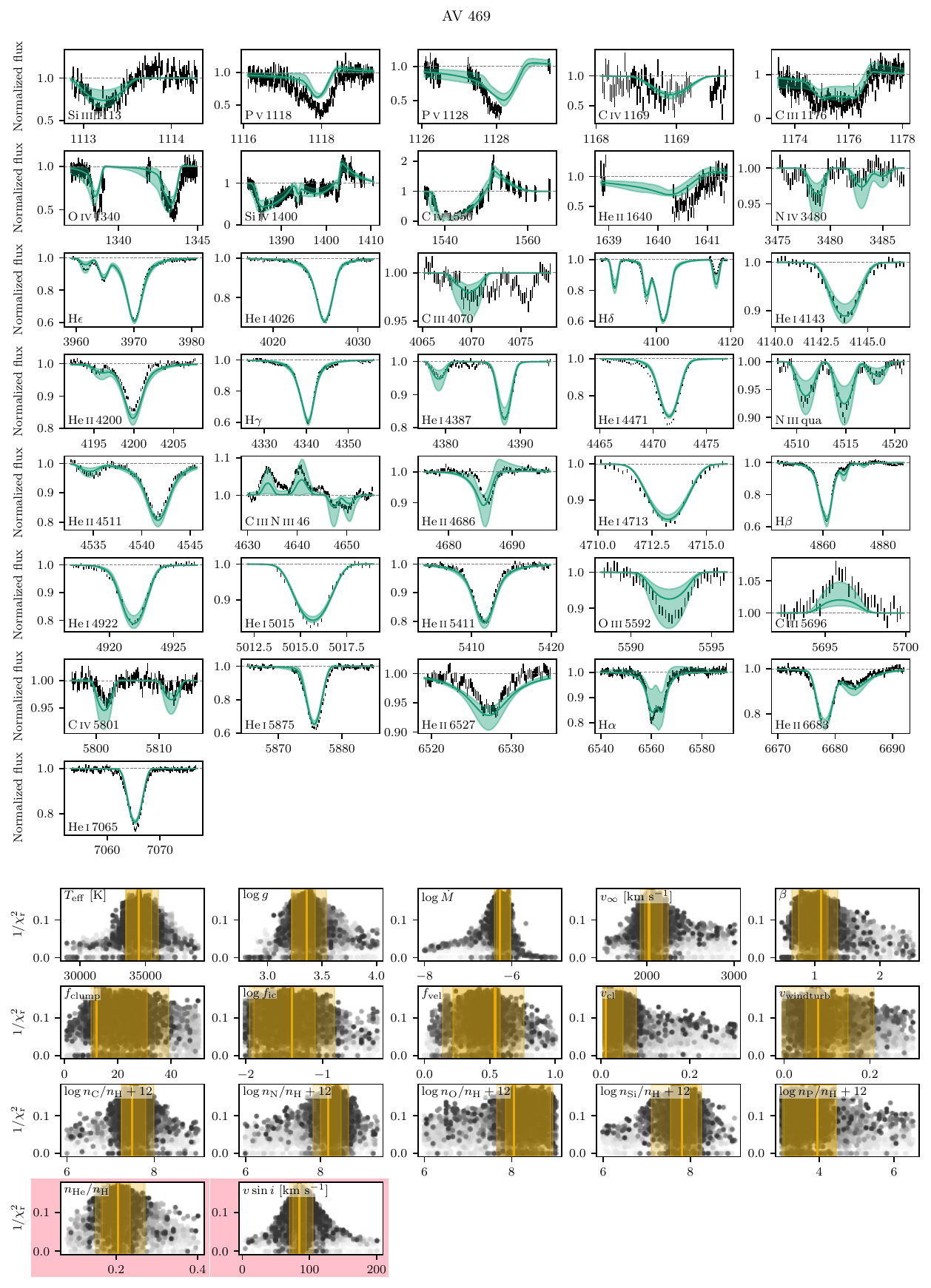}
    \caption{Same as Figure\,\ref{p2:fig:fit_summary}, but for AV\,469.}
\end{figure*}

\begin{figure*}
    \centering
    \includegraphics[width=0.95\textwidth]{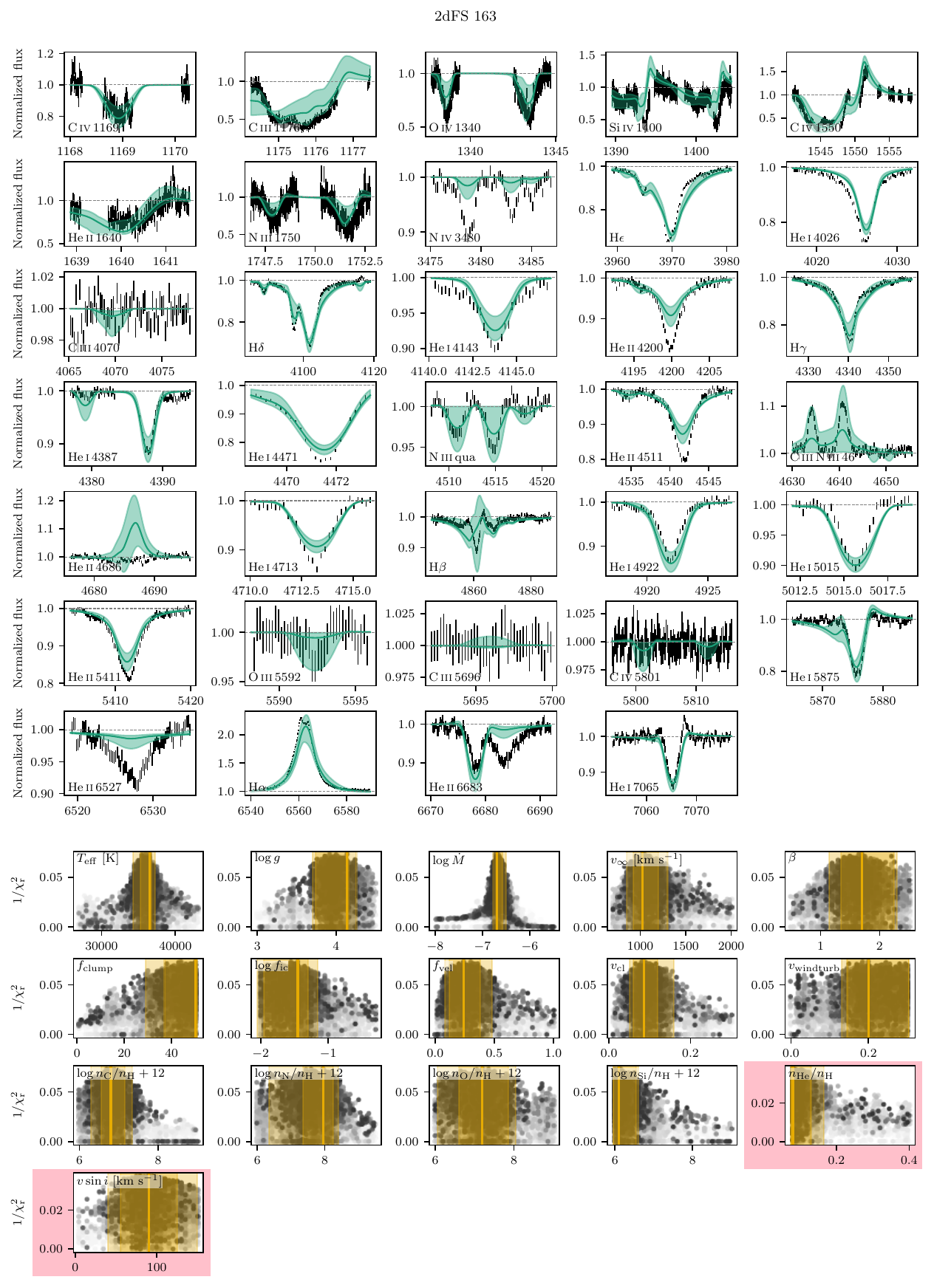}
    \caption{Same as Figure\,\ref{p2:fig:fit_summary}, but for 2dFS\,163.}
\end{figure*}

\begin{figure*}
    \centering
    \includegraphics[width=0.95\textwidth]{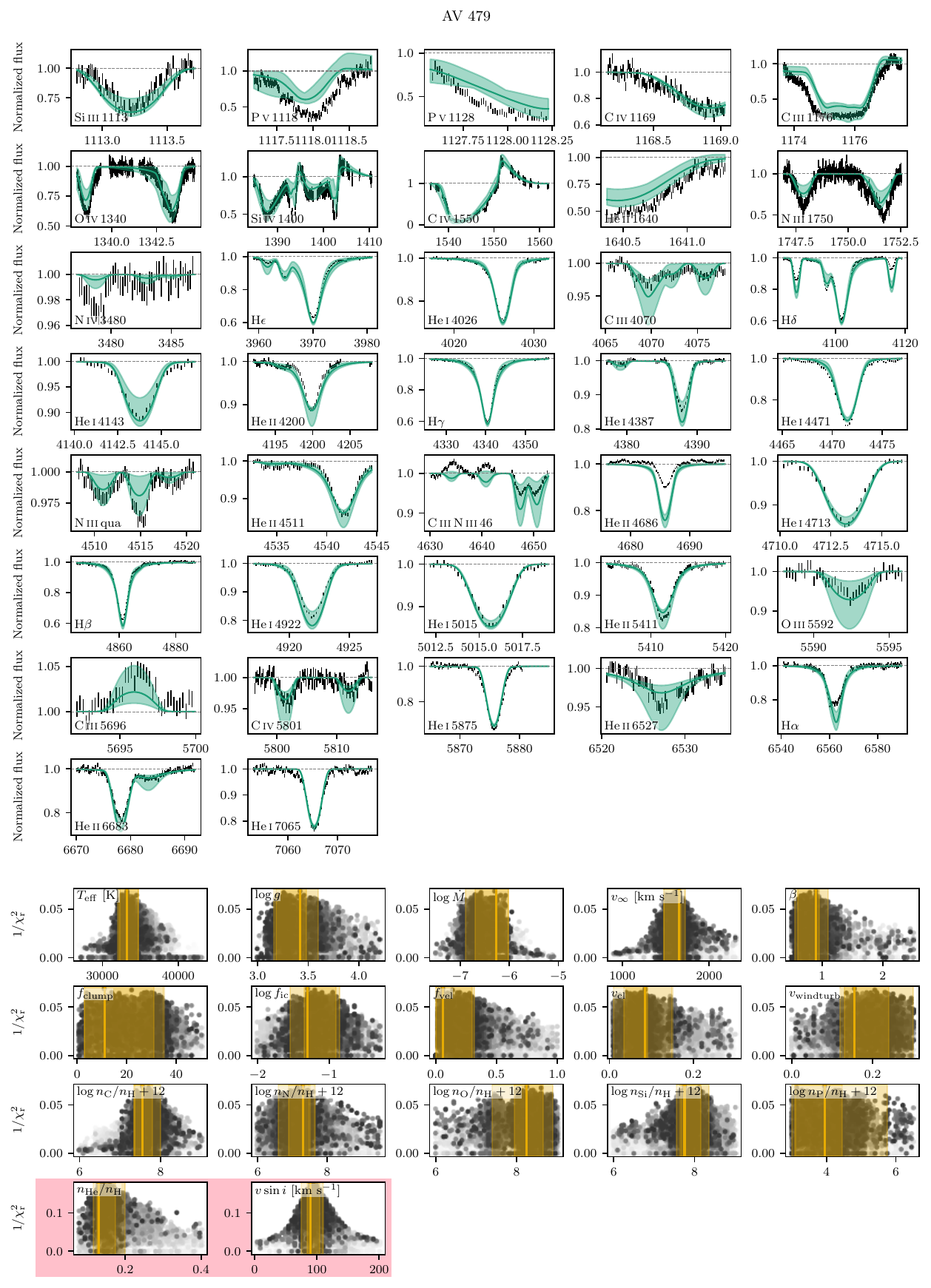}
    \caption{Same as Figure\,\ref{p2:fig:fit_summary}, but for AV\,479.}
\end{figure*}

\begin{figure*}
    \centering
    \includegraphics[width=0.95\textwidth]{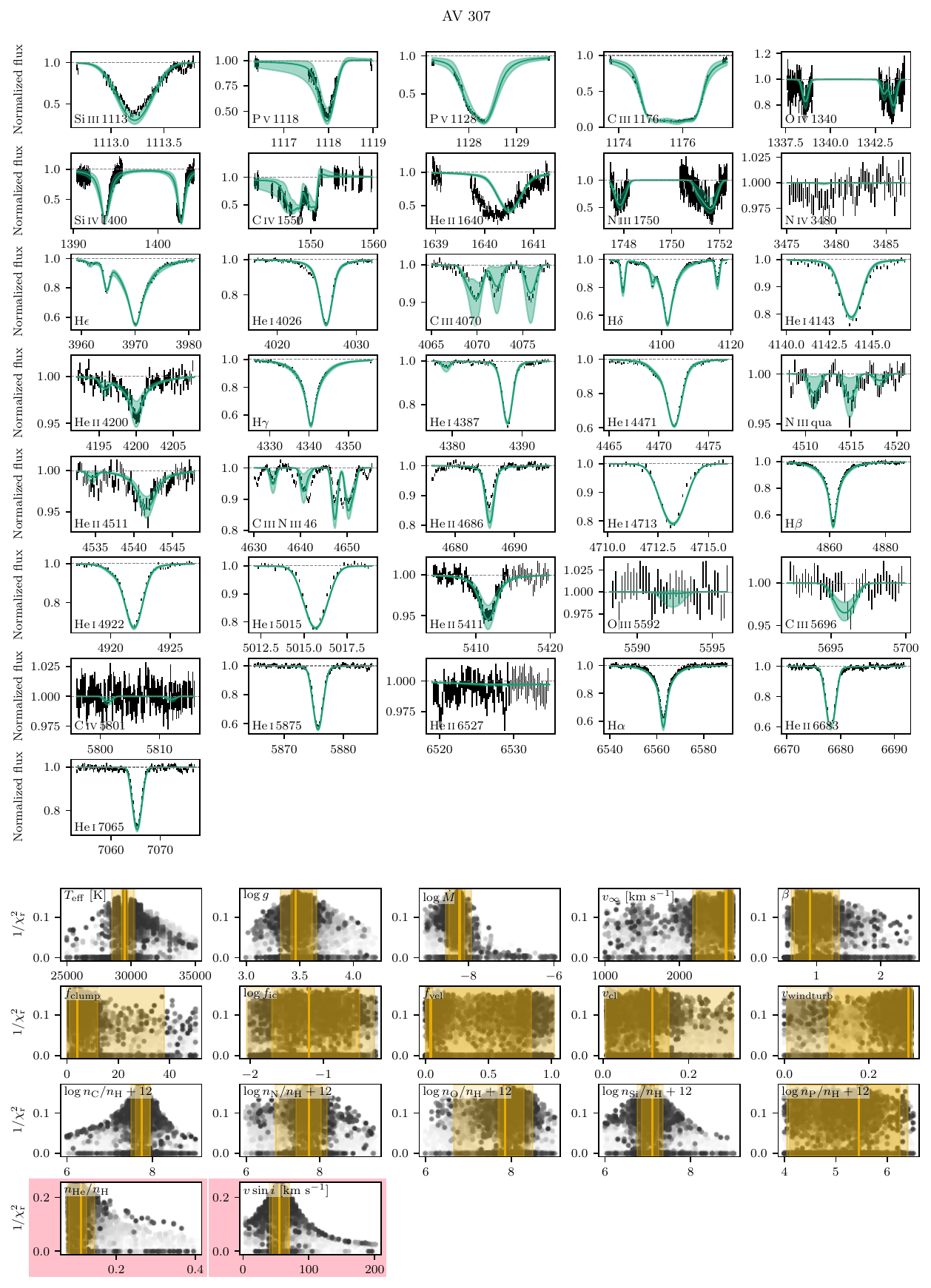}
    \caption{Same as Figure\,\ref{p2:fig:fit_summary}, but for AV\,307.}
\end{figure*}

\begin{figure*}
    \centering
    \includegraphics[width=0.95\textwidth]{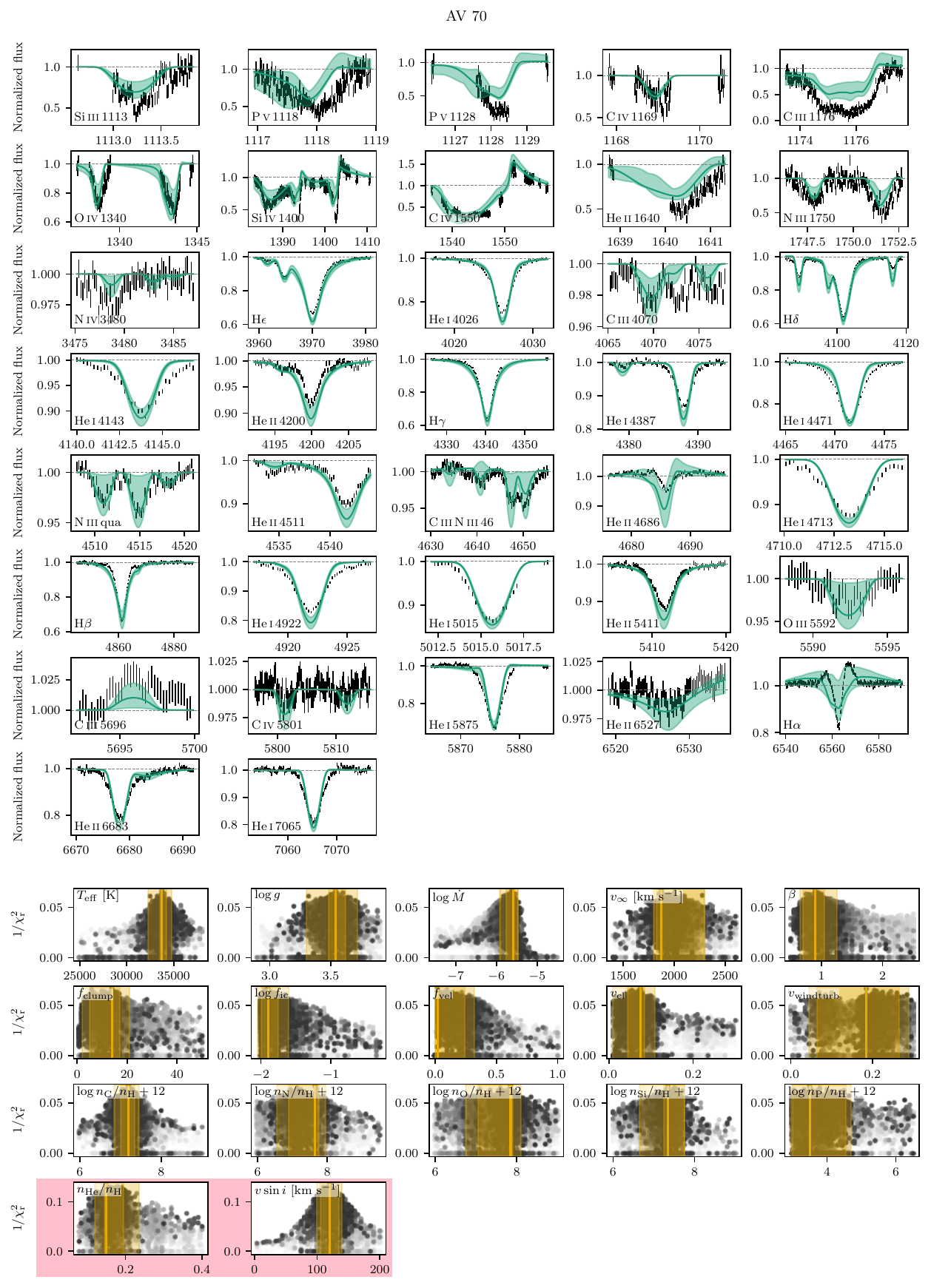}
    \caption{Same as Figure\,\ref{p2:fig:fit_summary}, but for AV\,70.}
\end{figure*}

\begin{figure*}
    \centering
    \includegraphics[width=0.95\textwidth]{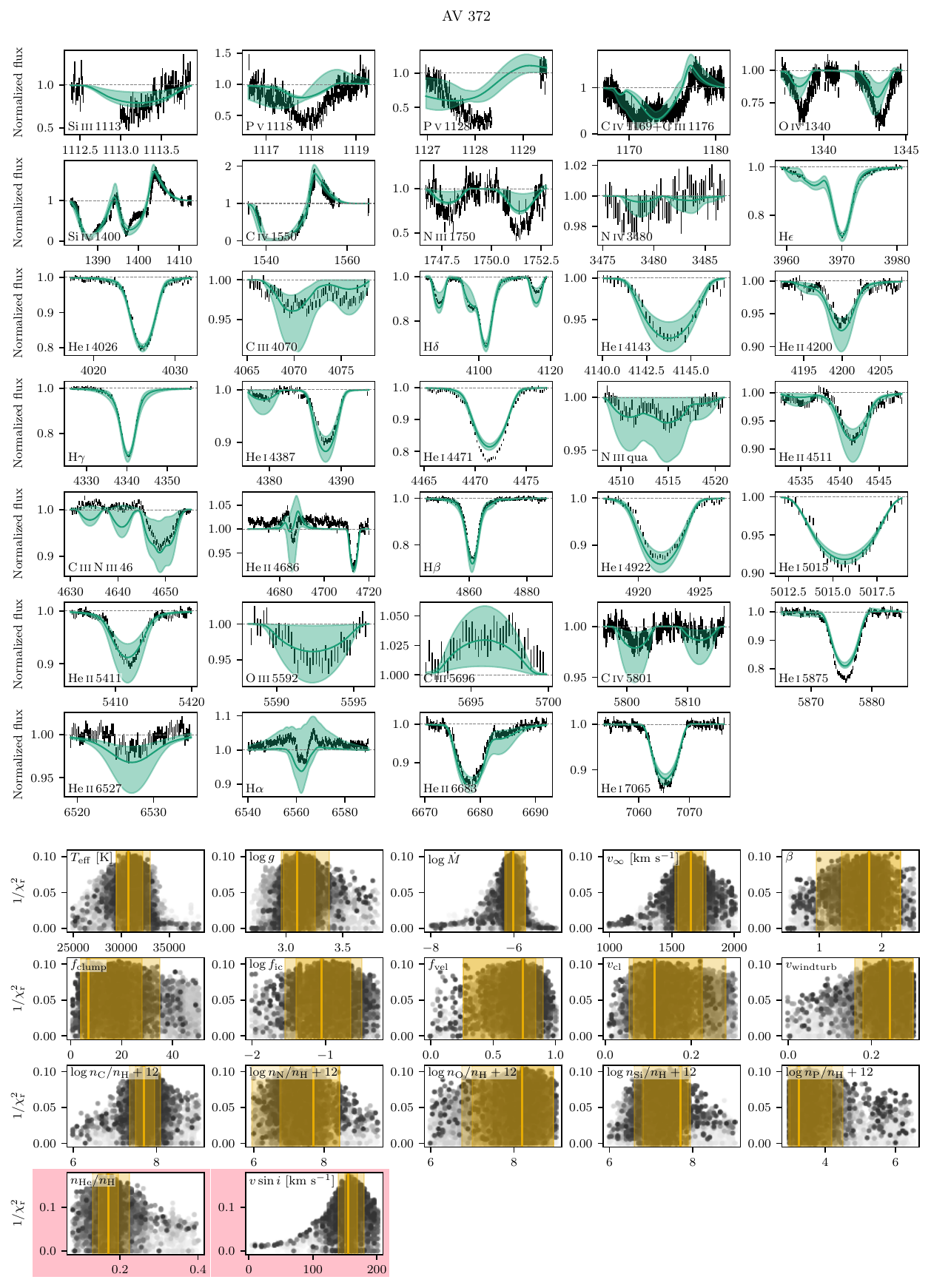}
    \caption{Same as Figure\,\ref{p2:fig:fit_summary}, but for AV\,372.}
\end{figure*}

\begin{figure*}
    \centering
    \includegraphics[width=0.95\textwidth]{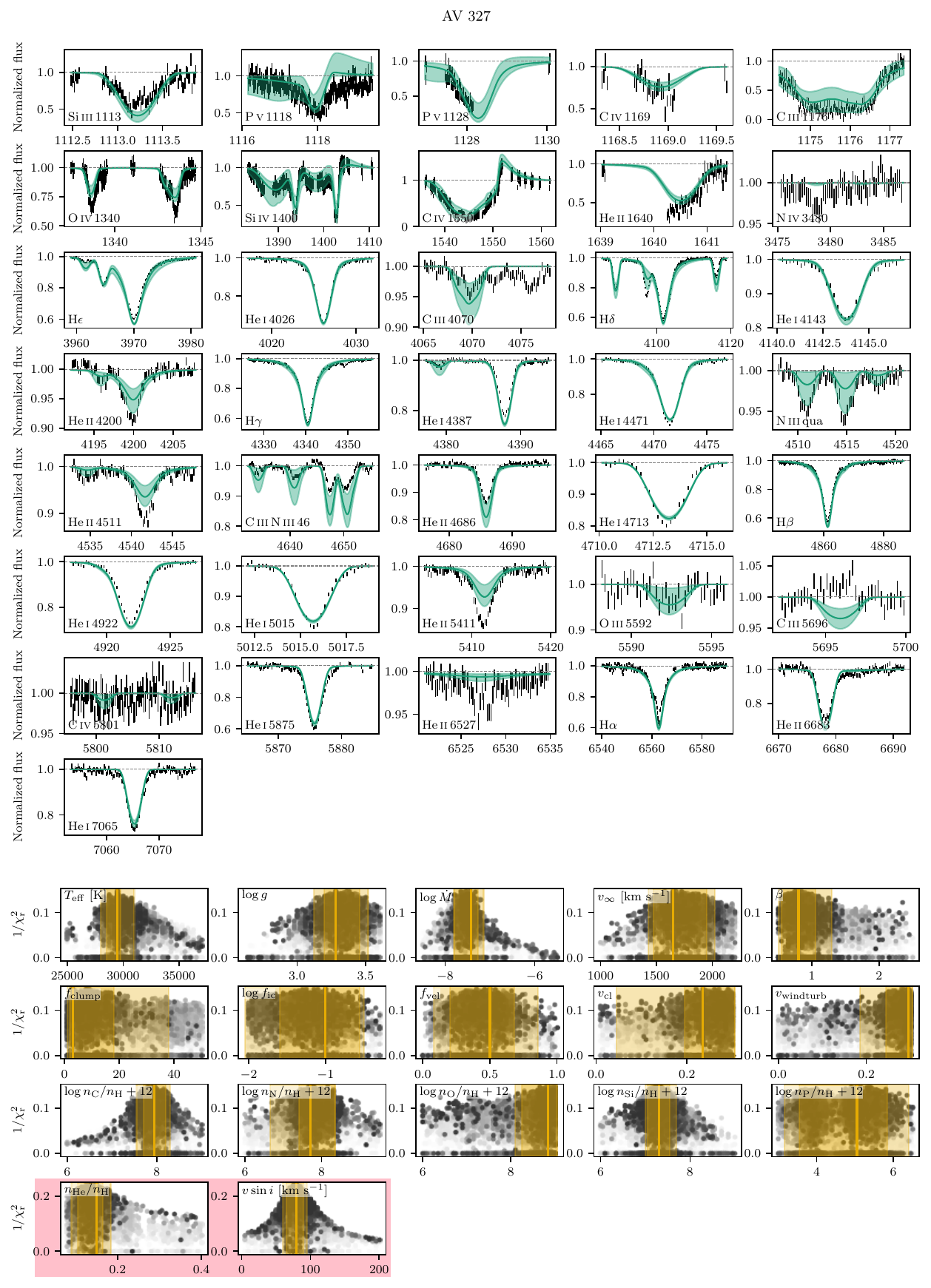}
    \caption{Same as Figure\,\ref{p2:fig:fit_summary}, but for AV\,327.}
\end{figure*}

\end{document}